\documentclass[12pt]{article}

\usepackage{scicite}

\usepackage{times}

\usepackage{amssymb,amsmath,epsfig,makecell,listings, amsthm, mathtools, paralist, enumitem, mathrsfs,gensymb,url}
\usepackage{lineno}
\usepackage{booktabs}
\usepackage{todonotes}
\newcounter{todocounter}

\presetkeys{todonotes}{inline, color=yellow!30}{}
\usepackage{nameref}
\usepackage[font=small, labelfont=bf]{caption}
\usepackage[font=footnotesize, labelfont=bf]{subcaption}

\newcommand{\Z}{\ensuremath{\mathbb{Z}}}
\DeclareMathOperator{\Ima}{im}

\usepackage{nameref,hyperref}
\usepackage{caption}
\usepackage{subcaption}
\usepackage{setspace}
 \usepackage{booktabs,colortbl}
 \definecolor{LRed}{rgb}{.8,.8,1}
\definecolor{grey}{rgb}{0.9,0.9,0.9}

\newenvironment{sciabstract}{%
\begin{quote} \bf}
{\end{quote}}

\title{Multiscale Topology Characterises Dynamic Tumour Vascular Networks}

\author{Bernadette J. Stolz$^{1\ast}$, Jakob Kaeppler$^2$, Bostjan
	Markelc$^{2,3}$, Franziska Braun$^4$,\\
	Florian Lipsmeier$^5$, Ruth J. Muschel$^2$, Helen M. Byrne$^1$, Heather A. Harrington$^{1,6\ast}$\\
\\
\normalsize{$^{1}$Mathematical Institute, University of Oxford, Oxford, UK}\\
\normalsize{$^{2}$Oxford Institute for Radiation Oncology, University of Oxford, Oxford, UK}\\
\normalsize{$^{3}$Department of Experimental Oncology, Institute of Oncology Ljubljana,}\\
\normalsize{Ljubljana, Slovenia}\\
\normalsize{$^{4}$Data Science,  pRED Informatics, Pharma Research \& Early Development, }\\
\normalsize{ Roche Innovation Center Munich, Germany}\\
\normalsize{$^{5}$Digital Biomarkers, pRED Informatics, Pharma Research \& Early Development,}\\
\normalsize{ Roche Innovation Center Basel, Switzerland}\\
\normalsize{$^{6}$Wellcome Centre for Human Genetics, University of Oxford, Oxford, UK}\\
\\
\normalsize{$^\ast$To whom correspondence should be addressed; E-mail:  stolz@maths.ox.ac.uk or} \\ 
\normalsize{harrington@maths.ox.ac.uk.}
}

\date{}

\begin{document} 


\baselineskip24pt


\maketitle

\begin{sciabstract}
Advances in imaging techniques enable high resolution 3D visualisation of vascular networks over time and reveal abnormal structural features such as twists and loops, and their quantification is an active area of research. Here we showcase how topological data analysis (TDA), the mathematical field that studies `shape' of data, can characterise the geometric, spatial and temporal organisation of vascular networks. We propose two topological lenses to study vasculature,
which capture inherent multi-scale features and vessel connectivity, and surpass the single scale analysis of existing methods. We analyse images collected using intravital and ultramicroscopy modalities and quantify spatio-temporal variation of twists, loops, and avascular regions (voids) in 3D vascular networks.
This topological approach validates and quantifies known qualitative trends such as dynamic changes in tortuosity and loops in response to antibodies that modulate vessel sprouting; furthermore, it quantifies the effect of radiotherapy on vessel architecture. 
\end{sciabstract}

\section*{Introduction}

The advent of high resolution imaging techniques has driven the development of reconstruction algorithms, which generate exquisitely detailed 3D renderings of biological tissues, such as tumour vascular networks~\cite{Bates2015,Bates2019}. Tumour vasculature is highly dysfunctional as vessels tend to be very leaky,
the direction of blood flow can change over time, and the structure of the vessel network looks markedly different than normal vessels ~\cite{Goel2011}.
Visualisation of tumour vasculature in 3D and over time offers a detailed picture of abnormal structural changes such as twists and loops~\cite{Nagy2009,Goel2011,Pittet2011,ritsma2013,Dodt2007,Dobosz2014}. The quantification of the 3D architecture is important because vessel structure affects vessel function (i.e., delivery of oxygen, nutrients, and therapies). 
Existing analyses have quantified structural features, including vessel density, number of vessels and branching points~\cite{Vilanova2017}, fractal dimension~\cite{Gazit1997}, and lacunarity~\cite{Gould2011}, and highlighted their relevance for predicting disease progression~\cite{Ehling2014,Bullitt2005} and response to treatment~\cite{Dobosz2014}. Such spatially averaged summaries lose information from detailed 3D renderings and do not account for vessel connectivity, or higher order topological features such as loops and voids; 
the latter correspond to avascular tumour regions
characterised by hypoxia and necrosis, and associated with reduced patient survival and poor responses to therapy~\cite{Goel2011}. 
Therefore, more detailed, automated and reproducible methods for quantifying vessel networks are needed, which may provide future benefit to basic research, clinical assessments, treatment planning, and monitoring. 

For large studies~\cite{Hoffmann2019}, machine learning algorithms are excellent at quantifying 3D microscopy features
(e.g., images obtained from adipose tissue~\cite{geng20213}). 
Here we use existing image processing methods, based on machine learning, to reconstruct 3D vascular networks from high resolution spatial data. 
We then use these 3D segmented networks to
quantify, compare and interpret the spatial organisation of tumour vasculature and responses to treatment. 
The novelty of our approach lies in the deep quantification of the vascular networks, and not the collection and segmentation of the experimental data.
In more detail, we present a topological framework that quantifies different notions of connectivity in 
reconstructed 3D vascular networks (e.g., for the first time quantifying loops and voids), and complements, extends and surpasses existing descriptors (see Figs.~\ref{Fig:MC38Cor},~\ref{Fig:RocheCor}) by providing a multi-scale summary of these topological features.

Mathematically, one can describe tumour vasculature as a spatial network, i.e. nodes embedded in three-dimensional space, connected by edges that represent blood vessel segments. 
An emerging mathematical field that uses topological and geometric approaches to quantify the ``shape" of data is \emph{topological data analysis (TDA)} \cite{Edelsbrunner2002,Carlsson2009}. A central method in TDA is \emph{persistent homology} (PH)~\cite{Edelsbrunner2002, Edelsbrunner2008, Carlsson2009,Edelsbrunner2010,Otter2017}. PH computes features called topological invariants of the data at different spatial scales; features that persist over a wide range of spatial scales are generally considered better to represent robust topological signals in the data. TDA has been successful in neuroscience, specifically analysing functional brain network data (for a small selection of examples see \cite{stolz2017persistent,stolz2018topological,sizemore2018,Giusti2015,
lord2016insights,reimann2017cliques}). Improved computations in PH~\cite{Otter2017} have increased the scope of its applications to include structural and spatial data, such as brain arteries~\cite{Bendich2014},  neurons~\cite{Kanari2018}, airways~\cite{Belchi2018}, stenosis~\cite{nicponski2019}, zebrafish patterns~\cite{mcguirl2020}, contagion dynamics~\cite{taylor2015}, and spatial networks~\cite{Feng2020,Stolz2019}.
More recently, interest has emerged for applications of TDA to patient-specific data in oncology \cite{bukkuri2021} and
TDA has been successfully applied to classify synthetic data from mathematical models of angiogenesis, the process in which tumour blood vessels form from existing ones~\cite{Nardini2021}.
The characteristics of tumour vascular networks that we study here using PH features are tortuosity~\cite{Nagy2009} (or `bendiness'), loops~\cite{Nagy2009}, and size of avascular regions~\cite{Goel2011}.

Tortuosity has been quantified previously using standard measures in tumour vessels \cite{Kannan2018,jain2013normalizing}, and using TDA in ageing vasculature \cite{Bendich2014}. Here, we propose a normalised topological tortuosity descriptor. To our knowledge, this is the first time loops and voids (which may correspond to avascular tumour regions) have been quantified in vasculature.
The leap from a single 2D slice to 3D reconstruction presents opportunities for quantifying 3D connectivity that would otherwise be impossible. As we describe later, these topological approaches are inherently multi-scale and quantify global connectivity of the data that surpasses standard descriptors. 
Such quantification could serve as a biomarker for charateristics of vascular networks and their response to vascular targeting treatments.

\begin{figure}[ht!]
 \centering
\includegraphics[width=\textwidth]{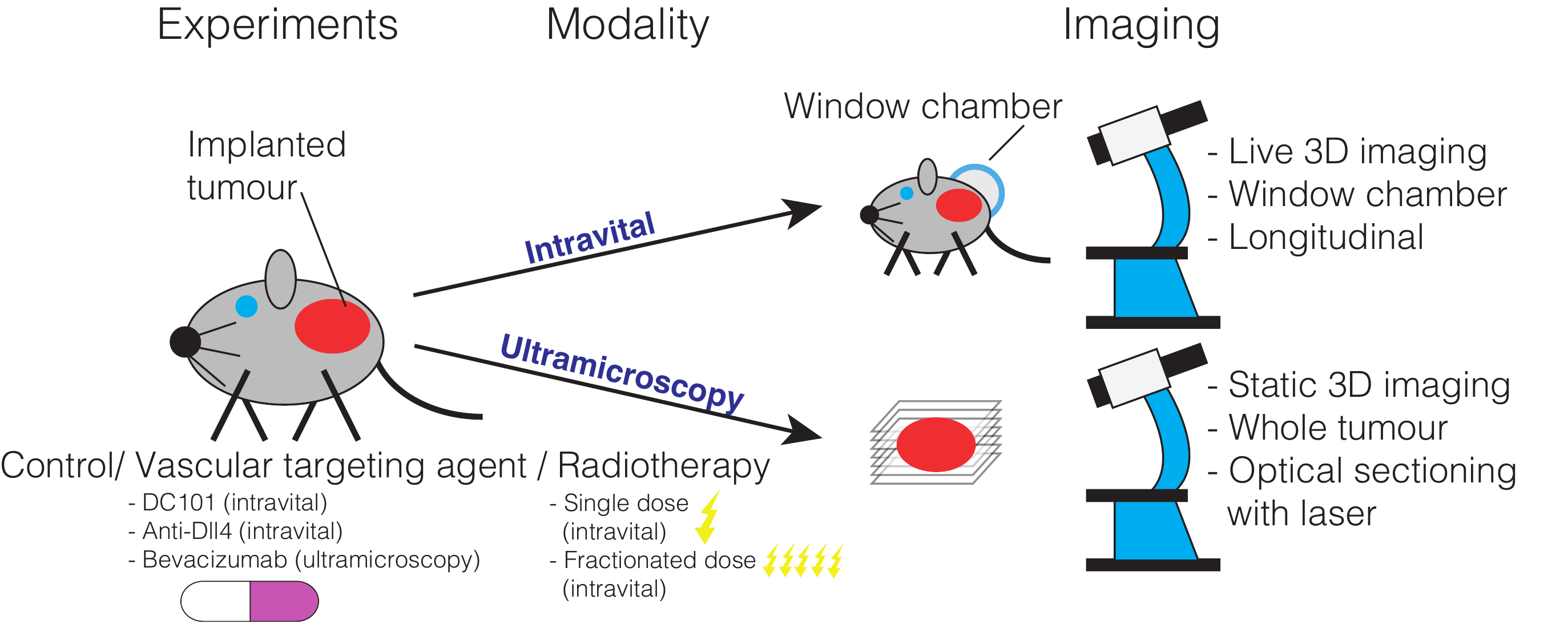}
\caption{{\bf Description of datasets.} We illustrate the treatments and imaging techniques used to generate the experimental data that we analyse. Both datasets consist of 3D stacks of tumour vasculature images from mice undergoing different treatments (vascular targeting agents and radiotherapy). Intravital data were collected from live animals observed over several days. Ultramicroscopy data~\cite{Dobosz2014} were obtained from multiple tumours excised at different times after treatment (one time point per tumour). These data are not directly comparable since they were generated from two different mouse models (see Table \ref{tab:data}), using different experimental setups (see Methods).} \label{Fig:Data}
\end{figure}

\begin{table}
\begin{tabular}{ p{2.4cm} p{2cm} p{1.7cm}  p{8.2cm}}
\toprule
Dataset & Type & Model & Experimental conditions \\
\midrule
\midrule
  Multiphoton {\it intravital} 3D microscopy & Dynamic (over multiple days) & Mouse colorectal cancer in mice & 1. Control ($n = 7$) \newline 2. DC101 (decrease sprouting, $n = 5$) \newline 3. Anti-Dll4 (increase sprouting, $n = 3$) \newline 4. Irradiated (single dose 15 Gy, $n = 5$)\newline 5. Irradiated (fractionated dose $5 \times 3$ Gy, $n = 4$ )  \\
 \midrule
  Multispectral fluorescence {\it ultramicroscopy} & Static & Human breast cancer in mice & 1. Control ($n = 18$) \newline 2. Bevacizumab ($n = 13$) \\
\bottomrule
\end{tabular}
\caption{\label{tab:data} {\bf Summary of datasets.} Summary of datasets analysed in this study including the number of mice $n$ (see also Figure 1). For information on the size of the images and extracted networks, see Table S2 in Supplementary Information.}
\end{table}

\clearpage

We showcase our topological approach by analysing three-dimensional vascular networks reconstructed from microscopy images from two different studies: intravital data and ultramicrospy data~\cite{Dobosz2014} (see Fig.~\ref{Fig:Data} and corresponding table).
In the intravital dataset, the same vascular networks are observed over time, providing a time course. The low penetration depth of intravital imaging means that only part of the vasculature can be imaged. 
The intravital dataset contains control (untreated) tumours and tumours subjected to either vascular targeting agents or radiation therapy. The agents consist of 
antibodies DC101~\cite{Kannan2018} and anti-Dll4~\cite{Liu2011}, 
which decrease and increase vessel sprouting, respectively~\cite{Ehling2014,Folkman1971,Hanahan2011,Vilanova2017}.  
The irradiated tumours receive either a single dose ($1 \times 15$ Gy) or fractionated doses ($5 \times 3$ Gy) of radiation therapy. 
Even though radiation therapy is commonly used to treat tumours, observations of structural changes in the vasculature have remained inconsistent~\cite{Park2012}. 
The second dataset, ultramicroscopy ~\cite{Dobosz2014}, gives three-dimensional reconstructions of the entire tumour vasculature. The dataset includes multiple time points (snapshot data), where we obtain one time point per tumour. The data include control tumours and tumours treated with bevacizumab~\cite{Ferrara2004}, a drug that inhibits angiogenesis and is thought to (transiently) normalise~\cite{Goel2011} tumour vasculature, i.e. reduce structural and functional abnormalities. See Section~\nameref{Sec:DataPreProcessing} in Methods for details on network binarization, skeletonisation, pruning, and testing to reconstruct the networks we analyse \cite{unetRuss,Bates2015,
Bates2017Thesis, Bates2019, Dobosz2014}.

\paragraph*{Standard measures and existing descriptors.} 
In order to describe the architecture of abnormal tumor vasculature, several morphological characteristics have been used. The most common one, microvascular density (MVD), is often used to compare 2D tumor sections. A high MVD has been shown to independently predict death from several types of cancer \cite{weidner1995intratumor}. Other descriptors often used in the literature include vessel volume, number of branching points, vessel diameter, vessel length, and vessel tortuosity \cite{jain2013normalizing}. However, none of these features can recapitulate the complexity of the entire vascular network. With the advent of personalized medicine, different imaging modalities such as magnetic resonance imaging (MRI), dynamic contrast enhanced – MRI (DCE –MR) and computed tomography are often used to aid with patient diagnosis and treatment presonaliisation. A common feature to these approaches is that the resulting images are 3D volumes \cite{Kannan2018}. As such, the morphological descriptors should capture the complexity of the 3D vascular network, not just the single-vessel-level characteristics. 

Existing analyses of blood vessel networks have quantified structural features and shape, including vessel density, number of vessels (i.e. number of edges) and branching points (i.e. number of nodes)~\cite{Vilanova2017}. 
To highlight the additional insight generated by our TDA descriptors, we calculate existing descriptors (at each time point); specifically the number of branching points, mean vessel diameter, mean vessel length, and length to diameter ratio for both the intravital data and ultramicroscopy. For the intravital data, we report two existing tortuosity measures (for details, see Methods Fig.~\ref{Fig:Tortuosity}b,c). The first tortuosity descriptor is the chord-length-ratio (clr)~\cite{Bates2017Thesis,
Bates2019}, which is the ratio of the distance between the branching or end points of the vessel and the path length of the vessel, where a value of one corresponds to a straight vessel and zero is tortuous. The second tortuosity descriptor is the sum-of-angles-metric (SOAM)  \cite{Bullitt2005}, which is computed by summing the angles of regularly sampled tangents along a blood vessel skeleton, where a value of zero corresponds to a straight vessel and tortuous vessels correspond to larger values.  For the ultramicroscopy data, we report the number of vessel segments (both as computed by \cite{Dobosz2014} and \cite{unetRuss,
Bates2019}), number of branching points (both as computed by \cite{Dobosz2014} and \cite{unetRuss,
Bates2019}),   necrotic tumour volume as computed by \cite{Dobosz2014}, tumour volume as computed by \cite{Dobosz2014}, and vital tumour volume as computed by \cite{Dobosz2014}.
All existing descriptors are normalised by day 0 of observation/treatment, and computed from the freely available {\sc python} code package {\sc unet-core}~\cite{unetRuss}. Note that to compare with existing descriptors, we compress our TDA descriptors and therefore lose information.

\paragraph*{Topological Data Analysis.} 
\begin{figure}[ht!]
 \centering
\includegraphics[width=\textwidth]{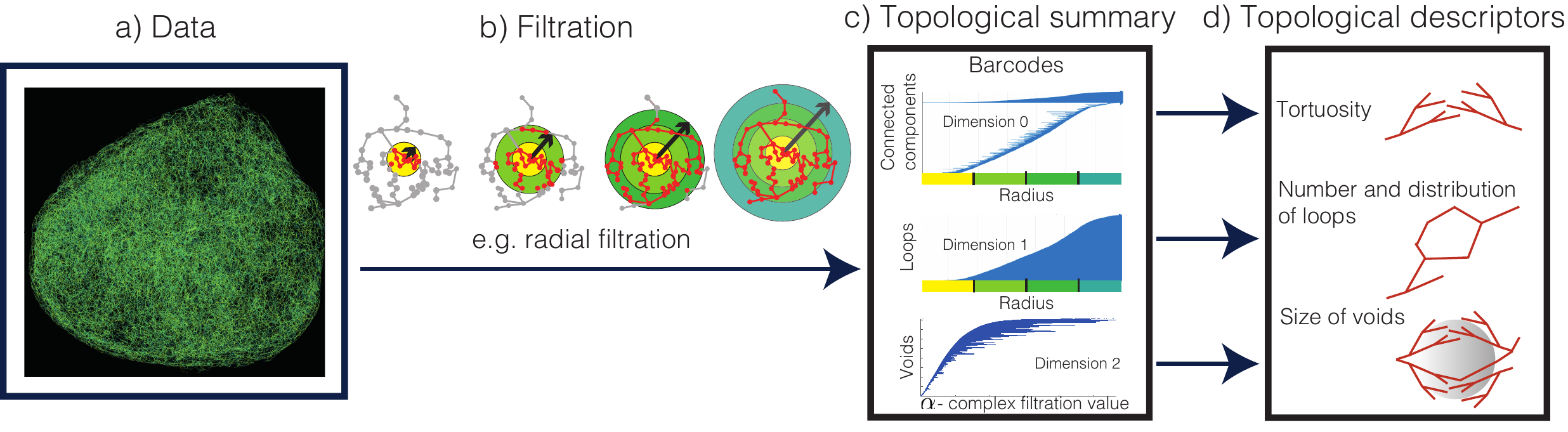}
\caption{{\bf Schematic illustration of topological data analysis (TDA) for vascular network data.} {\bf a)} We reconstruct the three-dimensional vascular network from image stacks. 
{\bf b)} We apply the radial filtration and the $\alpha$-complex filtration. {\bf c)} We compute the topological summary of the data, 
which consists of a collection of barcodes~\cite{Ghrist2008}. The horizontal axis of a barcode represents a spatial parameter such as radial distance to the tumour centre (radial filtration) or the scale at which we view the data ($\alpha$-complex filtration). Every line in a barcode corresponds to a topological feature, i.e. a connected component, loop, or void, in the data. In the radial filtration, we analyse the network within the sphere (highlighted in red) and compute connected components and loops as the sphere grows from the tumour centre outwards. 
In the barcodes the bars start at the radius (measured from the tumour centre) where the corresponding connected component or loop first enters the sphere. For a connected component, its corresponding bar ends at the radius at which it merges with another component, i.e. it connects to another part of the vascular network within the growing sphere. A bar representing a loop finishes at the final radius of the filtration.
For voids, we study the data at different scales using the $\alpha$-complex filtration (see Methods Section~\nameref{Sec:Methods}) and the range of a bar represents the scale values where the void is detectable. Its length is a proxy for the volume of the void. {\bf d)} We extract interpretable topological descriptors of the data from barcodes.
}\label{Fig:TDAVessels}
\end{figure}

Here we present topological descriptors to quantify vascular network characteristics across different spatial scales, and over time. We first explore appropriate multi-scale lenses of the data, called filtrations, which feed into persistent homology computations. 

We propose two filtrations for tumour vascular networks: the radial filtration quantifies topological features with respect to distance from the tumour centre; the $\alpha$-complex~\cite{edelsbrunner1993union} filtration (see Methods Section~\nameref{Sec:Methods}) quantifies avascular tumour regions that are devoid of blood vessels. Recall the data is embedded in 3D space. The network nodes are branching points (i.e. points where vessels branch) and vessel nodes (i.e. other points sampled along vessels).
In the radial filtration we determine the centre of mass of the 3D nodes and grow a sphere from the centre outwards in uniform steps. Inspired by a filtration to analyse neuronal tree morphologies radially from the root of a neuron \cite{Kanari2018}, the radial filtration differs from the well-known \v{C}ech filtration since we only consider one ball from the centre of the tumour (rather than many balls, for example, grown from points sampled on the network).
At each step we determine the nodes located inside the growing sphere and connect two nodes when there is a vessel between them, resulting in a growing network-- the radial filtration. We then compute the connected components and loops. The radial filtration depends on the choice of tumour centre whereas the $\alpha$-complex filtration does not.
For the $\alpha$-complex filtration~\cite{edelsbrunner1993union,Otter2017}, we construct a sequence of nested simplicial complexes (i.e., collections of nodes, edges, triangles, and tetrahedra) on the 3D nodes of the vessel network. 
Each edge, triangle, or tetrahedron can be assigned a filtration value $\alpha^2$, which can be thought of as a proxy for volume. The filtration value is increased to obtain the filtration on the data until the Delaunay triangulation~\cite{Delauney1934} of the 3D nodes of the vessel network is constructed. We then compute voids in this filtration.

PH computes topological features such as connected components (dimension 0), loops (dimension 1), and voids (dimension 2) and how they change across different scales. 
These multi-scale and multi-dimensional topological features are summarised in a barcode~\cite{Ghrist2008}  (see Fig.~\ref{Fig:TDAVessels}).  From these barcodes, we compute interpretable topological descriptors in the Results. These topological calculations extend the toolbox of existing descriptors by quantifying connectivity across spatial and temporal scales.

\paragraph*{Loss of information.}
Given the sheer size of these data (see Supplementary Information Section  \nameref{Sec:ComputData} for the variation in size), they cannot be processed, analysed and summarised without some loss of information. At processing stage, the reconstruction of 3D networks from 2D slices in the ultramicroscopy dataset \cite{Dobosz2014} had original image stacks taken with resolution of 5.1$\mu$m, thus limiting the loss of information in the z-direction while reconstructing the image stack. We used existing segmentation and skeletonisation algorithms \cite{Bates2015,
Bates2017Thesis,Bates2019} and existing code that computes skeletonisation and standard metrics \cite{unetRuss,
Bates2019} (see Methods for details on data, processing, and testing to minimise segmentation errors). Data analysis quantities depend on the segmentation and preprocessing of the data. For biological networks, noise contamination and its consequences on data analysis is an active area of research~\cite{Kavran2021}.
A strength of TDA is that its output has been proven to be robust to small amounts of noise in data, which are given by stability theorems \cite{Cohen-Steiner2005}. However, the TDA output will change if the resulting network changes significantly, as will existing morphological descriptors.  The TDA filtration step sizes are discrete and coarse to ensure computations are feasible for this dataset (e.g., 500 filtration steps for the radial filtration); therefore small features may be ``stepped over" and some fine information may be lost between filtration steps. As described later (see Results and Supplementary Information Section \nameref{Sec:ComputData}), the topological tortuosity measure depends on the short bars in the barcode. For the ultramicroscopy data we may have required more filtration steps, or finer data resolution in the x-y plane to compute topological tortuosity; however we were limited by computational resource and processing of experimental data. Furthermore, due to the size of the reconstructed ultramicroscopy networks (e.g., ranging from 12,500 to 118,000 branching points, see Supplementary Information Section \nameref{Sec:ComputData}), we had to subsample points from the network (see Methods Section~\nameref{Sec:DataPreProcessing} for details). 
All these factors may affect the computation of small connected components (see discussion of tortuosity descriptor in the Results and Supplementary Information). 

Topological summaries, such as barcodes, can be equipped with a metric that is stable with respect to small perturbations to the data \cite{Cohen-Steiner2005}; however, the space of barcodes is complex, with arbitrarily high curvature \cite{turner2014frechet}. This metric alone is not suitable for integration with machine learning, which has lead to the development of stable vectorisation methods \cite{bubenik2015statistical,Adams2015,carriere2015stable} to reduce loss of information. Combining TDA and machine learning is an active research area, ranging from segmentation \cite{wu2017optimal,hu2019topology} to statistical analyses of topological summaries (see a tiny selection here \cite{turner2020same,chazal2018density,
robinson2017hypothesis,
kusano2016persistence}). 

\paragraph*{Gain of information.}
The persistent homology algorithm used for the topological data analysis computations is underpinned by stability theorems
\cite{Cohen-Steiner2005}, which ensure that the computed topological features are stable with respect to small perturbations to the data. Moreover, the algorithm will output the same topological barcode (summarising the multiscale descriptors) even if it is re-run or computed multiple times on the same dataset; therefore, it is both accurate and reproducible. In contrast, manual counting is prone to human errors and is often limited to 2D slices, as done by Shayan et al \cite{Shayan}, limiting detection to features (e.g. vessel segments, branching points and loops) in the plane.
Rather than manual counting or standard descriptors, the mathematical framework that we employ (theory and algorithms) enables quantification of loops and voids in 3D. Furthermore, this topological quantification is automated, systematic and performed across spatial scales. To compare this multiscale topological data analysis to standard descriptors (which are single scale) requires its compression; TDA gives additional information that surpasses manual counting and existing descriptors. Therefore, topological descriptors offer a significant improvement on both manual counts and standard descriptors.

\section*{Results} 

\begin{figure}[ht!]
 \centering
\includegraphics[width=\textwidth]{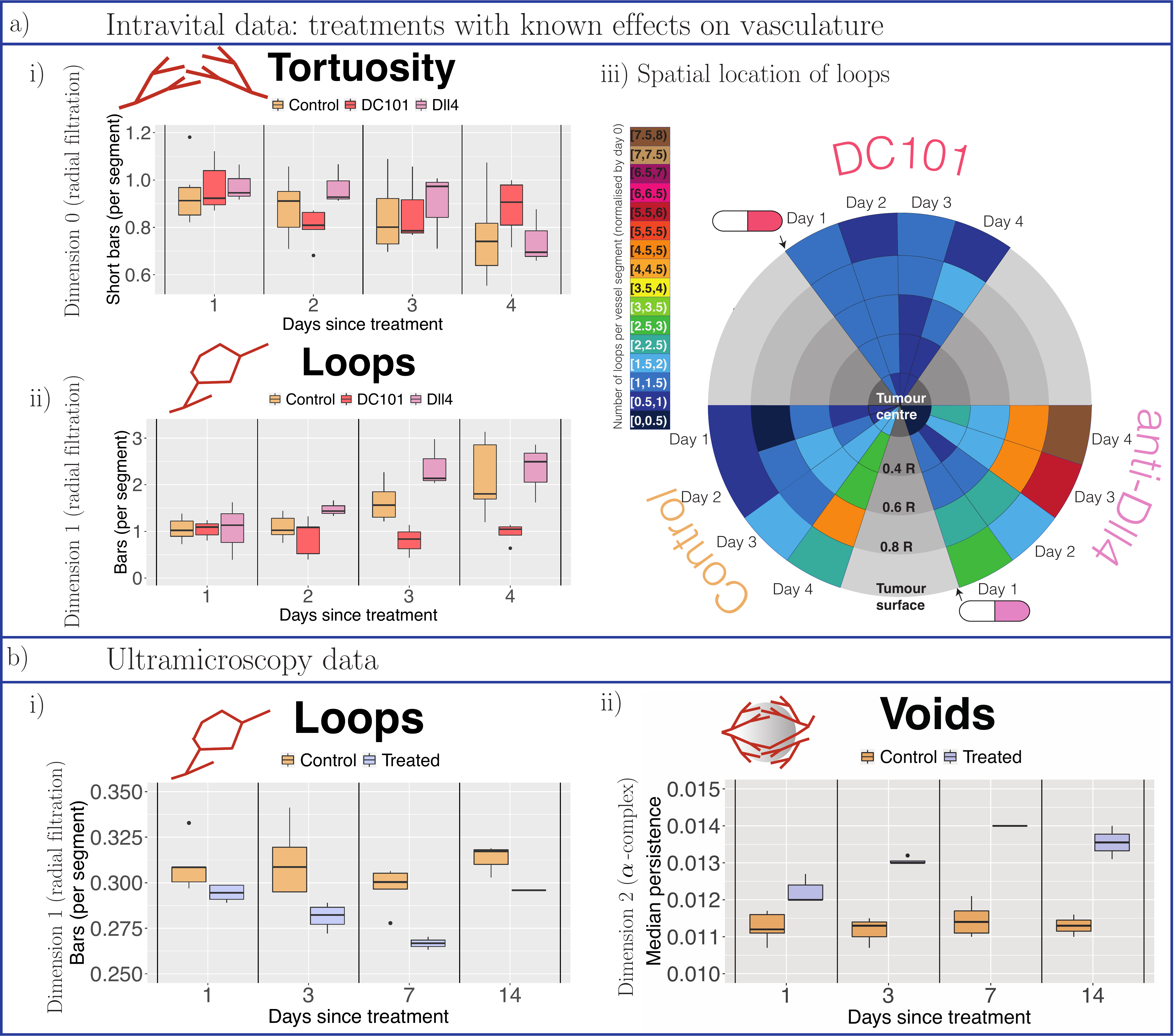}
\caption{{\bf Topological descriptors extracted from tumour blood vessel networks treated with vascular targeting agents with known effects.} {\bf a)} Intravital data results. We normalised all descriptors with respect to values on the day on which treatment is administered or, for controls, the day on which observations commence (day 0). 
Data was collected from controls (beige), and tumours treated with vascular targeting agent DC101~\cite{Kannan2018} (dark pink),
or vascular targeting agent anti-Dll4~\cite{Liu2011} (light pink). 
(i) Tortuosity was computed as the ratio of short bars in dimension 0 barcodes of the radial filtration ($\leq 10\%$ of maximal radius used) to the number of vessel segments. (ii) Loops are the number of bars in dimension 1 barcodes of the radial filtration per vessel segment. (iii) Spatio-temporal resolution of the number of loops per vessel segment. We illustrate the changes in the median number of loops (normalised by day 0) in radial intervals around the tumour centres over the days of observation. We point to the day following treatment with vascular targeting agents with a cartoon drug.
{\bf b)} Ultramicroscopy data results. Due to the snapshot nature of the data (one time point per tumour), all reported topological descriptors are raw values.
Data was collected from controls (beige) and tumours treated with bevacizumab (purple). (i) We computed the number of vessel loops per vessel segment. 
(ii) We determined the size of voids (avascular regions) by computing the median length of bars in the dimension 2 barcodes of the $\alpha$-complex filtration.
}\label{Fig:Results}
\end{figure}

\paragraph*{Topology gives descriptors of tortuosity, loops and voids.}
We developed interpretable, quantitative descriptors of tortuosity~\cite{Bendich2014} (``bendiness''), loops and  voids (see Figures \ref{Fig:TDAVessels}d and \ref{Fig:Results})
based on the calculated topological summaries of 3D tumour vasculature. The connected components (dimension 0) of the radial filtration characterise the tortuosity: a vessel with high tortuosity will intersect the growing sphere multiple times and generate many small components which quickly connect as the sphere radius increases and manifest in the barcode as multiple short bars. The topological tortuosity measure proposed for brain arteries \cite{Bendich2014} was based on analysing data with a simpler tree structure, whereas tumour vessel networks may have multiple components. To ensure the tortuosity descriptor did not mistakenly count separate vessels as a single tortuous vessel, (i.e., that the descriptor captured topological connectivity), we normalised the descriptor. Specifically the tortuosity descriptor proposed here was defined as the ratio of the number of short bars ($\leq 10\%$ of maximal radius used in the radial filtration) in dimension 0 barcodes to the number of vessel segments (see Methods and Fig.~\ref{Fig:Tortuosity}). 
The loops descriptor was computed from the number of bars in dimension 1 barcodes of the radial filtration and divided by the number of vessel segments. 
PH of the $\alpha$-filtration allowed us to identify voids, i.e. avascular tumour regions, and their volume in the vessel networks. Long bars in the corresponding barcodes (dimension 2) represent large voids, while short bars represent small voids. The void descriptor measures the median persistence value or bar length in the barcodes.
 We further used the radial filtration to determine how the number of loops per vessel segment changes over time in annuli at different distances from the tumour centre. 

\paragraph*{Validation of topological descriptors on two datasets.}
We validated the topological descriptors on data from studies in which tumours were treated with different agents with known effects on tumour vasculature: vascular targeting agents DC101 and anti-Dll4 in the intravital data and bevacizumab in the ultramicroscopy data (see Fig.~\ref{Fig:Results}). We found significant differences in our topological descriptors 
between control and treatment groups of both datasets despite that 
1) the biology in the two studies was different, e.g. treatments, tumour types, and mouse models (see Section~\nameref{seq:data} in Methods for description), which can influence the degree of tumour vascularisation and blood vessel structure~\cite{Konerding1999},
2) the imaging modalities are not straightforward to compare (intravital is timecourse data and can be normalised, has high spatial resolution in the $xy$-plane but low penetration depth, whereas ultramicroscopy is snapshot data at lower spatial resolution but across the whole tumour). While these technical differences led to discrepancies in computational feasibility and interpretation (see Sections \nameref{Sec:ComputData} and \nameref{Sec:Tortuosity}), we successfully completed computations and showed that our topological descriptors are interpretable for both datasets (see also Sections \nameref{Sec:ResStats} and \nameref{Sec:MoreResStats} for statistical analysis).

Our tortuosity descriptor and the number of loops per vessel segment succeeded in
capturing increased sprouting in the vascular networks induced by anti-Dll4 (see Fig.~\ref{Fig:Results}ai -- ii, \ref{Fig:ResultsPValues}a, \ref{Fig:KruskalMC38TortuosityDC101Dll4}, and \ref{Fig:KruskalMC38LoopsDC101Dll4}) and confirmed the transient phenomenon of vascular normalisation~\cite{Goel2011} induced by DC101 (see Fig.~\ref{Fig:Results}ai -- ii). Specifically, the tortuosity descriptor captured vascular normalisation 2 days after treatment in agreement with the literature~\cite{Goel2011} and our loop descriptor showed vessel normalisation 2 -- 4 days after treatment for loops, which has not been reported before. 

For the ultramicroscopy data, care is needed when interpreting the proposed tortuosity descriptor since these networks are less resolved in the $x-y$ plane, information loss may occur due to computational limitations of filtration step size, and the number of vessel segments reduces markedly following treatment with bevacizumab. These three factors lead to a counter intuitive increase in tortuosity after treatment since small vessels are ``stepped over" by the radial filtration without a finer spatial resolution (either in data or computation). Visual inspection (see Fig.~\ref{fig:ExampleImagesRoche}) does not show tortuous vessels. If we consider the non-normalised tortuosity descriptor, we observe a decrease in tortuosity compared to control (see Figure~\ref{Fig:RocheTortuosityDiscussion}).

\begin{figure}[ht!]
\centering \includegraphics[width=.95\textwidth]{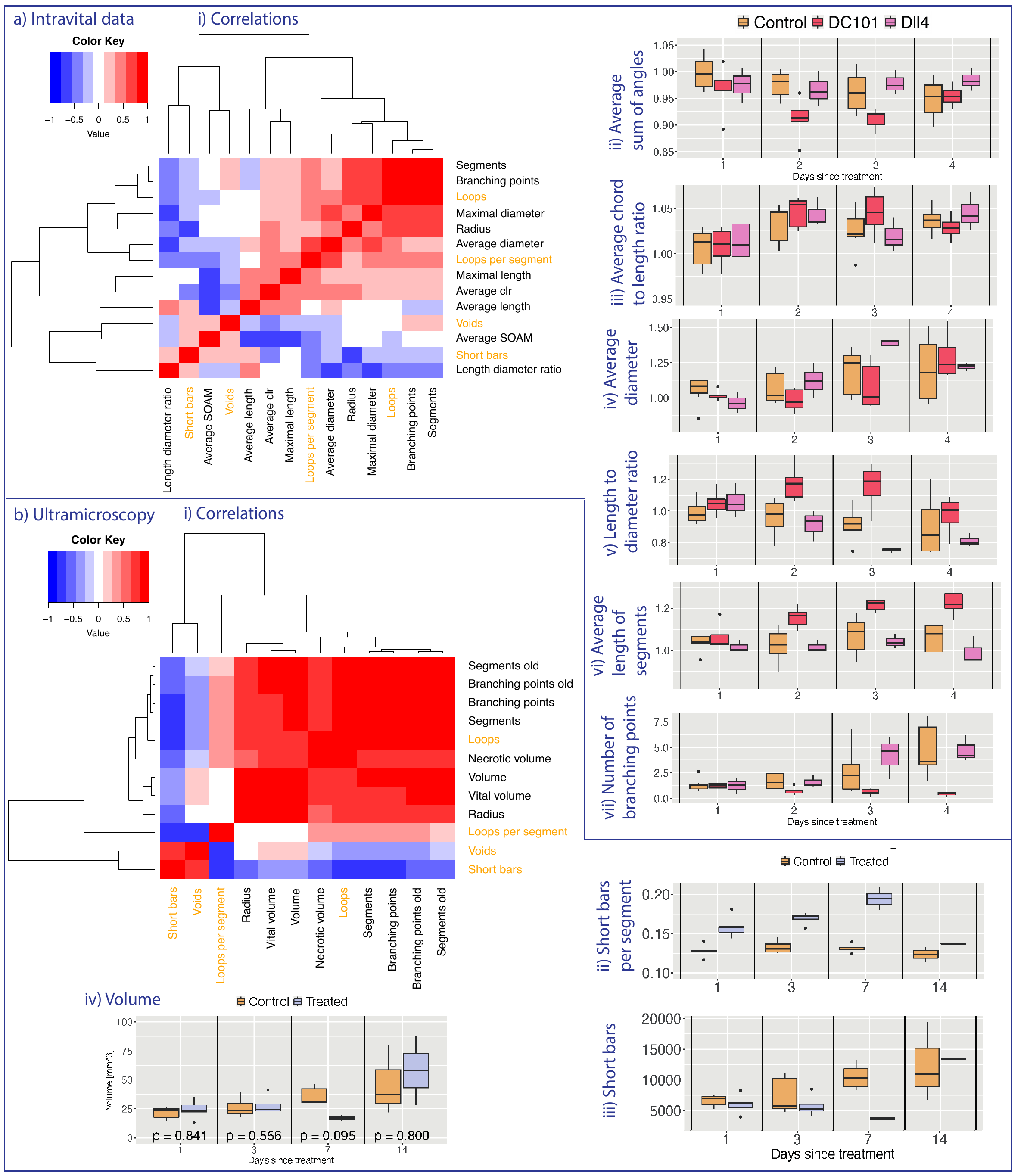}
\caption{
{\bf Heatmap displaying the pairwise Pearson correlation coefficients between different vascular descriptors (standard and topological) derived from the (a) intravital data and (b) ultramicroscopy. } (ai,bi) The dendrogramme represents complete linkage clustering using the Euclidean distance measure. We compute standard vascular descriptors (see text).
We highlight the topological measures in orange including both the number of loops and number of loops per vessel segment to highlight the effect of the normalisation. (aii-aiii) Existing tortuosity descriptors. (aiv-avii) Standard measures. (bii-biii) While the proposed topological tortuosity descriptor is a good measure for intravital data (see Figure 3), care must be taken with the ultramicroscopy data (see text for details). }\label{Fig:CorrCombined}
\end{figure}

\paragraph*{Topological loop and void descriptors surpass standard measures.} 
Throughout our analysis, we computed the topological descriptors indexed by a filtration value, which corresponds to tracking the evolution of connectivity at different spatial scales. Therefore, standard (spatially averaged) morphology descriptors are not directly comparable with topological (spatially resolved) descriptors. To perform a comparison, required us to compute the topological descriptor for the entire network, losing spatial information encoded in the barcode (see Fig.~\ref{Fig:Results}).  We report a comparison between standard and topological descriptors and their correlations in Fig.~\ref{Fig:CorrCombined}.
Topological descriptors provided complementary information to standard statistical measures and surpassed them by providing multiscale information of spatial location of tortuosity as well as connectivity information captured with the loop descriptor (see correlations in Fig.~\ref{Fig:CorrCombined}ai and Fig.~\ref{Fig:MC38Cor}).  
The tortuosity of the intravital dataset appeared qualitatively consistent with the conventional tortuosity measure mean sum of angles metric (SOAM) across the network (see Fig.~\ref{Fig:CorrCombined}aii and Fig.~\ref{Fig:KruskalMC38SOAMDC101Dll4}). Our results suggested that, the discriminatory power of the tortuosity descriptor for this dataset lies between SOAM and mean chord length ratio (clr), another measure for tortuosity (see Fig.~\ref{Fig:Tortuosity} and Figures~\ref{Fig:CorrCombined}aii-iii, \ref{Fig:KruskalMC38TortuosityDC101Dll4}, \ref{Fig:KruskalMC38SOAMDC101Dll4}, and \ref{Fig:KruskalMC38CLRDC101Dll4}). Compared to standard measures calculated on the intravital vascular networks (see Figures~\ref{Fig:CorrCombined}aiv-vii and Figures~\ref{Fig:KruskalMC38SegmentsDC101Dll4}-\ref{Fig:KruskalMC38MaxLengthDC101Dll4}), the effect of the treatments on the number of loops highlighted either more significant and discriminatory differences from day 2 after treatment onwards (average vessel diameter, maximal vessel diameter, and maximal vessel length) or higher significance on day 3 (number of vessel segments, number of branching points, and average vessel length). In comparison to the length-diameter ratio, the number of loops captured a more prolonged change in network structure which was still discernible on day 4 after treatment.

In the ultramicroscopy data, our loop descriptor further confirmed transient normalisation effects of bevacizumab visible 1 -- 7 days after treatment (see Fig.~\ref{Fig:Results}bi), whereas the void descriptor captured sustained effects of bevacizumab on angiogenesis (see Fig.~\ref{Fig:Results}bii). 
While the topological descriptors showed known effects, these trends could not be explained by changes in standard measures, such as tumour volume (see Fig.~\ref{Fig:CorrCombined}b and \ref{Fig:RocheVolume}) and, therefore, represented genuine structural changes in the degree of vascularisation. The avascular regions captured by the voids descriptor did not correlate with any existing standard measures (Fig.~\ref{Fig:CorrCombined}bi), suggesting these topological descriptors provide additional quantification of network connectivity.
The differences between the treatment groups in the ultramicroscopy dataset were significant for all topological descriptors on day 1 and 3 after treatment (see Fig.~\ref{Fig:ResultsPValues}). The void descriptor was ideally suited for this dataset as it contains the full tumour rather than a slice as in the intravital data (see Fig.~\ref{Fig:KruskalMC38Voids}).

\paragraph*{Spatio-temporal variation of vascular networks captured by loop descriptor.} 
In contrast to the ultramicroscopy data (see Fig.~\ref{Fig:RocheIntervals}), we found spatio-temporal variation in the number of loops in response to different treatments in the intravital data (see Fig.~\ref{Fig:Results}aiii).
We divided the radial filtration into different spatial intervals (corresponding to spherical shells around the tumour centre) and observed the median number of vessel loops per vessel segment over time in each shell, normalised by day 0 of treatment.
We again confirmed known mechanisms of action for vascular targeting agents DC101 and anti-Dll4; anti-Dll4 increased sprouting predominantly from blood vessels close to the tumour periphery, thereby leading to the formation of loops (see orange/red/brown coloured sectors in Fig.~\ref{Fig:Results}aiii); whereas DC101  reduced the number of loops across the entire vessel network (see blue coloured sectors in Fig.~\ref{Fig:Results}aiii).  

 \begin{figure}[ht!]
 \centering
\includegraphics[width=\textwidth]{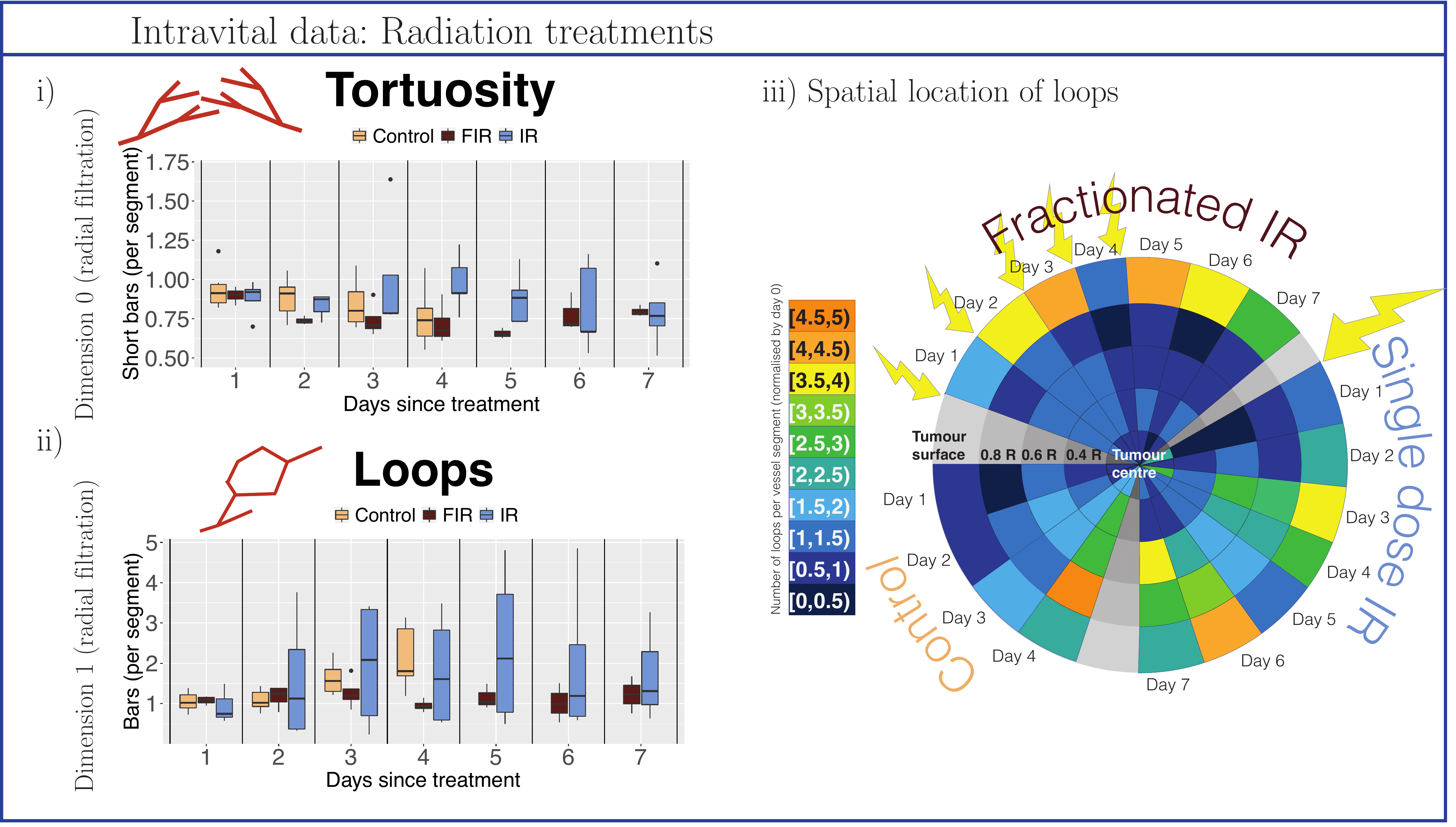}
\caption{{\bf Topological descriptors extracted from tumour blood vessel networks treated with radiation therapy.} We normalised all descriptors with respect to values on the day on which treatment is administered (day 0) or, for controls, the day on which observations commence (day 0).
Data was collected from control mice (beige),
mice treated with fractionated irradiation (FIR, brown), 
and mice treated with single dose irradiation (IR, blue). 
(i)  Tortuosity was computed as the ratio of short bars ($\leq 10\%$ of maximal radius used in the radial filtration) in the dimension 0 barcodes of the radial filtration to the number of vessel segments. (ii) Loops are the number of bars in the dimension 1 barcodes of the radial filtration per vessel segment in the network. (iii) Spatio-temporal resolution of the number of loops per vessel segment. We illustrate the changes in the median number of loops (normalised by day 0) in different radial intervals around the tumour centres over the days of observation.
The yellow arrows highlight days for which the tumours have received treatment on the prior day (i.e. an arrow on day 1 signifies that (a dose of) treatment was administered on day 0). }\label{Fig:ResultsIR}
\end{figure}

\paragraph*{Topological descriptors quantified unknown effects of radiation therapies.} 
Our topological descriptors quantified and, furthermore, elucidated the unknown effects of single- and fractionated-dose irradiation treatments on vascular networks (see Figures~\ref{Fig:ResultsIR}, \ref{Fig:ResultsPValues}, \ref{Fig:KruskalMC38TortuosityIRFIR}, and \ref{Fig:KruskalMC38LoopsIRFIR}).
Reductions of tortuosity and the number of loops from single-dose irradiation were apparent only on day 1 after treatment and showed great variation across different tumours over time. 
Spatially, the effect of single-dose irradiation manifested in a decrease in the number of loops in the whole tumour only on day 1 after treatment and thereafter remained stable only very close to the tumour centre; by contrast, the number of loops increased again in most parts of the vessel network (see Fig.~\ref{Fig:ResultsIR}iii).
Beneficial effects of fractionated-dose irradiation became apparent after a time lag of 2 (tortuosity) or 3 days (loops, with statistically significant difference to controls on day 4, see Fig.~\ref{Fig:ResultsPValues}) after initial treatment. Spatially, the number of loops decreased below the tumour surface but increased in the tumour periphery from day 2 after initial treatment onwards (see Fig.~\ref{Fig:ResultsIR}iii after start of fractionated irradiation treatment).
Trends in tortuosity and number of loops revealed changes in network structure and differ from those seen for the approximate tumour radius (see Fig.~\ref{Fig:KruskalMC38Radii}).

\paragraph*{Vascular architecture evolution linked to increasing topological complexity.} 
Finally, when comparing all five treatment groups in the intravital data, we found significant differences for tortuosity on day 2 after treatment (see Fig.~\ref{Fig:KruskalMC38Tortuosity}), followed by significant differences in the number of loops on day 3 after treatment (see Fig.~\ref{Fig:KruskalMC38Loops}). We hypothesise that vascular targeting agents and radiotherapy first show effects on the level of tortuosity before changes manifest in more complex network structures such as vessel loops. Biologically, this can be explained by treatments having an immediate effect on individual vessels while visible changes in network connectivity associated with angiogenesis take longer to occur.

\section*{Discussion} 
In the present work, we showcased the application of an interpretable and powerful, multi-scale topological method to analyse highly resolved images. Our approach represents a much needed paradigm shift in the analysis of images of biological tissues, closing the current gap between the level of detail in data from modern imaging modalities, which are highly resolved over space and time, and  coarse quantitative descriptors commonly extracted from these images. We quantified, validated, and uncovered aspects of network connectivity in tumour vasculature by exploiting the three-dimensionality of state-of-the-art data across different scales, from small vessel loops to large voids, information which is inaccessible using standard summaries. Our topological descriptors characterise tortuosity and vessel loops (radial filtration), and tumour vascularisation ($\alpha$-complex filtration) in a novel way, giving  unprecedented quantification, in terms of spatial location and connectivity, of dynamic changes in the network architecture of tumour blood vessels during disease progression and treatment. In addition to validating the known dynamic effects of vascular targeting agents on vessel density, we also provided novel quantification of their spatial location effects on the vasculature. Hitherto, we offered a multiscale topological characterisation of the effects of radiotherapy on vasculature. 

 When performing TDA, the choice of filtration and its interpretation can reveal different information about a system. For example, the plane-sweeping filtration is better suited than the $\alpha$-filtration where there is a clear direction in the data (e.g., brain arteries \cite{Bendich2014}).  Here we used points sampled from the vessels in 3D space and then constructed the $\alpha$- filtration, which is robust with respect to rotation. However, the barcode will change with deformation (e.g. loops being stretched) as persistent homology computes information about both geometry and topology. In this study, we made the simplifying assumption that the tumours are spherical, as is often done in modelling; however, this does not hold if the tumour is ellipsoid in shape. For different tumour geometries, the vessels may not be radially oriented, so a future extension could be to consider an ellipsoid filtration \cite{breiding2018learning}. See \cite{Stolz2019} for an exploration of different filtrations for applications. The filtrations we employed in this work cannot be applied to all tumours, for example, a recent study predicting survival of glioblastoma motivated a new topological statistic for analysing shape that effectively analyses multiple directions of sublevel sets \cite{crawford2020predicting}.
We also found that the utility of different topological descriptors (i.e. dimensions) may depend on imaging resolution. Our results indicated that the high planar resolution of intravital data better captures tortuosity, resulting in more short connected components in the radial filtration, whereas it is less appropriate for measuring voids due to the shallow imaging depth. Conversely, the deep z direction of ultramicrosopy enables quantification of voids (i.e., higher dimensional homology features), while comparably low planar resolution may not suffice to generate the small (0-dimensional) features needed to quantify tortuosity. We demonstrated that loops were well quantified for both modalities. In the future, we will investigate more computationally efficient topological invariants, such as the Euler Characteristic and Betti numbers, 
which also provide quantitative information about vessel connectivity. In other future work, we will extract the spatial locations of the tumour and immune cells from the images and apply similar topological analyses to these data to compare control and treated tumours. 

Our topological descriptors provide global and multiscale quantification of vascular connectivity, and represent a first step towards understanding the relationship between structure and function of the vasculature. For example, high tortuosity of vessels has been observed to reduce blood flow \cite{penta2015role}, and a future extension will be to develop directed topology approaches for tackling such directed vascular networks. Even with the state-of-the-art data used here, ethical constraints preclude the collection of more data, thereby limiting the strength of the biological conclusions that can be drawn. If more data were available, our topological descriptors could be fed directly into machine learning algorithms and analyses.
Since our datasets were obtained from two different studies, we were unable to directly compare them. The data were generated on different scales and, therefore, the values differ significantly, with respect to regions of interest and spatial resolution. In future work, it would be informative to cross-validate the descriptors by performing intravital imaging followed by ultramicroscopy imaging on the same mice with small tumours, to validate the method on the same vessel networks, exclude any influences from different imaging modalities, and work towards topological data integration. At this time, analysis of the same tissue with both imaging modalities is logistically impossible.
We propose that the topological descriptors be tested with different imaging modalities used in the clinic to determine their practical use for monitoring the response of tumours to therapy.

We conclude by noting that the topological perspective for analysing and preserving the multiscale nature of data is broadly applicable to other spatial networks~\cite{Feng2020} and biological systems, where it can also be used to quantify perturbations to network topology.
Such networks arise across many different biomedical applications but are also relevant in other biological settings ranging from leaf vessel networks to collagen fibres and signalling networks.

\section*{Methods}

\subsection*{Experimental procedures for intravital data}\label{Seq:Exp}

\paragraph*{Abdominal imaging window implantation.} This procedure was based on a previously described method~\cite{ritsma2013}. Mice were prepared in a surgical unit, administered with inhalational anaesthesia and pre-operative analgesics. Body temperature and respiration rates were monitored throughout the procedure. A one-cm cut was made along the abdominal midline approximately 5  mm underneath the sternum followed by blunt dissection around the cut to separate the connective tissue from the skin. A custom-made imaging window frame (Workshop at the Department of Oncology, Oxford University) was fitted underneath the skin. Continuous sutures were used to secure the skin to the window frame. Approximately $2.5 \times 10^5$ MC38 cells stably expressing eGFP in 5 $\mu$L containing 30\% of Matrigel and 10\% of Evan’s blue dye were injected under the connective tissue and above the abdominal muscle layer. The chamber was then flushed with water to lyse non‑injected cells by osmotic shock, tapped dry with sterile cotton swabs and flooded with saline. A cover glass glued on the chamber’s lid was secured onto the window frame. The animals were then placed onto a heat mat for post‑operative recovery, and their health and tumour growth was monitored by visual examination. 

\paragraph*{Treatment regimes.} Animals with tumours approximately 100 mm$^3$ of the chamber were administered with either anti-mouse VEGFR2 antibody (clone DC101~\cite{Kannan2018}, 27 mg/kg, BioXCell), anti-mouse Dll4 antibody~\cite{Liu2011} twice per week at a dose of 5 mg/kg (in two doses on the initial day of imaging and three days later), or one of two radiation treatments. For the radiation treatments, mice were anaesthetised under inhalation with isoflurane and placed in an imaging-guided small animal radiation research platform (SARRP) irradiator (Xstrahl Ltd). A Cone Beam CT scan (computerised tomography) of each mouse was obtained and the treatment was planned using Muriplan (Xstrahl Ltd). The SARRP was used to deliver 15 Gy of X-rays (220 kVp copper filtered beam with HVL of 0.93 mmCu) to the tumour at ~2 Gy per minute. This was given either in a single dose or at 5 daily fractionations of 3 Gy X-ray radiation to the tumour. Dosimetry of the irradiator was performed as previously described~\cite{Hill2017}.

\paragraph*{Intravital two‑photon imaging.} 
In order to visualize the tumor vasculature, we used a transgenic mouse model in which the fluorescent protein tdTomato is expressed in both normal and tumor endothelial cells (EC). We used transgenic mice bearing a Cre recombinase-tamoxifen receptor fusion protein (Cre-ERT2) driven by the VE cadherin promoter. These mice were crossed with Gt(ROSA)26Sortm9(CAG-tdTomato)Hze mice so that activation of Cre by tamoxifen resulted in EC expression of tdTomato (schematic shown in Figure 1A). Similarly tumor ECs (TEC), identified as CD31 positive cells in allografted tumors, were rendered generally over 90\% tdTomato positive (Figure 1B-D, Figure S1). For imaging purposes we only used mice with greater than 95\% fluorescent EC.

Mice were imaged for four days following initial treatment for vascular targeting agents and seven days for radiation treatment with a Zeiss LSM 880 microscope equipped with an aesthetic vaporiser and respiratory monitoring system. Stage and atmosphere were heated to 37 \degree C. To label perfused vessels,  Quantum dot‑705 solution (1 $\mu$M, Invitrogen) was infused intravenously using a motorised pump at a rate of 0.84 $\mu$L$\cdot$min$^{-1}$. A mode‑locked MaiTai laser tuned to 920 nm was used to simultaneously excite eGFP, tdTomato and Qdot705. The Qdot705 signal was acquired through a BP700/100 filter with a non-descanned detector. GaAsP detectors were used to acquire the signal of tdTomato selected by a BP 650/45 filter and the eGFP selected by a BP525/50 filter. 
Images were acquired in Z-stack tile scans with a pixel size of 0.823 $\mu$m and an image size per tile of 512 $\times$ 512 $\times$ 5 in x, y and z, respectively. A water immersion 20 $\times$ objective made for UV-VIS-IR transmission with a numerical aperture of 1 was used.
The segmentation of tumor blood vessels was based on the TECs expressing tdTomato. We used intravenous injection of Qdots to distinguish perfused from non-perfused tumor vessels, i.e. vessels labelled with the infused Qdots and vessels not labelled with it.  As further evidence, we note that no Qdot positive, endothelial negative vessels were identified. If the Qdots were in the lymphatics then they would have identified vessels not lined by vascular endothelium; this did not happen. 
All animal experiments were conducted in accordance with the United Kingdom Animals (Scientific Procedures) Act 1986 as amended (Amendment Regulations 2012 [SI 2012/3039]), under the authority of a UK Home Office Project Licence (PPL 30/2922 and PCDCAFDE0), with local ethical approval from the University of Oxford Animal Welfare and Ethical Review Panel.

\subsection*{Datasets} \label{seq:data}
We analysed two different tumour blood vessel datasets: data obtained by multiphoton intravital 3D imaging~\cite{Pittet2011} (see above for description of experimental procedures) and data obtained by ultramicroscopy~\cite{Dodt2007}. Both datasets consist of 3D stacks of images of tumour blood vessels subjected to different experimental conditions.

\paragraph*{Dataset I: Multiphoton intravital 3D imaging.} The \emph{intravital dataset} consists of tumour vasculature images that were obtained from multiphoton intravital 3D imaging~\cite{Pittet2011} of transgenic mice injected with murine colon adenocarcinoma cells (cell line MC38). The animals were imaged alive and over several days using the experimental procedures described in Section \nameref{Seq:Exp}. The mice were divided into groups that were subjected to different experimental conditions: 
\begin{enumerate}
\item Controls (7 mice).
\item Anti-Dll4 treated tumours (3 mice): The mice were treated using anti-Dll4 antibodies~\cite{Liu2011} which block Dll4 signalling and, thereby, increase vessel sprouting. The resulting networks are very dense and complex. 
\item DC101 treated tumours (5 mice): The mice were treated using DC101 antibodies~\cite{Kannan2018} which block VEGFR-2 signalling and, thereby, reduce vessel sprouting. 
\item Single-dose irradiated tumours (5 mice): The mice were treated with a single dose of 15 Gy on the first day of imaging. 
\item Dose-fractionated irradiated tumours (4 mice): The mice were treated with five doses of 3 Gy over 5 consecutive days followed by two days of rest starting on the first day of imaging. 
\end{enumerate} 
In each case, we refer to the start of treatment or observation as day 0. 

\paragraph*{Dataset II: Multispectral fluorescence ultramicroscopy data.} The \emph{ultramicroscopy dataset} consists of multispectral fluorescence ultramicroscopy~\cite{Dodt2007} images of blood vessels of human breast cancer tumours (cell line KPL-4, HER2 positive) that were implanted into 31 immunodeficient mice.
The experiments were carried out by Dobosz \emph{et al.}~\cite{Dobosz2014}, \emph{Roche Diagnostics/Institute for Biological and Medical Imaging, Helmholz Zentrum, Munich}.
The mice were divided into a control group and a treatment group:
\begin{enumerate}
\item Controls (18 mice). 
\item Anti-VEGF-A treated tumours (13 mice): The mice were treated with bevacizumab~\cite{Ferrara2004}, an antibody which binds to VEGF-A and, thereby, induces normalisation~\cite{Goel2011} of the vessel networks, i.e. reduces some of their structural and functional abnormalities, and lowers their permeability~\cite{Dobosz2014}. 
\end{enumerate}
Treatment was administered once the tumours reached a volume of approximately 60~mm$^3$, controls were observed accordingly.
To test the effect of treatment on drug delivery at different time points, both controls and anti-VEGF-A treated mice were also treated with trastuzumab~\cite{Hudis2007} (anti-HER2 antibody) six hours before the tumour was extracted and prepared for imaging. Different subgroups of tumours were imaged on day 1 (5 controls, 5 treated), day 3 (5 controls, 4 treated), day 7 (5 controls, 2 treated), and day 14 (3 controls, 2 treated) after administration of bevacizumab. For more details on experimental conditions see reference~\cite{Dobosz2014} (note that the dataset in reference~\cite{Dobosz2014} was created under the same conditions and overlaps with the data used in this work, but the two are not identical, e.g. the dataset in reference~\cite{Dobosz2014} consists of 5 controls and treated mice for day 1, 3, and 7 after treatment each but does not include day 14 after treatment). Imaging was performed \emph{ex vivo} at a spatial resolution of 5.1$\mu$m on the $xy$-plane with images taken every 5.1$\mu$m in the $z$-direction.
Skeletonisations of the images were produced by Dobosz \emph{et al.}~\cite{Dobosz2014} using a custom \emph{Definiens} Developer script. For details of how the ultramicroscopy data was skeletonised, please see Section \nameref{Sec:DataPreProcessing} in Methods.

\subsection*{Data preprocessing}\label{Sec:DataPreProcessing}

\paragraph*{ Intravital data.} Skeleton files were extracted from the imaging data by combining two segmentation models and taking their geometric average. The skeletons were then pruned (see reference~\cite{Bates2017Thesis}, p. 165, for a full description). The segmentation method used for the intravital dataset was extensively tested against synthetic datasets as well as against manually segmented intravital microscopy images \cite{
Bates2017Thesis}. This method achieved a Dice score of 0.97. Moreover, a Skeleton Error (given in $\mu m$) - the distance between skeletons which was computed using the Modified Hausdorff Distance was determined. This skeleton error can be interpreted as the average shortest distance between any point on the ground truth skeleton to some point on the target skeleton and vice versa. With our method this skeleton error was ~5 $\mu m$ compared to ground truth in the synthetic dataset as well as in the intravital dataset, where manually segmented images were considered as ground truth. Lastly, the method used also achieved coverage of 0.96-0.99 both in the synthetic datasets and intravital microscopy datasets. This shows, that the errors introduced by the segmentation method were relatively small.

We extracted blood vessel networks from skeleton files using the method {\tt VesselTree} from {\sc unet\_core.vessel\_analysis} in the {\sc python} code package {\sc unet-core}~\cite{unetRuss}. The extracted networks consist of points on vessel branches (multiple points per vessel branch including branching points) which represent the network nodes, and the vessels that connect them which constitute the edges of the network. {\tt VesselTree} also enables us to extract network features such as number of vessel segments (i.e. edges of the network), number of branching points (i.e. nodes of the network), vessel diameters, vessel lengths, and measures of tortuosity (chord-length-ratio and sum of angles metric) for every point. We account for the difference in resolution between the $z$-axis and the $xy$-plane by rescaling the coordinates in the $z$-direction using a factor of $\frac{0.83}{5}$ on the $z$-coordinates before further analysis.

We excluded the following data from our analysis due to imaging and/or segmentation quality: control tumour 24\_2C, day 4; fractionated-dose irradiated tumour 60\_1E, day 5 onwards. 
For the radial filtration, due to the very high number of points in some of the blood vessel networks, we reduced the data size by including all branching points but sampling only every second point from every branch in the following networks: control tumour 18\_4E, day 3; control tumour 18\_4E, day 4; control tumour 29\_1B, day 3; control tumour 29\_1B, day 4; control tumour 34\_2A, day 4; control tumour 60\_2A, day 4; DC101 treated tumour 51\_2C, day 1; DC101 treated tumour 54\_2D, day 2; anti-Dll4 treated tumour 24\_2A, day 3; anti-Dll4 treated tumour 24\_2A, day 4. The days listed refer to the days after tumour treatment.
For the $\alpha$-complex filtration we used the full set of nodes as input.

\paragraph*{Ultramicroscopy data.}
We preprocessed grey scale skeletonisation files provided in the ultramicroscopy dataset from individual {\tt .tif} files (one for every $xy$-plane slice of the vessel network) to .tif stacks in {\tt uint8} format using the software {\sc ImageJ}~\cite{ImageJ}. We then converted the {\tt .tif} stacks to {\tt .nii} format using the function {\tt tiff2nii.m} from the {\sc Matlab} toolbox~\cite{niiMatlabToolbox}. We used the {\tt .nii} files as input for our {\sc unet-core}~\cite{unetRuss} in our {\sc python} scripts.
Even though {\sc unet-core} was originally trained on multiphoton intravital 3D imaging, we justify our approach by the fact that the skeletonisations are clear, high-contrast images (see Fig.~\ref{fig:ExampleImagesRoche} for extracted networks). Any imaging specific effects were removed by the skeletonisation process which was developed specifically for this dataset~\cite{Dobosz2014}. We compared the number of branching points and the number of vessel segments extracted by {\sc unet-core} with similar measurements extracted previously by Dobosz \emph{et al.}~\cite{Dobosz2014} and found that these are highly correlated (see Figures~\ref{Fig:CorrCombined}b, \ref{Fig:RocheCor}).  

We note that we obtained 3D coordinates for network nodes. The distances between these nodes scale linearly with the true distance in $\mu$m. Since we were only interested in features with respect to their relative distance to the tumour centre, this was sufficient for our purposes. A coordinate set true to distance could be obtained by comparing an exemplary output network closely to microscopy images. 

For the radial filtration, due to the very high number of points in these blood vessel networks (on the order of millions of nodes in comparison with on the order of thousands of nodes in the intravital data), we reduced the point clouds for all tumours by including all branching points but sampling only every fourth point from every branch. Despite our reduction approaches, we were not able to run our codes on one of the treated tumours from day 14 networks. For the $\alpha$-complex filtration we used the full set of nodes as input.

\subsection*{Topological data analysis}\label{Sec:Methods}

Topological data analysis (TDA) is an umbrella term used for methods that allow the study of potentially high dimensional data using mathematical concepts from topology~\cite{Munkres2000}. \textit{Persistent homology} (PH)~\cite{Edelsbrunner2002, Edelsbrunner2008, Carlsson2009,Edelsbrunner2010} quantifies global topological structures (e.g., connectedness, loops, and voids) in data. More details on TDA and PH are in the Supplementary Materials.

\paragraph*{Homology and simplicial complexes.} To compute (persistent) homology from data, we first constructed simplicial complexes, which can be thought of as collections of generalized triangles. From the constructed simplicial complexes, we quantified and visualised the datasets' connected components (dimension 0), loops (dimension 1), and voids (dimension 2) at different spatial scales in the data.

Persistent homology is based on the topological concept of \emph{homology} (for intuitive introductions, see, for example, references~\cite{
SimplexMinded,sizemore2018}; for more formal introductions see references~\cite{Kosniowski1980, Hatcher2002,munkres1984}). 
To compute topological invariants, such as connected components (dimension 0) and loops (dimension 1), we used homology. To obtain homology from a simplicial complex, $X$, we constructed vector spaces whose bases are the $0$-simplices, $1$-simplices, and $2$-simplices, respectively, of $X$.  There is a linear map between $2$-simplices and $1$-simplices, called the boundary map $\partial_2$, which sends triangles to their edges. Similarly, the boundary map $\partial_1$ sends edges to their points and $\partial_0$ sends points to 0. The action of the boundary map $\partial_1$ on the simplices is stored in a binary matrix where the entry $a_{i,j}$ denotes whether the $i$-th $0$-simplex forms part of the boundary of the $j$-th $1$-simplex. If so, then $a_{i,j} = 1$; otherwise, $a_{i,j}=0$. We computed the kernel $\text{Ker}(\cdot)$ and image $\text{Im}(\cdot)$ of the boundary maps to obtain the vector spaces $H_0(X) = \text{Ker}(\partial_{0}) / \text{Im}(\partial_{1})$ and
$H_1(X) = \text{Ker}(\partial_{1}) / \text{Im}(\partial_{2})$. These vector spaces are also referred to as homology groups and their dimensions define the topological invariant we studied called the \emph{Betti numbers} of $X$, $\beta_0$ and $\beta_1$, which give the number of connected components and loops, respectively.
We studied a vascular network at multiple scales in different ways as described later in this section. The multiple scales of the data can be encoded by a \emph{filtration}, which is a sequence of embedded simplicial complexes $X_0\subseteq X_1 \subseteq \dots \subseteq X_\text{end}$ built from the data. 

\paragraph*{Persistent homology.} Persistent homology is an algorithm that takes in data via a filtration and outputs a topological summary, which visualises changes in topological features such as connectedness (dimension 0) and loops (dimension 1) across the filtration. The simplicial complexes are indexed by the scale parameter of the filtration. The inclusion of a simplicial complex $X_i\subseteq X_j$ for $i\leq j$ gives a relationship between the corresponding homology groups $H_p(X_i)$ and $H_p(X_j)$ for $p = 0,1,2$. This relationship allowed us to track topological features such as loops along the simplicial complexes in the filtration. Intuitively, a topological feature is born in filtration step $b$ when it is first computed as part of the homology group $H_p(K_b)$ and dies in filtration step $d$ when that feature no longer exists in $H_p(K_d)$, \emph{i.e.}, when a connected component merges with another component or when a loop is covered by 2-simplices. The output from persistent homology is a multiset of intervals $[b,d)$ which quantifies the persistence of topological features. Each topological feature is said to persist for the scale $d-b$ in the filtration. 

\paragraph*{Method I: Radial filtration.} We applied a \emph{radial filtration}~\cite{Stolz2019} 
to the 3D vessel networks, i.e. the collection of nodes (both branching points and points along vessel branches), their spatial coordinates, and the edges between them. We built the filtration starting in the tumour centre, which we approximated by the centre of mass of the points sampled from the tumour blood vessels, e.g. the nodes of our networks. 
We then proceeded in the following way. We divided the maximal distance of a node in the network to the centre of mass into 500 steps and from this constructed a sequence of uniformly increasing radii. By increasing the radial distance stepwise, each filtration step we included all nodes within the specified radius. If two nodes that were connected by an edge were also within the given radius, we added that edge to our filtration. 
In the barcodes from this filtration we could capture tortuosity (from connected components in dimension 0 barcodes with persistence $\leq$ 10 \% of the maximal radius used, see Fig.~\ref{Fig:Tortuosity}a), loops (dimension 1) and their spatial distribution.
We note that for the intravital (shallow) imaging data, the approximated tumour centre is defined by the image (vessel nodes) viewed through the window chamber. The approximated tumour centre was calculated based on the vasculature in this small segment and, as such, does not represent the true tumour centre.

\paragraph*{Method II: $\alpha$-complex.} On 3D data, the $\alpha$-complex~\cite{Otter2017,gudhi:AlphaComplex} filtration builds a sequence of nested simplicial complexes (collections of nodes, edges, triangles, and tetrahedra) whose final element $K_{end}$ is the Delaunay triangulation~\cite{Delauney1934}, i.e. the triangulation of the 3D convex hull of the data points by tetrahedra. We built the filtration on the 3D nodes of the vessel networks. Inductively, starting with the highest dimension (i.e. first tetradedron, then edges), each simplex $\sigma$ in $K_{end}$ was assigned a filtration value given by the square of its circumradius $\alpha$ in the case that the circumsphere contains no other vertices than the vertices of $\sigma$; otherwise, its filtrations value was given by the minimum of the filtration values of the higher-dimensional simplices of which $\sigma$ was a face. To construct the filtration, edges, triangles, and tetrahedra were included up to a set filtration value which increased stepwise. The effect of the assignment of the filtration values was, for example, that in 2D the long edge of a slim triangle was only included when the whole triangle was included. This avoided the formation of cycles for slim triangles. 
In the barcodes from this filtration we could capture the degree of tumour vascularisation (from voids, dimension 2).

\subsection*{Existing descriptors}

The standard morphological descriptors that we computed from the segmented intravital microscopy images were the number of vessel segments (i.e. number of edges), number of branching points (i.e. number of nodes), maximal vessel diameter, average vessel diameter, maximal vessel length, average vessel length, average chord length ratio (clr), average sum of angles measure (SOAM), and vessel length/diameter ratio.
The standard descriptors that we included for the ultramicroscopy dataset were the
number of vessel segments (both as computed by \cite{Dobosz2014} and {\sc unet-core}~\cite{unetRuss}), number of branching points (both as computed by \cite{Dobosz2014} and {\sc unet-core}~\cite{unetRuss},   necrotic tumour volume as computed by \cite{Dobosz2014}, tumour volume as computed by \cite{Dobosz2014}, and vital tumour volume as computed by \cite{Dobosz2014}.

\begin{figure}[ht!]
 \centering
\includegraphics[width=\textwidth]{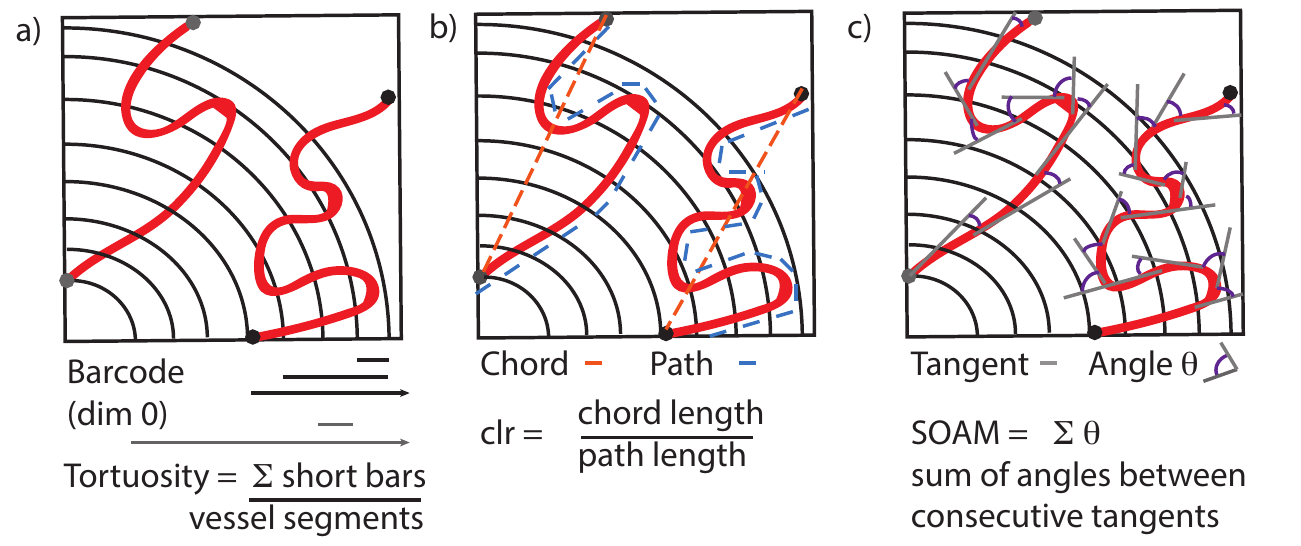}
\caption{{\bf Schematic of tortuosity descriptors.} (a) The topological descriptor is defined as the number of short bars in the barcode (connected components in dimension 0 barcodes with persistence $\leq$ 10 \% of the maximal radius) normalised by the number of vessel segments. In this schematic, there are two vessels. This normalisation ensures the connected components in the tortuosity measure do not also count different vessel segments. This descriptor is in contrast to the topological tortuosity reported in \cite{Bendich2014}, which did not have multiple vessel segments. (b) The chord-length-ratio  (clr)~\cite{Bates2017Thesis}  is the ratio between the chord connecting two ends of a curve (orange) and the path length of the curve (blue). Clr measures the deviation from a straight line. The sum-of-angles-measure (SOAM) measures the sum of angles between consecutive tangents of a curve, so a high score is given to a curve rapidly changing direction.} \label{Fig:Tortuosity}
\end{figure}

\paragraph*{Tortuosity: Sum-of-angles-metric (SOAM).}
The sum-of-angles-metric (SOAM) was applied as a measure of tortuosity in blood vessels by Bullit \emph{et al.}~\cite{Bullitt2005}. It is the sum of the angles of regularly sampled tangents along a blood vessel skeleton and can take values from zero (straight vessel) to infinity. For tortuous vessels, the metric increases monotonically with vessel length. See Fig.~\ref{Fig:Tortuosity}b for a schematic.

\paragraph*{Tortuosity: Chord-length-ratio (clr).}
The chord-length-ratio (clr)~\cite{Bates2017Thesis} of a blood vessel is defined as the ratio of the distance between the branching/end points of the vessel and the length of the vessel. The measure can take a value of at most one (straight vessel) and tends to zero for very tortuous vessels. See Fig.~\ref{Fig:Tortuosity}c for a schematic.

\paragraph*{Statistical analysis.} We analysed the statistical significance of differences between treatment groups in the tortuosity values, number of loops per vessel segment, and median persistence of voids. We performed a pairwise Wilcoxon's rank sum test on the ultramicroscopy data for each day separately to determine the statistical significance of our topological measures (see Fig.~\ref{Fig:ResultsPValues}b). We tested at significance level 0.05.
For the intravital data, we performed a Kruskal-Wallis test to determine whether at least one treatment group differs significantly from the others for the topological descriptors (see Figures~\ref{Fig:KruskalMC38TortuosityDC101Dll4}, \ref{Fig:KruskalMC38LoopsDC101Dll4}, \ref{Fig:KruskalMC38TortuosityIRFIR}, \ref{Fig:KruskalMC38LoopsIRFIR}, Figures~\ref{Fig:KruskalMC38Tortuosity} -- \ref{Fig:KruskalMC38Voids}) as well as for standard vasculature measures (see Figures~\ref{Fig:KruskalMC38TortuosityDC101Dll4} -- \ref{Fig:KruskalMC38LambdaDC101Dll4}). 
We further applied a pairwise Wilcoxon's rank sum test between the control group and the different treatment regimes for the topological descriptors (see Fig.~\ref{Fig:ResultsPValues}a). We again tested at significance level 0.05 and did not correct for false discovery rate. 
To explore correlations between different types of summary descriptors for vascular networks, we computed pairwise Pearson correlation values for the different descriptors in both datasets separately (see Figures \ref{Fig:MC38Cor} and \ref{Fig:RocheCor}). We performed all statistical analyses in R Studio~\cite{RStudio-Team:2016aa}, all our tests described above were by default two-sided.

\subsection*{Implementation}

We implemented the radial filtration in {\sc Matlab} and used the software package {\sc javaPlex}~\cite{Javaplex} to compute PH on our filtration. We divided the distance from the tumour centre (centre of mass) to the farthest away point in the blood vessel network into 500 steps to build the radial filtration. We implemented the $\alpha$-complex using the {\sc GUDHI} library~\cite{gudhi:AlphaComplex}. All code is freely available at the following repository: \\ \url{https://github.com/stolzbernadette/TDA-Tumour-Vasculature}.

\bibliography{Vessels}

\bibliographystyle{Science}

\section*{Acknowledgments}
We are very grateful to Gesine Reinert for advice on statistical analysis of our results and Ulrike Tillmann for valuable feedback. We thank Russell Bates, Mike Brady, Uli Bauer, James Grogan, Almut Koepke, Nina Otter, and Nicola Richmond for helpful discussions. BJS gratefully acknowledges EPSRC and MRC grant
	(EP/G037280/1) and F. Hoffmann-La Roche AG for funding her
	doctoral studies. BM acknowledges funding from the People Programme (Marie Curie Actions) of the European Union’s Seventh Framework Programme (FP7/2007- 2013) under REA grant agreement No 625631. HAH gratefully acknowledges funding from an EPSRC Fellowship (EP/K041096/1) and a
	Royal Society university research fellowship. BJS, HMB, and HAH are
	members of the Centre for Topological Data Analysis, funded by
	the EPSRC grant (EP/R018472/1).

\noindent [Author Information] BJS contributed to the design of the work, the investigation and statistical interpretation of results, creation of software, visualisation, and writing the original draft. 
	BM, FL, FB, JK, and RJM contributed to the acquisition of data. HMB and HAH contributed to the conception of the project, contextualisation of the work, and supervision. BJS, JK, BM, FB, HMB, and HAH contributed to writing the original draft.
	
\noindent [Competing Interests] The authors declare that they have no competing financial interests.

\noindent [Data availability] The datasets used in this study can be provided by the corresponding author, RJM, or BM pending scientific review and a completed material transfer agreement. Requests for the datasets should be submitted to: RJM or BM. All other data needed to evaluate the conclusions in the paper are present in the paper and/or the Supplementary Materials.

\noindent [Code availability] Computer code used to generate results is available on \url{https://github.com/stolzbernadette/TDA-Tumour-Vasculature}.

\noindent [Correspondence] Correspondence and requests for materials
	should be addressed to Bernadette J Stolz~(email: stolz@maths.ox.ac.uk) or Heather A Harrington~(email: harrington@maths.ox.ac.uk).

\clearpage

\section*{Supplementary materials}
Topological data analysis\\
Computational differences between datasets \\
Supporting Experimental Information \\
Tortuosity in the ultramicroscopy data \\
Alternative results figures and statistical analysis \\
Additional results and statistical analysis \\

\setcounter{page}{1}

\newcommand{\hbAppendixPrefix}{S}
\renewcommand{\thefigure}{\hbAppendixPrefix\arabic{figure}}
\renewcommand{\thesection}{\hbAppendixPrefix\arabic{section}}   
\renewcommand{\thepage}{\hbAppendixPrefix\arabic{page}}   
\renewcommand{\thetable}{\hbAppendixPrefix\arabic{table}}

\subsection*{Topological data analysis}

\paragraph*{Homology and simplicial complexes.}
Persistent homology is based on the topological concept of \emph{homology} (for intuitive introductions, see, for example, references~\cite{
SimplexMinded,sizemore2018}; for more formal introductions see references~\cite{Kosniowski1980, Hatcher2002,munkres1984}). Homology allows one to study shapes and forms disregarding any changes caused by stretching or bending. One can study the properties of a topological space by partitioning it into smaller, topologically simpler pieces, which when reassembled include the same aggregate topological information as the original space. Topological spaces can be very simple. Two trivial examples are the empty set $X = \emptyset$ or a space that consists of one single point $X = \{ x \}$.
 If we want to capture the topological properties of the second example $X = \{ x \}$, we could simply choose a single node to represent it. However, a node or even a collection of nodes does not allow one to capture the topological properties of more complicated spaces, such as a $2$-sphere or the surface of the earth. In such cases, one needs a simple object that carries the information that the space is connected but also encloses a hole. Consider, for example, a collection of triangles glued together to form a hollow tetrahedron; this is an example of a mathematical object called a \emph{simplicial complex}. The building blocks that one uses to approximate topological spaces are called \emph{$n$-simplices} which one can think of as generalised triangles. The parameter $n$ indicates the dimension of the simplex. Every $n$-simplex contains $n+1$ independent nodes: a point 
\tikz
\tikzstyle{every node}=[circle, draw, fill=black!50,
                        inner sep=0pt, minimum width=3pt]
                        \path[draw] (0,0) node (5ex){} ;
is a $0$-simplex, an edge 
\tikz
\tikzstyle{every node}=[circle, draw, fill=black!50,
                        inner sep=0pt, minimum width=3pt]
                        \path[draw] (0,0) node (5ex) {} -- (0.5,0) node {} ; 
is a $1$-simplex, a triangle \includegraphics[width=0.5cm]{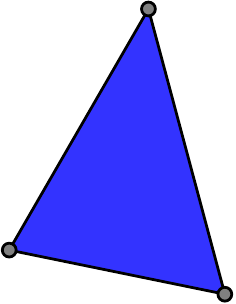}  
is a $2$-simplex, and a (solid) tetrahedron \includegraphics[width=0.7cm]{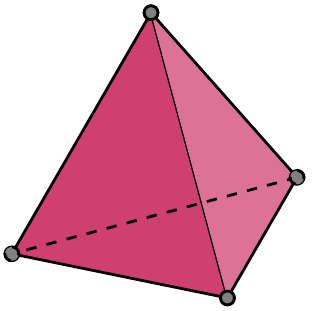} 
is a $3$-simplex. By using a numbering $x_i$ of vertices, we can write a $0$-simplex as $[x_0]$, a $1$-simplex as $[x_0,x_1]$, a $2$-simplex as $[x_0,x_1,x_2]$, and a $3$-simplex as $[x_0,x_1,x_2,x_3]$.
The lower-dimensional simplices form so-called \textit{faces} of the associated higher-dimensional objects. One combines different simplices into a \emph{simplicial complex} $X$ to capture all different aspects of a topological space. Two simplices that are part of a simplicial complex are allowed to intersect only in common faces.
The \emph{dimension} of a simplicial complex is defined to be the dimension of its highest-dimensional simplex. A subcollection of a simplicial complex $X$ is called a \emph{subcomplex} of $X$ if it forms a simplicial complex itself.

For every simplicial complex $X$ we can define a vector space $C_n(X)$ that is spanned by its $n$-simplices with coefficients in the field $\Z/2\Z$. The elements of the vector space $C_n(X)$ are called \emph{$n$-chains}. We can now define a linear map, the so-called \emph{boundary operator}, between vector spaces $C_n(X)$ and $C_{n-1}(X)$ which takes every $n$-simplex $x$ to the (alternating) sum of its faces, i.e. its boundary:
\begin{align}
\partial_n: C_n(X) &\rightarrow C_{n-1}(X), \nonumber \\
x &\mapsto \sum_{j = 0}^n (-1)^j[ x_0, \dots, x_{j-1},x_{j+1}, \dots, x_n ],\label{eq:boundaryOp}
\end{align}
i.e. in the $j$-th summand we omit the vertex $x_j$ from the vertices spanning the $(n-1)$-simplex. Note that the sum in Equation~(\ref{eq:boundaryOp}) is over the field $\Z/2\Z$ where $(-1) = 1$, i.e. we can omit the $(-1)^j$ term in the above equation. We can use the boundary operator to connect all $n$-chains of a simplicial complex $X$ in a sequence, the so-called \emph{chain complex} $\mathcal{C} = \{C_n, \partial_n\}$: 
\begin{align}
\dots \overset{\partial_{n+2}}{\longrightarrow} C_{n+1} \overset{\partial_{n+1}}{\longrightarrow} C_n &\overset{\partial_{n}}{\longrightarrow} C_{n-1} \overset{\partial_{n-1}}{\longrightarrow} \dots \overset{\partial_{1}}{\longrightarrow} C_{0}\\
c &\longmapsto \partial_n c.
\end{align}
We can represent a collection of edges that are connected to form a loop in a simplicial complex as a 1-chain, for example, $[x_0,x_1]+[x_1,x_2]+\dots+[x_{j},x_0]$.
If we apply the boundary operator to this 1-chain, we obtain $\partial([x_0,x_1]+[x_1,x_2]+\dots+[x_{j},x_0])=[x_1]-[x_0]+[x_2]-[x_1]+\dots+[x_0]-[x_{j}] = 0$.
In contrast, for a collection of edges that does not form a loop this is not the case, e.g., $\partial([x_0,x_1]+[x_1,x_2]+\dots+[x_{j-1},x_j]) =[x_j] -[x_0] = [x_0] + [x_j]$ (for coefficients from $\Z/2\Z$). Chains that are in the kernel 
 of $\partial_n$, i.e. their boundary is zero, are called \emph{$n$-cycles}. 
One can compute that the composition of two boundary maps yields zero, i.e. \begin{equation}
\partial_n \partial_{n+1} c = 0,
\end{equation} 
since the boundary of a boundary is empty. The image $\Ima \partial_{n+1}$ of the boundary operator is therefore a subspace of the kernel $\ker \partial_n$ and its elements are called \emph{$n$-boundaries}.

One can associate a family of vector spaces known as \emph{homology groups} to a simplicial complex $X$ based on its cycles and boundaries. For every dimension $n \geq 0$ one defines the $n$th \emph{homology group} as:
\begin{align}
H_n(X) = \frac{\ker \partial_n}{\Ima \partial_{n+1}}.
\end{align}
 In dimension $2$, the elements of the homology group $H_2$ are called \emph{voids}; in dimension $1$,  the elements of the homology group $H_1$ are called \emph{loops}; in dimension $0$, the elements of the homology group $H_0$ are called \emph{connected components}.
Two elements in $H_n$ are considered to be different, if they differ by more than a boundary, i.e. if they represent different $n$-dimensional holes. We then say that they belong to different \emph{homology classes}.

We measure the number of $n$-dimensional holes of a simplicial complex by considering its $n$th \emph{Betti number} $\beta_n$:
\begin{align}
\beta_n = \dim H_n(X) = \dim \ker \partial_n - \dim \Ima \partial_{n+1}.
\end{align}
The first three Betti numbers, $\beta_0$, $\beta_1$, and $\beta_2$, represent, respectively, the number of connected components, the number of $1$-dimensional holes, and the number of $2$-dimensional holes (i.e. voids) in a simplicial complex.

\paragraph*{Persistent homology.}
While homology gives information about a single simplicial complex, PH allows one to study topological features across embedded sequences, so-called \emph{filtrations}, of simplicial complexes, which can be constructed from data. A filtration~\cite{Carlsson2009, Edelsbrunner2008, Ghrist2008} of a simplicial complex $X$ is a sequence of embedded simplicial complexes,
\begin{equation}
	\emptyset = X_0 \subseteq X_1 \subseteq X_2 \subseteq \dots \subseteq X_{end} = X\,,
\end{equation}
starting with the empty complex and ending with the entire simplicial complex $X$. The simplicial complexes in the filtration are connected by inclusion maps. One can now apply an important property of homology, \emph{functoriality}: any map between simplicial complexes $f_{i,j}: X_i \rightarrow X_j$ induces a map between their $n$-chains $\tilde{f}^n_{i,j}: C_n(X_i)~\rightarrow~C_n(X_j)$ which induces a map between their homology groups $f^n_{i,j}: H_n(X_i) \rightarrow H_n(X_j)$. In particular, this means that there exist maps between the homology groups of every simplicial complex in a filtration, e.g., there are maps that relate the voids, loops or connected components in simplicial complexes across a filtration. One can visualise topological features such as loops or connected components across a filtration in a summary diagram called a \emph{barcode}~\cite{Carlsson2005,Ghrist2008}. For an appropriate choice of basis~\cite{zomorodian2005} of the homology groups $H_n$, a barcode represents the information carried by the homology groups and the maps $f^n_{i,j}: H_n(X_i) \rightarrow H_n(X_j)$. A topological feature of dimension $n$ in $H_n(X_{b})$ is \emph{born} in $H_n(X_{b})$, if it is not in the image of $f^n_{b-1,b}$. For example, intuitively, a loop is born in filtration step $b$, if the loop appears closed in the simplicial complex $X_b$ for the first time. A topological feature from $H_n(X_i)$ \emph{dies} in $H_n(X_d)$, where $i<d$, if $d$ is the smallest index such that the feature mapped to zero by $f^n_{i,d}$. If the topological feature is a loop, intuitively it dies in the filtration step where it is first fully covered by triangles (or other higher-dimensional simplices). Note that some topological features never die in a filtration, for example, we always have one connected component in a non-empty simplicial complex that is never mapped to zero. In a barcode, topological features in the filtration of a simplicial complex are represented by half-open intervals $[b,d)$. The lifetime of a topological feature, the so-called \emph{persistence} $p$, is defined as 
$p = d-b $.
For topological features that persist until the last filtration step (and beyond), the persistence is said to be infinite.

\subsection*{Computational differences between datasets}\label{Sec:ComputData}

The differences in the biology and the imaging of our datasets led to discrepancy in computational feasibility (see Table \ref{tab:data2}). In particular, the network sizes and penetration depth of the imaging differed considerably,
which significantly affected the computations for the radial filtration. 
We first performed the majority of computations for the intravital data on a IBM System x3550 M4 16 core server with 768 GB RAM over 3 months but were not able to obtain all results. For the ultramicroscopy data as well as the remaining intravital data, we required a Dual Intel Xeon Gold 6240M 18 core processor system with 3TB of RAM to complete computations over further 3 months. While for the intravital data we were able to compute the radial filtration on all networks in the dataset (in some cases after reduction approaches for the number of nodes, see Data preprocessing in Methods description), in the ultramicroscopy data, we were not able to obtain results for one of the control tumours on day 14 of observation despite reducing the number of nodes (see Section Data preprocessing in Methods description).

\begin{table}
\begin{tabular}{ p{2.5cm} p{2.6cm} p{2.3cm} p{2.7cm} p{1.8cm} p{2.5cm}}
\toprule
Data set & Branching points & Segments (edges) & Tumour volume \newline (initial day of imaging)& Penetration depth & Radial \newline filtration \newline computation per network \\
\midrule
\midrule
 \rowcolor{grey} Intravital & 240 -- 10 025 & 260 -- 10 060 & 100 mm$^3$ & 300 $\mu$m  & Days to weeks \\
 \midrule
  Ultramicroscopy & 12 500 -- \newline 118 000 & 16 700 -- \newline 169 150 & 60 mm$^3$ &$\geq$ 5 mm  & Weeks to months\\
\bottomrule
\end{tabular}
\caption{\label{tab:data2} \textbf {Summary of data sets, experimental conditions, and computational time.}}
\end{table}

\clearpage

\subsection*{Supporting Experimental Information}
\begin{figure}[ht!]
 \centering
\includegraphics[width=.7\textwidth]{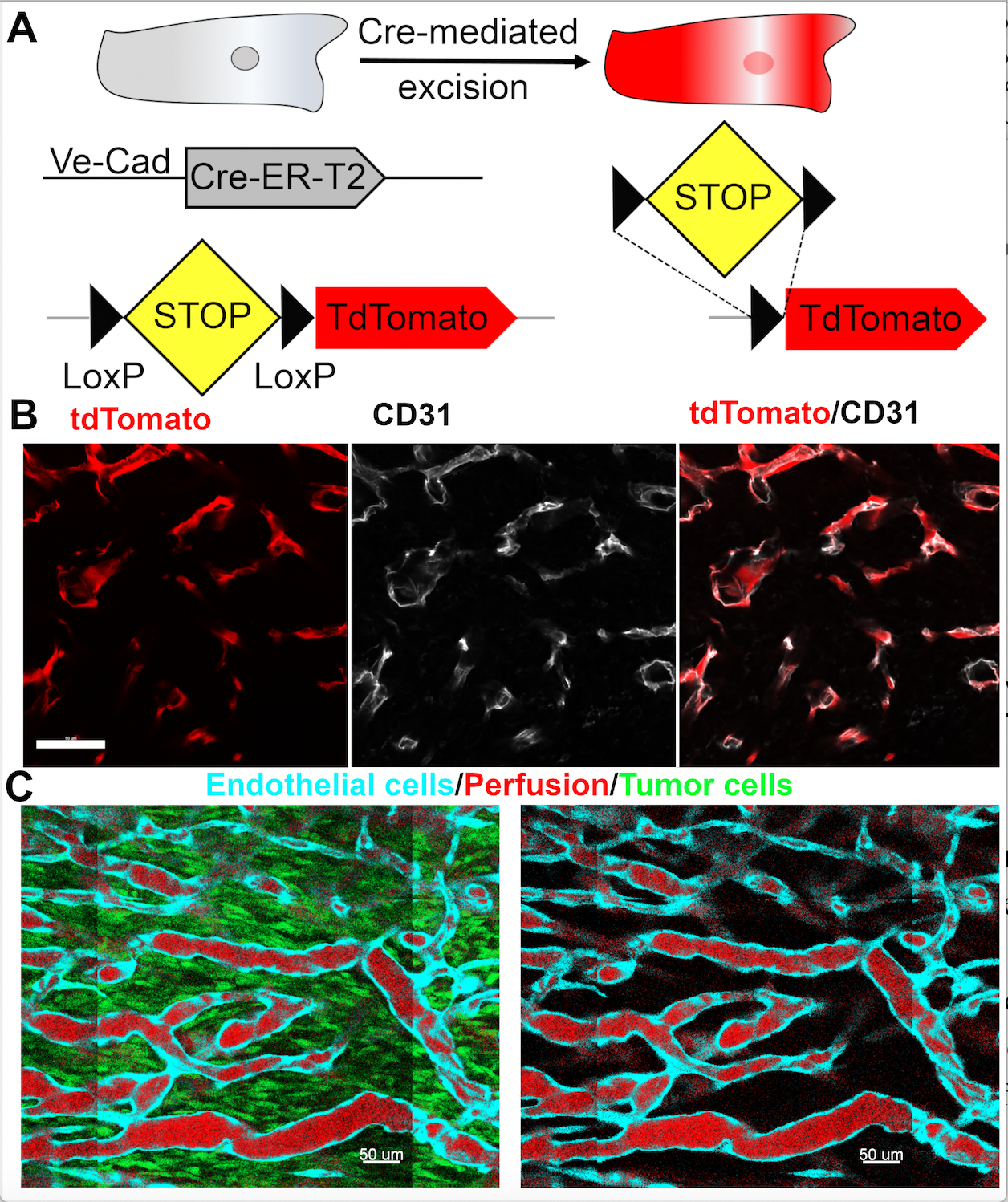}
\caption{{\bf tdTomato expression in ECs and TECs in VE-TOM mice.} {\bf A)} Schematic of the Ve-Cad (Cre-ERT2) system. Administration of tamoxifen by gavaging in adult VE-TOM mice activates the Cre-LoxP system in endothelial cells inducing tdTomato expression. 
{\bf B)} TECs expressing tdTomato (red), co-stained for CD31 (white). {\bf C)} TECs expressing tdTomato (red), co-stained for CD31 (white). Representative image of a MC38 tumor. GFP positive tumor cells (green), TECs (cyan), and infused Qdots (red) indicating perfused vessels. Scale bar in B and C is 50$\mu$m.}\label{Fig:EC}
\end{figure}

In order to visualize the response of the tumor vasculature, we used a transgenic mouse model in which the fluorescent protein tdTomato is expressed in both normal and tumor endothelial cells (EC). We used transgenic mice bearing a Cre recombinase-tamoxifen receptor fusion protein (Cre-ERT2) driven by the VE cadherin promoter. These mice were crossed with Gt(ROSA)26Sortm9(CAG-tdTomato)Hze mice so that activation of Cre by tamoxifen resulted in EC expression of tdTomato (schematic shown in Figure~\ref{Fig:EC}A). For imaging purposes we only used mice with greater than 95\% fluorescent EC. The segmentation of tumor blood vessels was based on the TECs expressing tdTomato. We used intravenous injection of Qdots 705 to distinguish perfused from non-perfused tumor vessels, i.e. vessels labelled with the infused Qdots and vessels not labelled with it.  As further evidence, we note that no Qdot positive, endothelial negative vessels were identified. If the Qdots were in the lymphatics then they would have identified vessels not lined by vascular endothelium; this did not happen.

\subsection*{Tortuosity in the ultramicroscopy data}\label{Sec:Tortuosity}

\begin{figure}[ht!]
 \centering
\subcaptionbox{}{\centering\includegraphics[height=.3\textwidth]{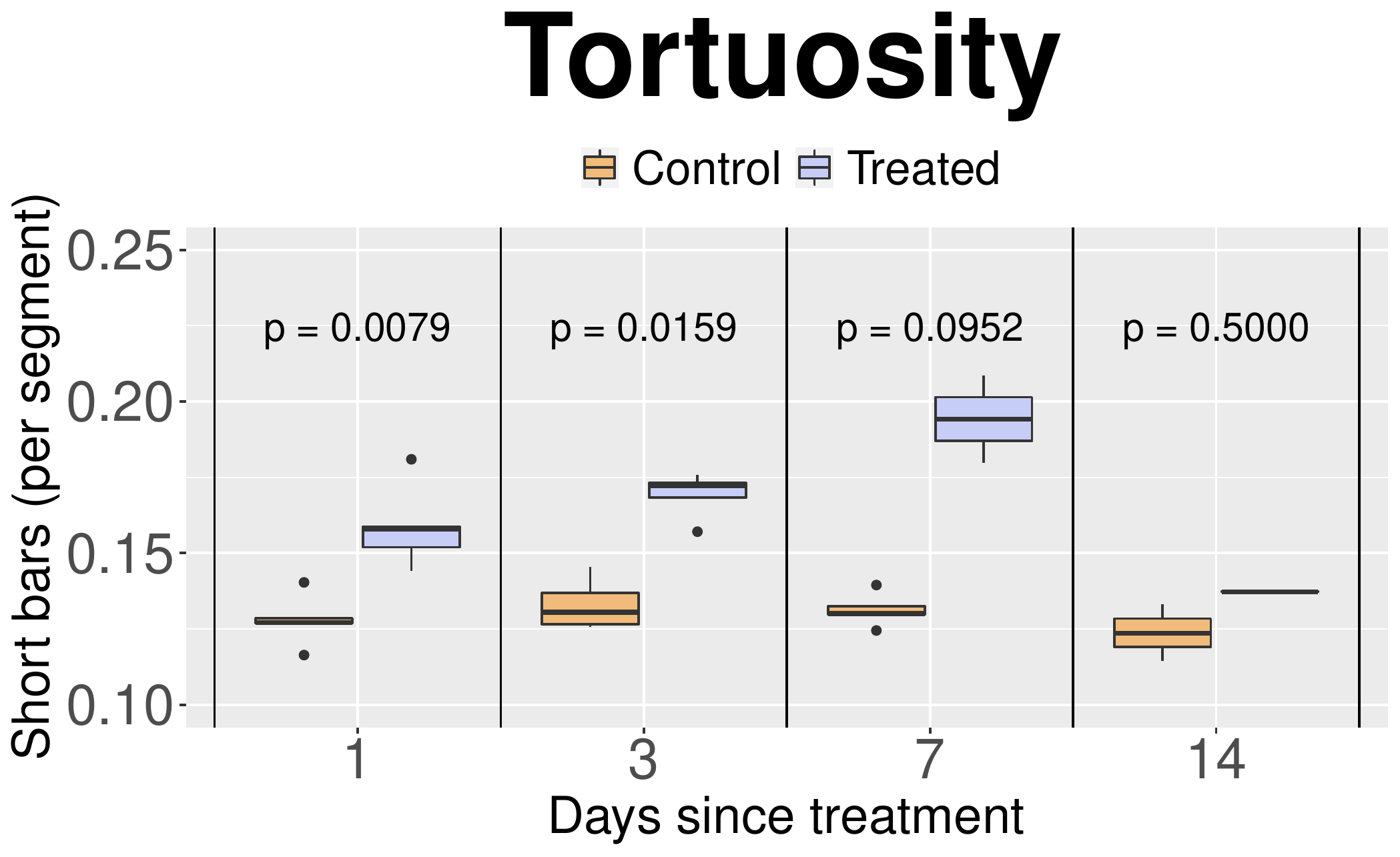}}%
\vspace{0.01\textheight}
\subcaptionbox{}{\centering\includegraphics[height=.3\textwidth]{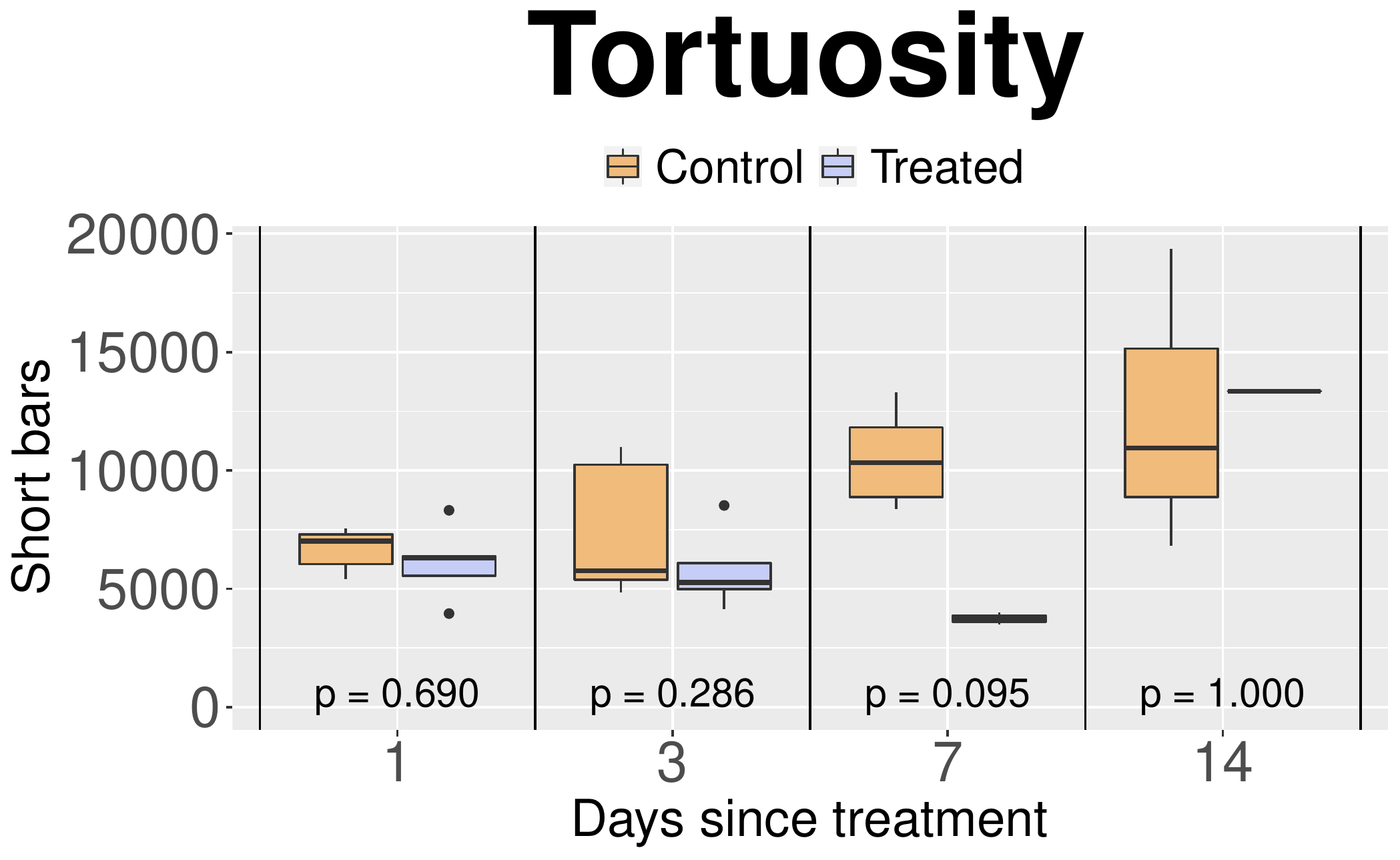}}%
\caption{{\bf Tortuosity in the ultramicroscopy dataset. a)} Tortuosity computed as number of short bars ($\leq 10\%$ of maximal radius used in the radial filtration) in dimension 0 barcode per vessel segment. {\bf b)} Tortuosity computed as number of short bars ($\leq 10\%$ of maximal radius used in the radial filtration) in dimension 0 barcode.}\label{Fig:RocheTortuosityDiscussion}
\end{figure}

The tortuosity descriptor is defined as the ratio of the number of short bars ($\leq 10\%$ of maximal radius used in the radial filtration) in dimension 0 barcodes to the number of vessel segments.
We divided by the number of vessel segments to ensure the contributions are topological and are not masked by an increase or decrease of vasculature. However, in the case of bevacizumab in the ultramicroscopy data, the significant decrease in  number of vessel segments~\cite{Dobosz2014}, which is also visually apparent when looking at examples of extracted vascular networks (see Fig.~\ref{fig:ExampleImagesRoche}), leads to a seemingly contradictory increase of tortuosity (see Figure~\ref{Fig:RocheTortuosityDiscussion}a). This is supported by a correlation between our tortuosity measure and the size of voids which we observed in this dataset (see Fig.~\ref{Fig:RocheCor}). When considering the raw number of short bars without dividing by the number of vessel segments, we observe the expected effect of bevacizumab on tortuosity (see Figure~\ref{Fig:RocheTortuosityDiscussion}b).

As discussed in Section \nameref{Sec:ComputData}, the vascular networks in the ultramicroscopy data are much larger than in the intravital dataset. However, they are less well resolved in the $xy$-plane (see Data description in Methods). This has two consequences on our analysis of tortuosity:
1) the tortuosity of vessels is likely to be captured to a lesser degree than in the intravital data, 2) the number of filtration steps needed to be able to capture tortuosity adequately would need to be significantly higher than the 500 used in the radial filtration.
Indeed, example images from this data (see Fig.~\ref{fig:ExampleImagesRoche}) do not appear to show strikingly tortuous vessels. Moreover, our computation of the radial filtration in 500 steps was already at the edge of computational feasibility (see Section \nameref{Sec:ComputData}). Thus further refinement of the filtration is not possible. Alternatively, we can observe the coarse trends change over
time without a normalisation by the number of vessel segments as shown in Figure~\ref{Fig:RocheTortuosityDiscussion}b.
While our topological descriptor therefore quantified a genuine and significant change in the vascular networks on the ultramicrocopy data, its interpretation here needs to be made with care.

\subsection*{Alternative results figures and statistical analysis}\label{Sec:ResStats}

\begin{figure}[ht!]
 \centering
\includegraphics[width=\textwidth]{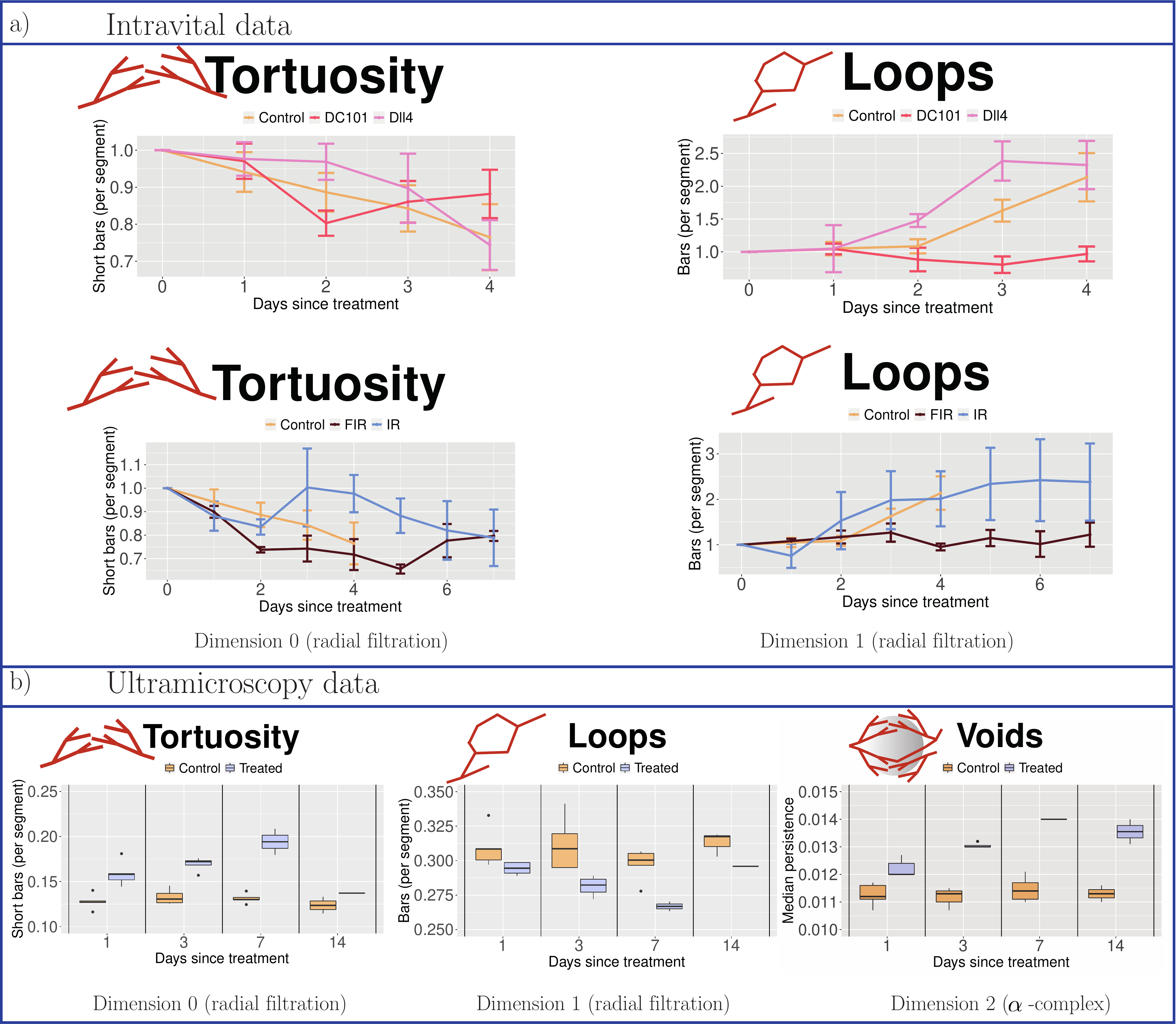}
\caption{{\bf Topological descriptors extracted from tumour blood vessel networks treated with vascular targeting agents with known effects II.} {\bf a)} Intravital data results. We normalised all descriptors with respect to values on the day on which treatment is administered (day 0) or, for controls, the day on which observations commence (day 0). 
Data was collected from control mice (beige), mice treated with the vascular targeting agent DC101~\cite{Kannan2018} (dark pink), mice treated with vascular targeting agent anti-Dll4~\cite{Liu2011} (light pink), mice treated with fractionated irradiation (FIR, brown), and mice treated with single dose irradiation (IR,blue).
Tortuosity was computed as the ratio of short bars ($\leq 10\%$ of maximal radius used in the radial filtration) in the dimension 0 barcodes of the radial filtration to the number of vessel segments. Loops are the number of bars in the dimension 1 barcodes of the radial filtration per vessel segment in the network. 
{\bf b)} Ultramicroscopy data results. Due to the snapshot nature of the data (one time point per tumour), all reported topological descriptors are raw values.
Data was collected from control mice (beige) and mice treated with bevacizumab (purple). We computed tortuosity values and the number of vessel loops per vessel segment, in the same way as for the intravital data. 
We also determined the size of voids (avascular regions) by computing the median length of bars in the dimension 2 barcodes of the $\alpha$-complex filtration.}\label{Fig:ResultsLinePlots}
\end{figure}

We show alternative representations of our results from the main text. In Fig.~\ref{Fig:ResultsLinePlots} we present the results for the intravital data as mean time series for each treatment group with error bars (standard error of the mean) to highlight that our data is dynamic over time.

\begin{figure}[ht!]
 \centering
\includegraphics[width=\textwidth]{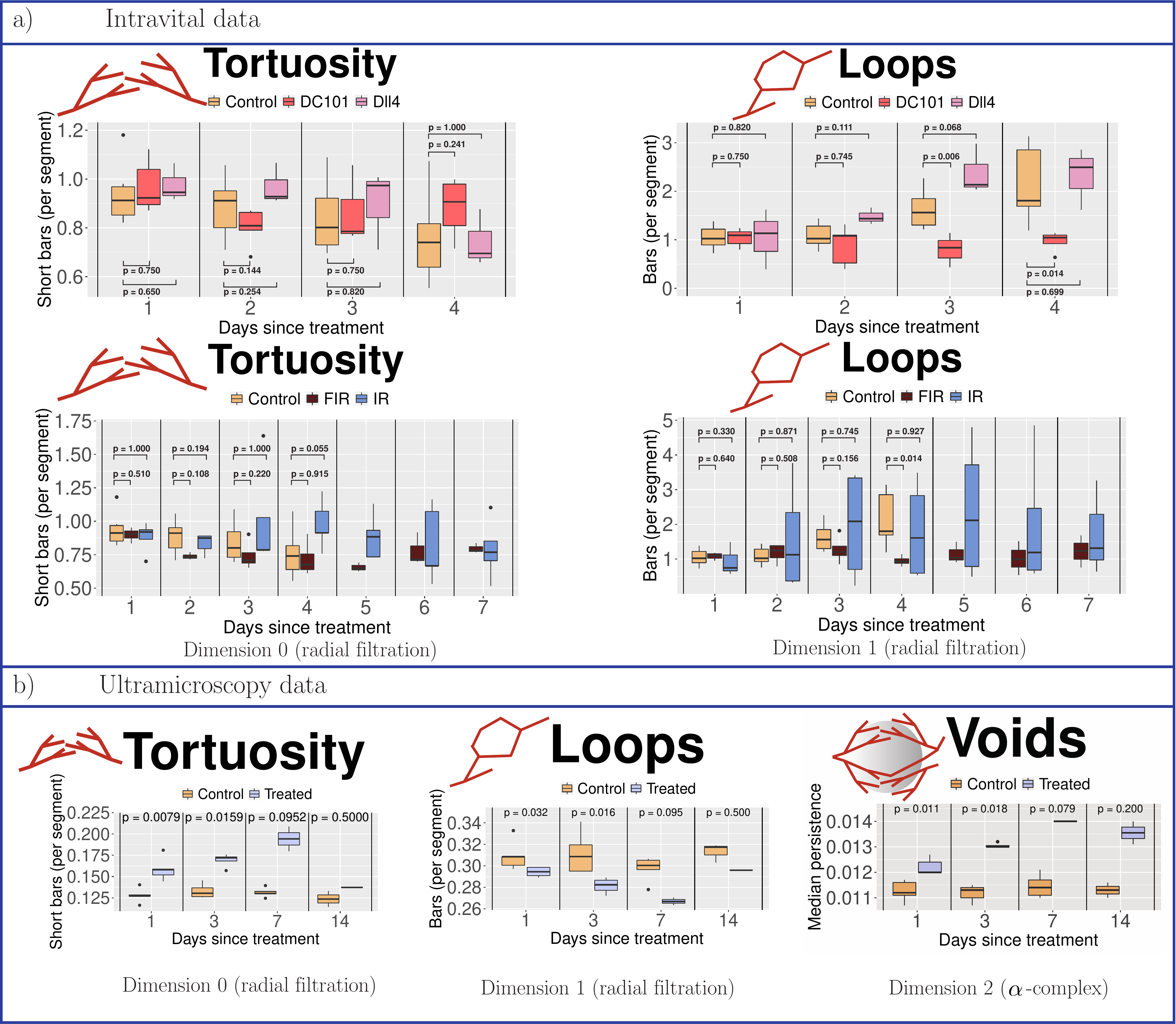}
\caption{{\bf Topological descriptors extracted from tumour blood vessel networks treated with vascular targeting agents with known effects III.} {\bf a)} Intravital data results. We normalised all descriptors with respect to values on the day on which treatment is administered (day 0) or, for controls, the day on which observations commence (day 0). 
Data was collected from control mice (beige), mice treated with the vascular targeting agent DC101~\cite{Kannan2018} (dark pink), mice treated with vascular targeting agent anti-Dll4~\cite{Liu2011} (light pink), mice treated with fractionated irradiation (FIR, brown), and mice treated with single dose irradiation (IR,blue).
Tortuosity was computed as the ratio of short bars ($\leq 10\%$ of maximal radius used in the radial filtration) in the dimension 0 barcodes of the radial filtration to the number of vessel segments. Loops are the number of bars in the dimension 1 barcodes of the radial filtration per vessel segment in the network. 
{\bf b)} Ultramicroscopy data results. Due to the snapshot nature of the data (one time point per tumour), all reported topological descriptors are raw values.
Data was collected from control mice (beige) and mice treated with bevacizumab (purple). We computed tortuosity values and the number of vessel loops per vessel segment, in the same way as for the intravital data. 
We also determined the size of voids (avascular regions) by computing the median length of bars in the dimension 2 barcodes of the $\alpha$-complex filtration.}\label{Fig:ResultsPValues}
\end{figure}

\noindent In Fig.~\ref{Fig:ResultsPValues} we present our results including $p$-values from our statistical analysis. We compute the (non-exact and unadjusted) $p$-values for the intravital data using the {\sc R} function {\tt pairwise.wilcox.test()} in {\sc RStudio}~\cite{RStudio-Team:2016aa} to perform a pairwise Wilcoxon's rank sum test between the control group and each of the treatment groups. For the ultramicroscopy data we use the function {\tt stat\_compare\_means()} from the library {\tt ggpubr} to perform Wilcoxon's rank sum test. All our tests are by default two-sided.

\begin{figure}[ht!]
\centering \subcaptionbox{Radial interval I.}{\centering\includegraphics[width=.36\textwidth]{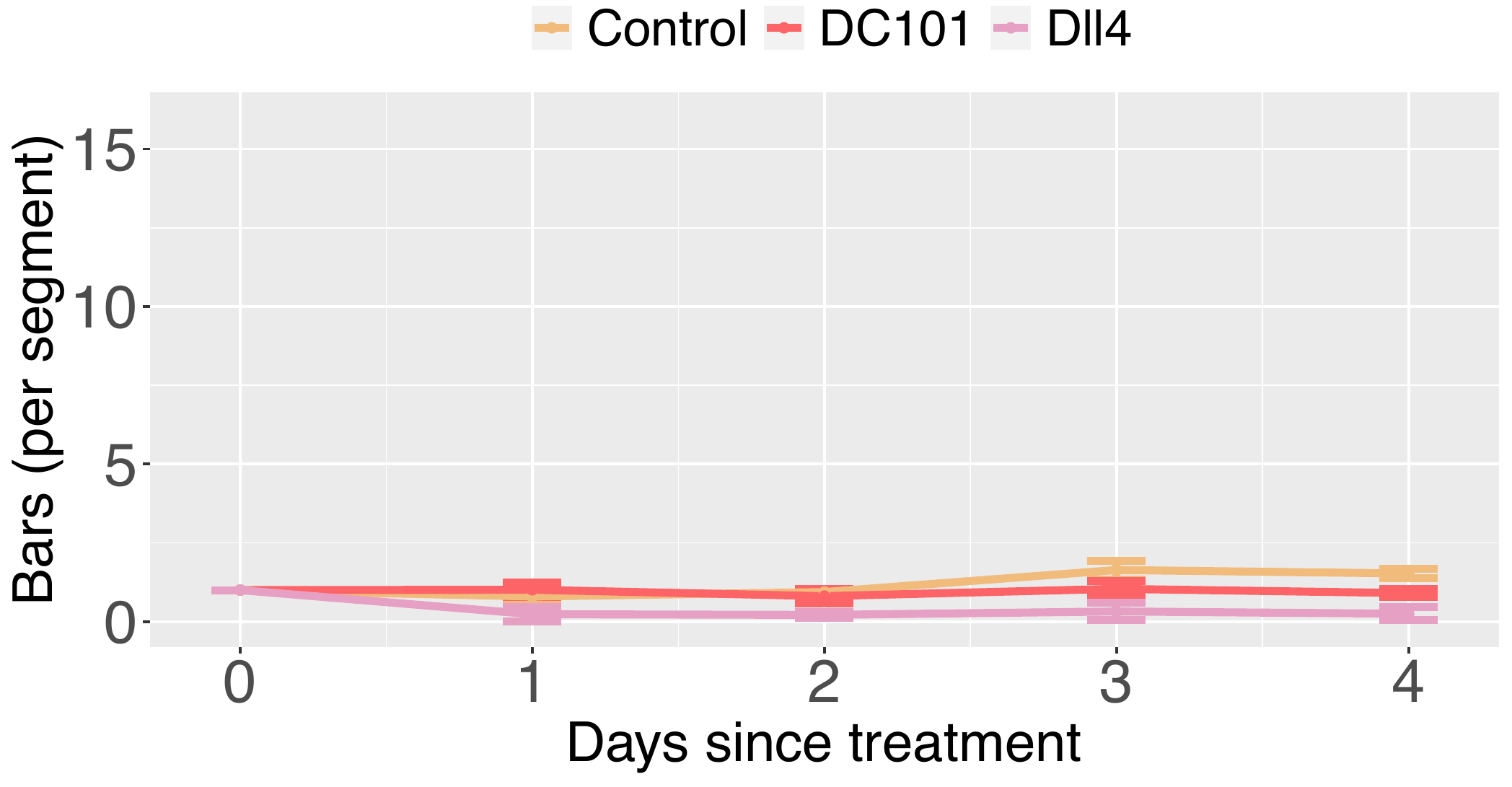}\hspace{1.5cm} \includegraphics[width=.36\textwidth]{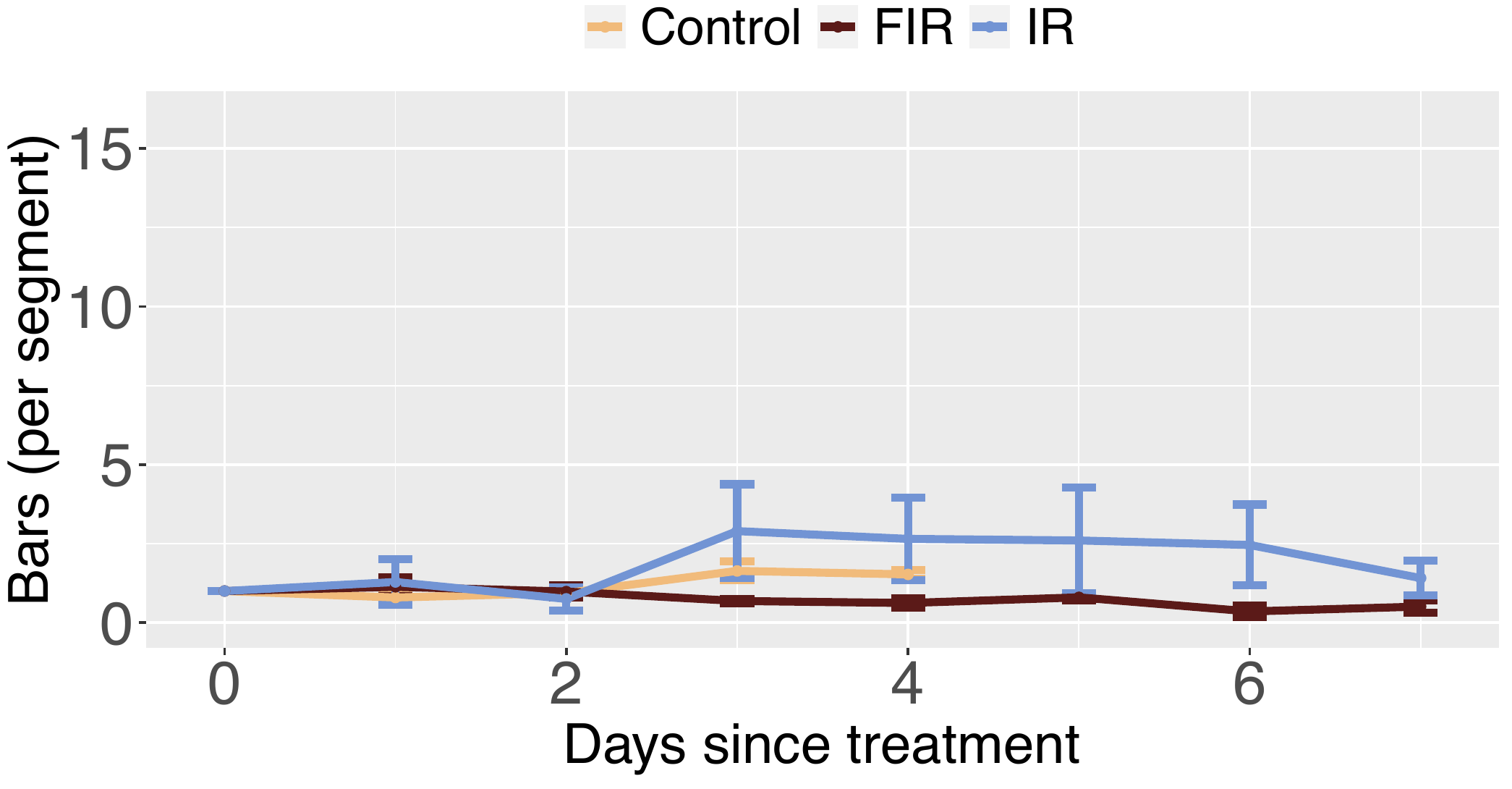}}%
\vspace{0.01\textheight}
\subcaptionbox{Radial interval II.}{\centering\includegraphics[width=.36\textwidth]{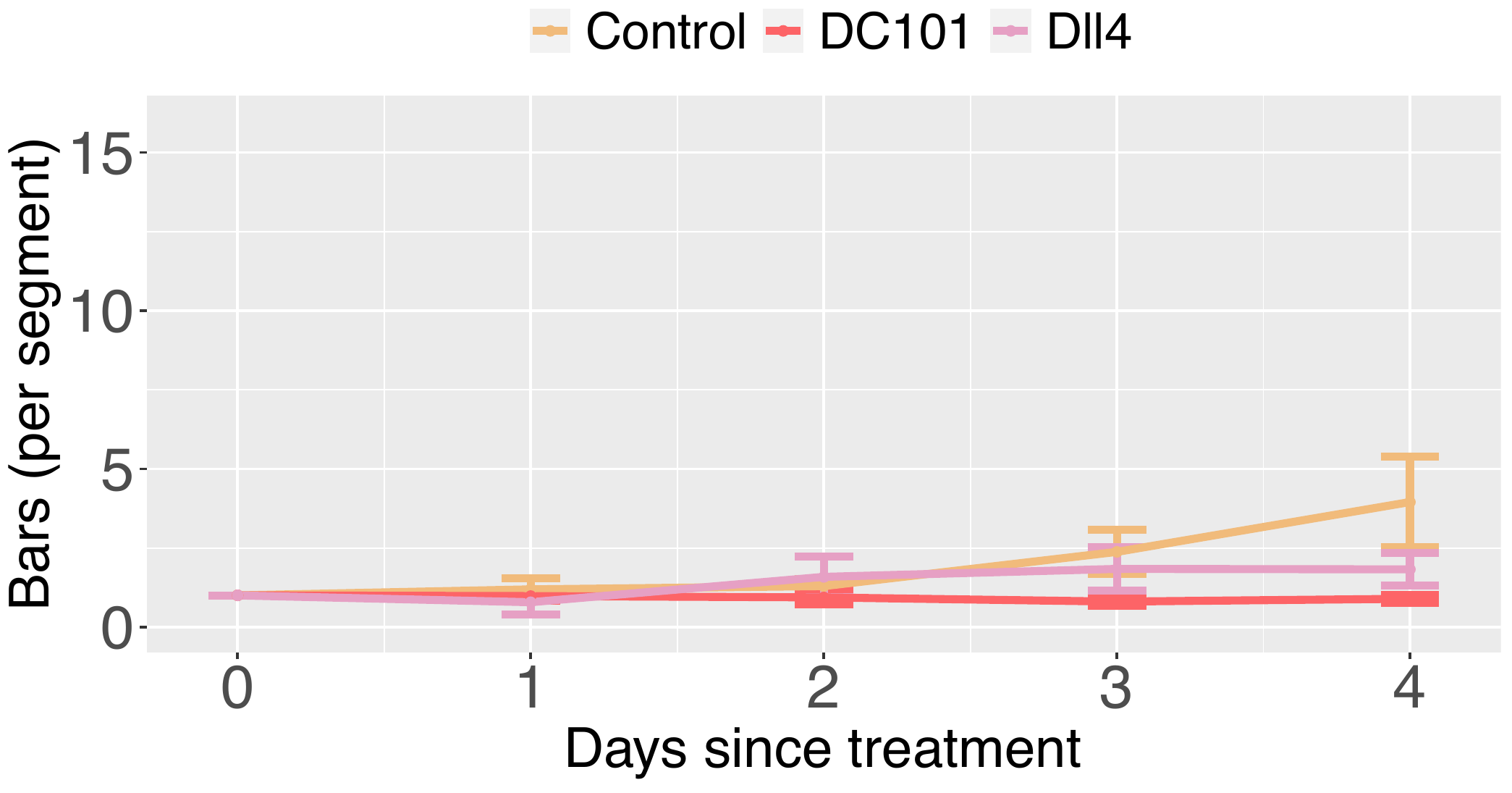}\hspace{1.5cm}\includegraphics[width=.36\textwidth]{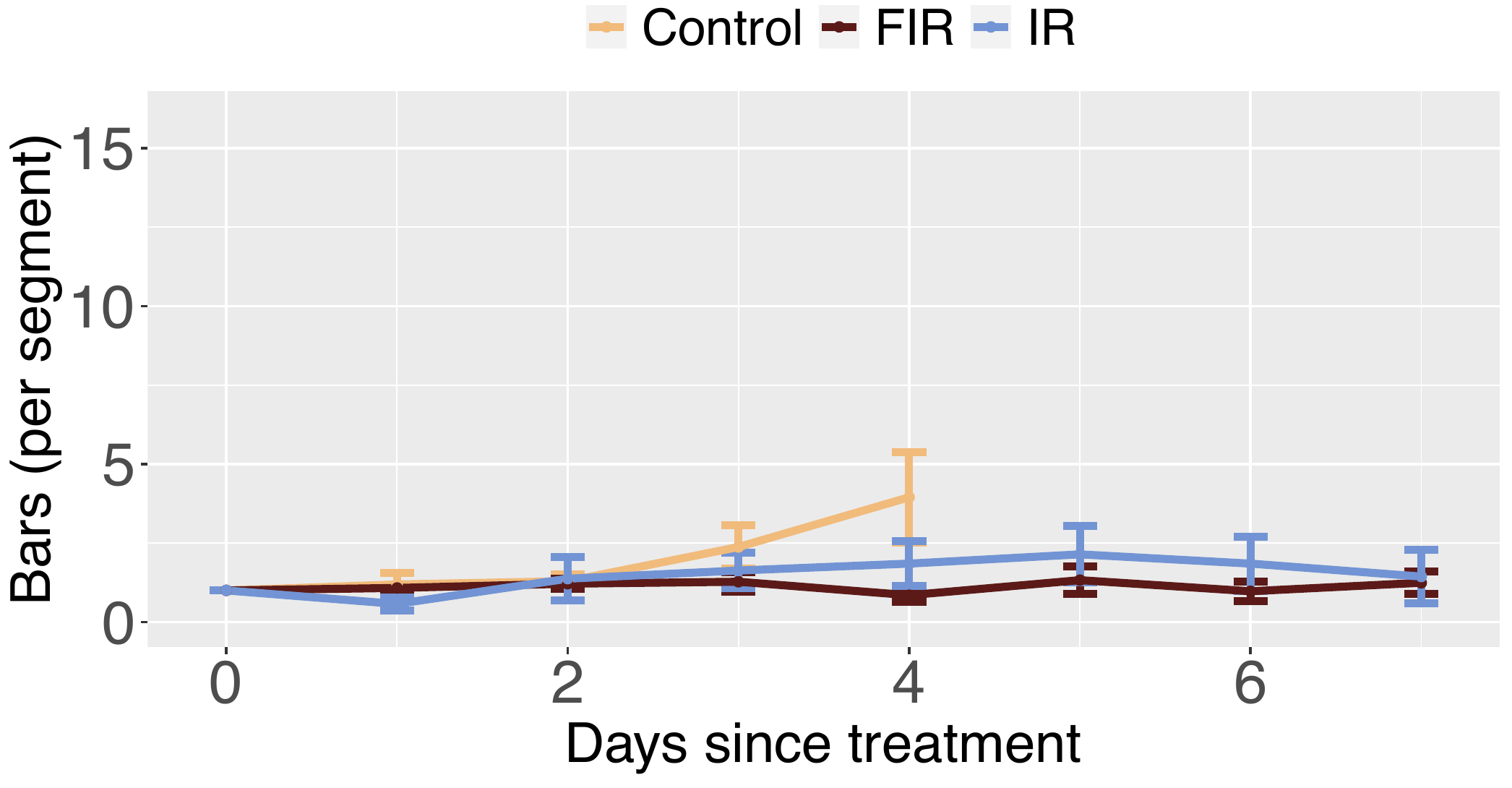}}%
\vspace{0.01\textheight}
\subcaptionbox{Radial interval III.}{\centering\includegraphics[width=.36\textwidth]{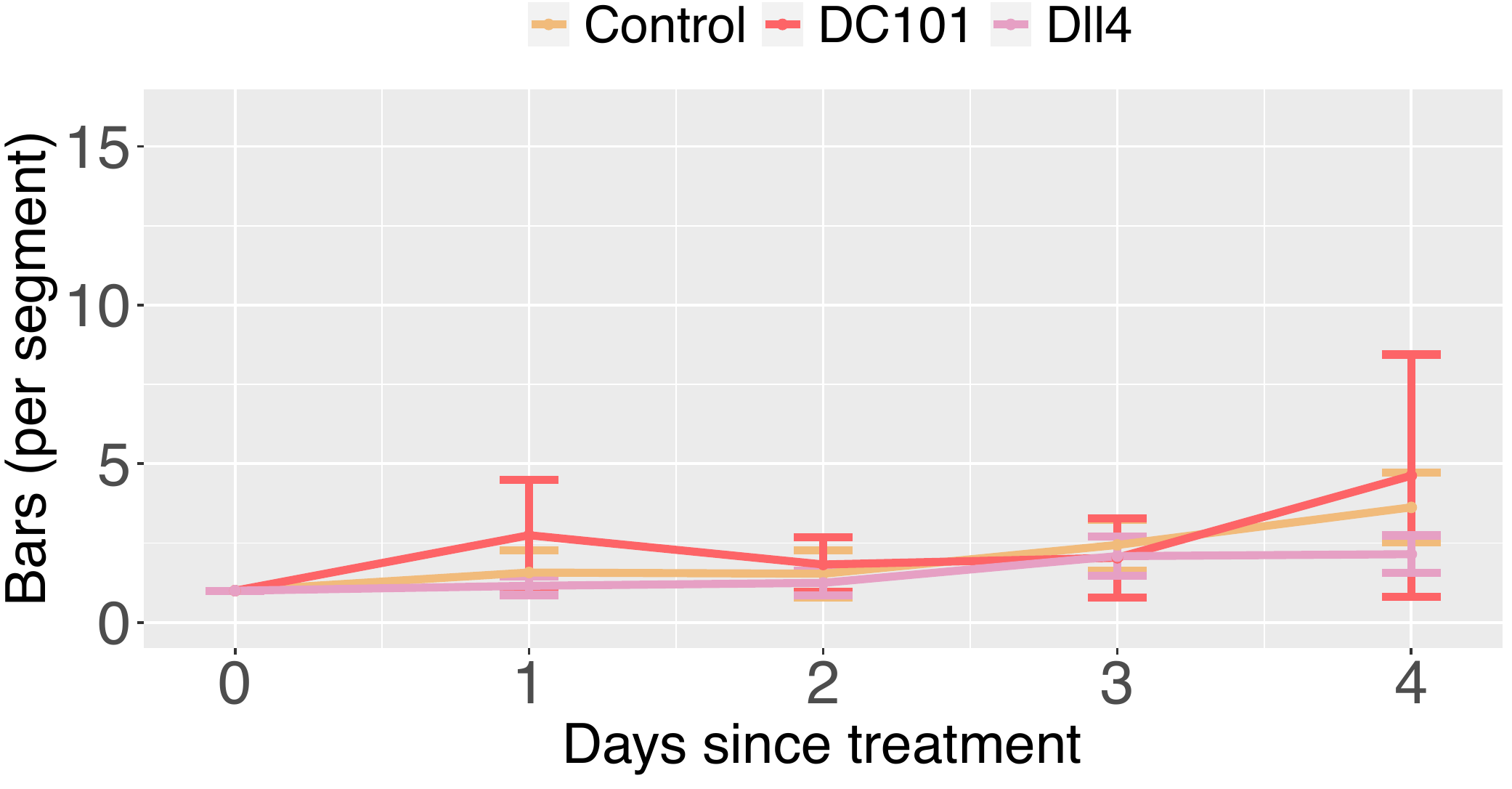}\hspace{1.5cm}\includegraphics[width=.36\textwidth]{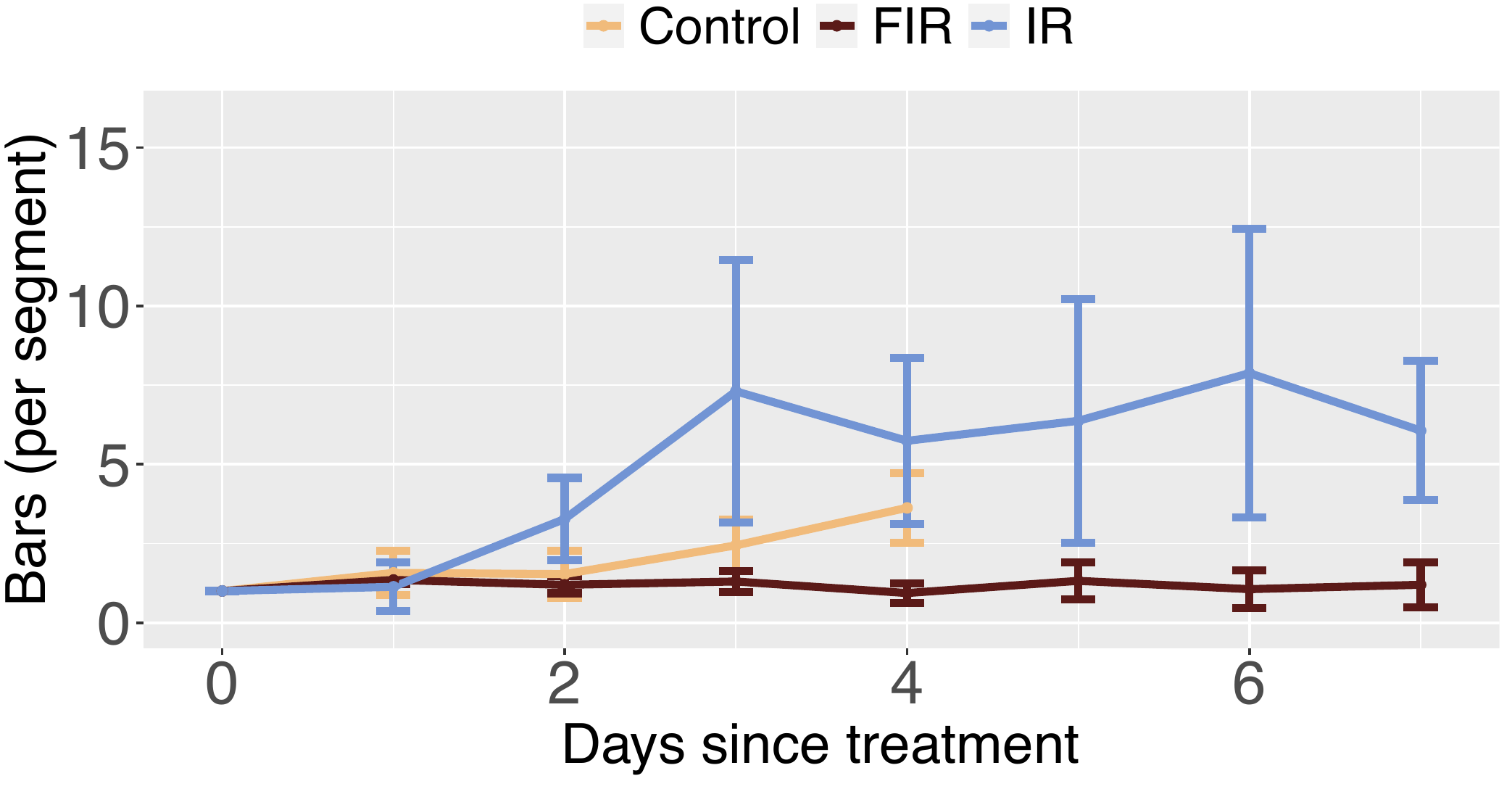}}%
\vspace{0.01\textheight}
\subcaptionbox{Radial interval IV.}{\centering\includegraphics[width=.36\textwidth]{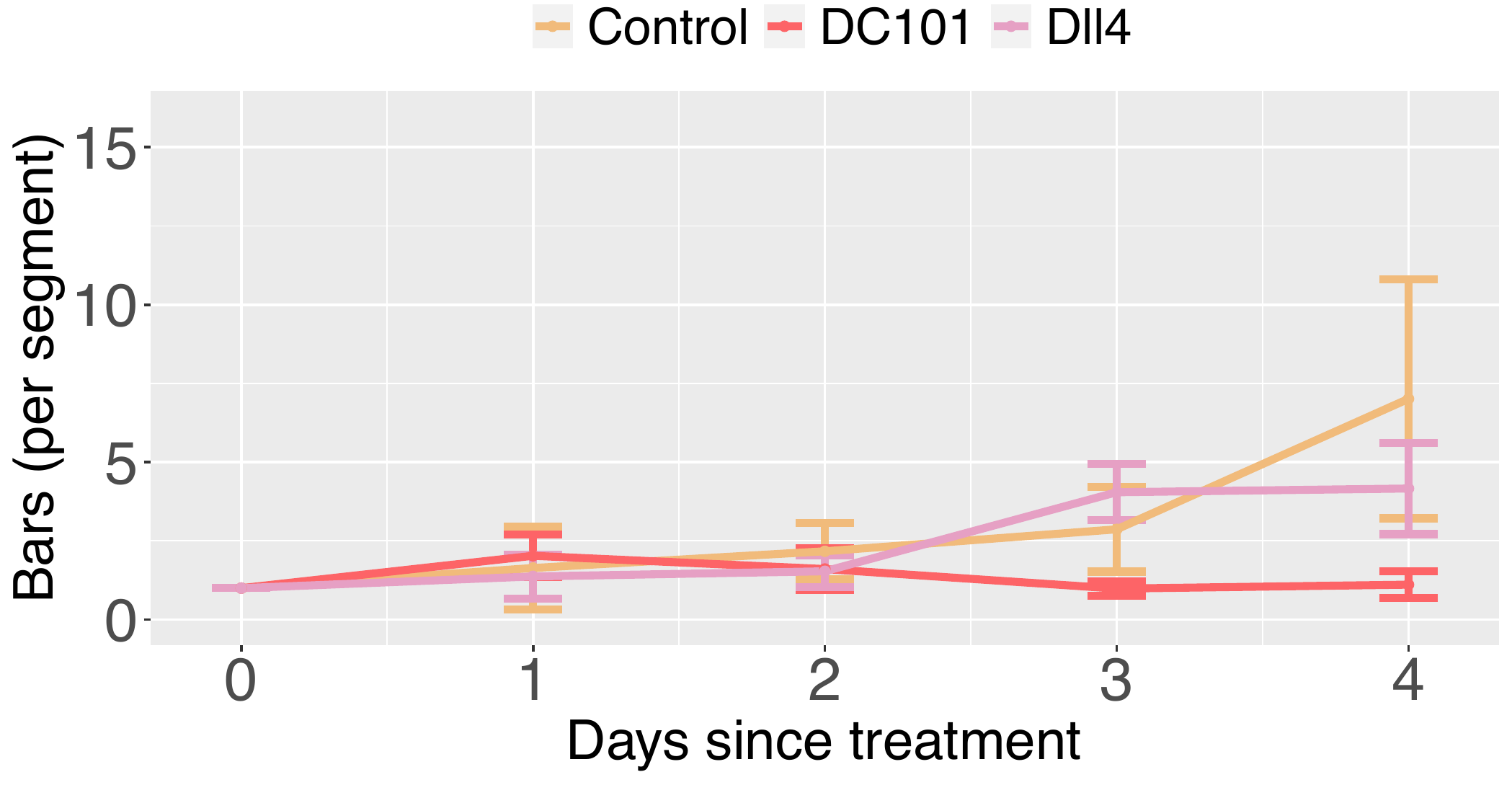}\hspace{1.5cm}\includegraphics[width=.36\textwidth]{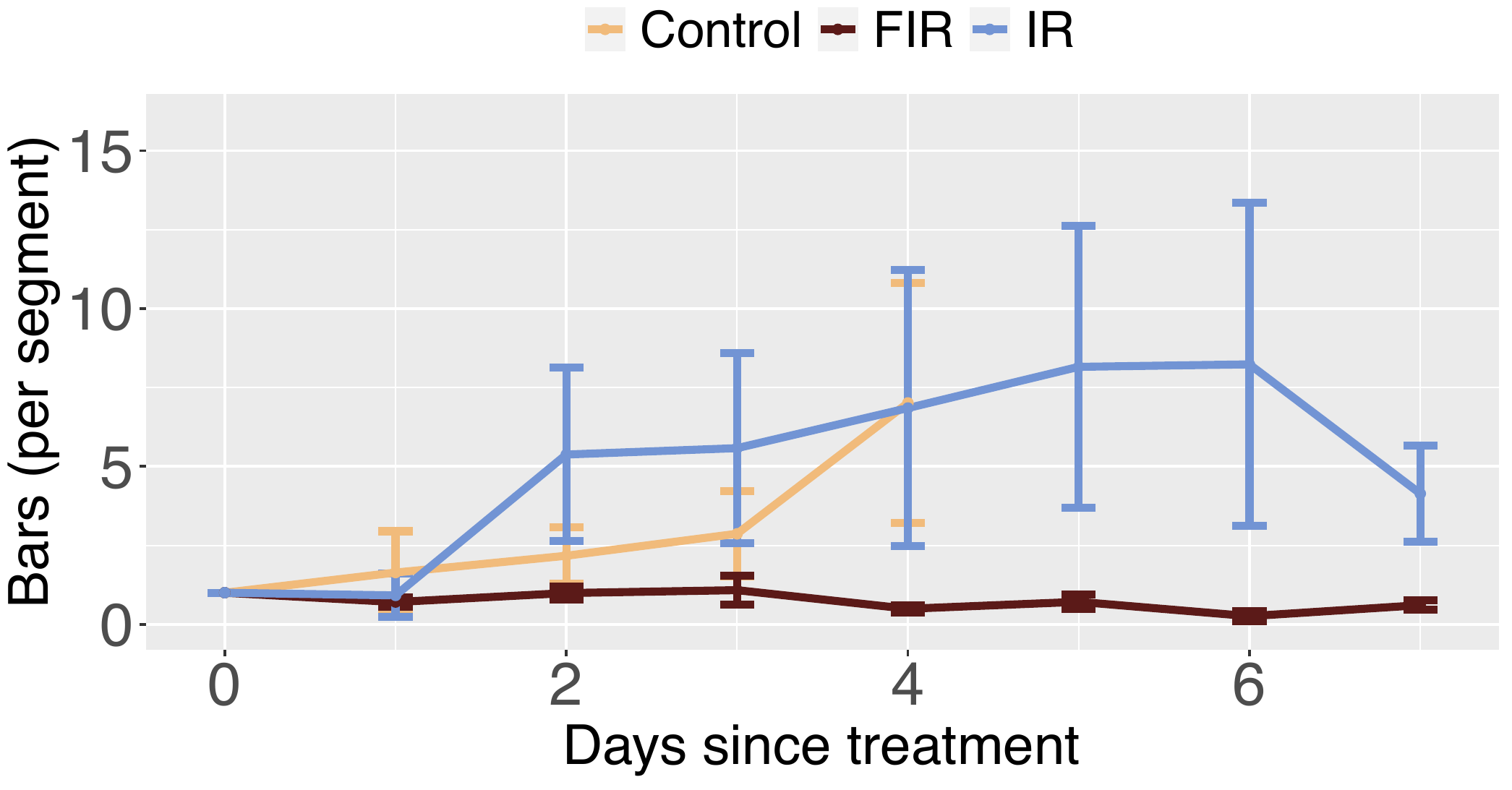}}%
\vspace{0.01\textheight}
\subcaptionbox{Radial interval V.}{\centering\includegraphics[width=.36\textwidth]{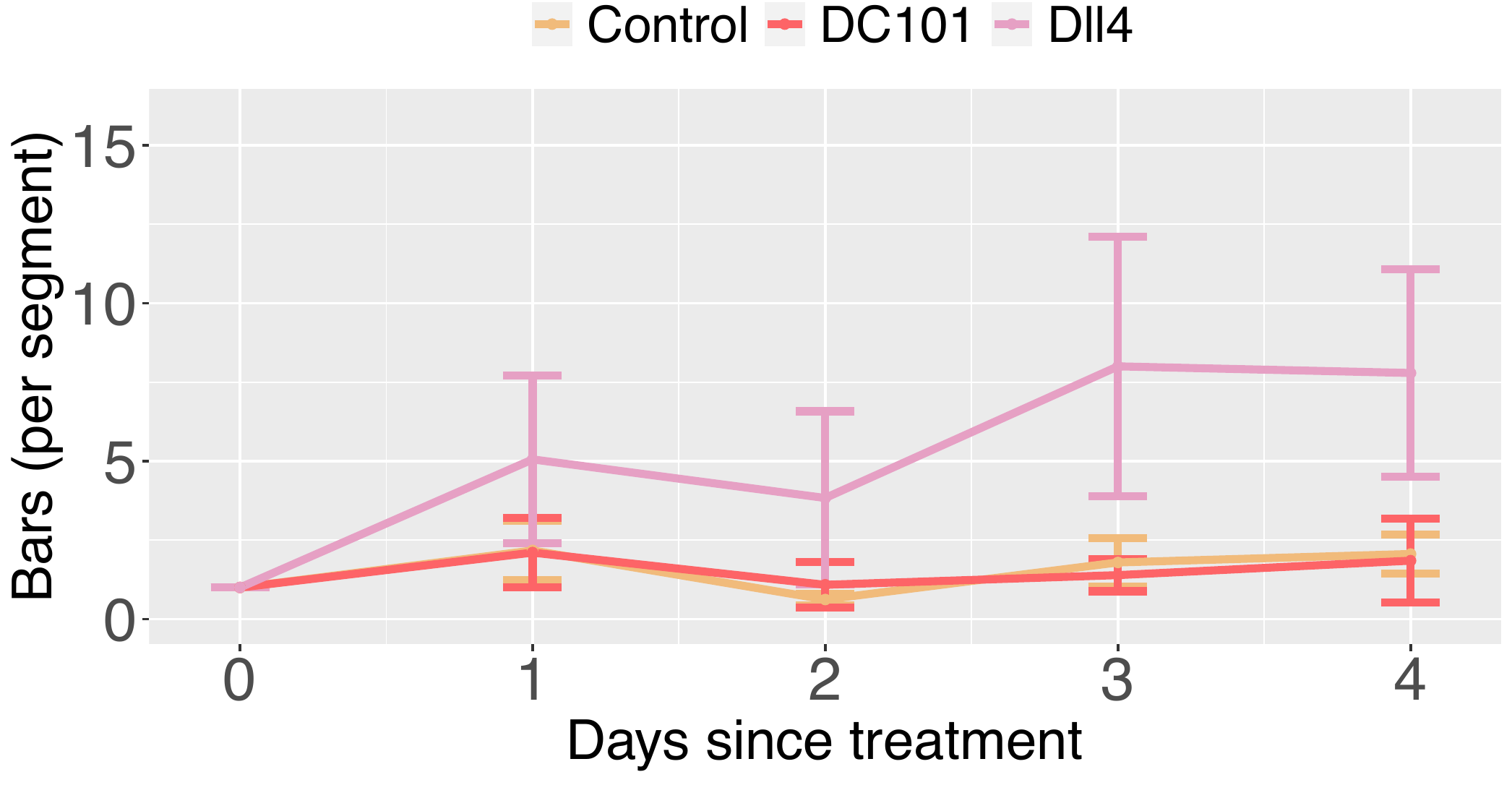}\hspace{1.5cm}\includegraphics[width=.36\textwidth]{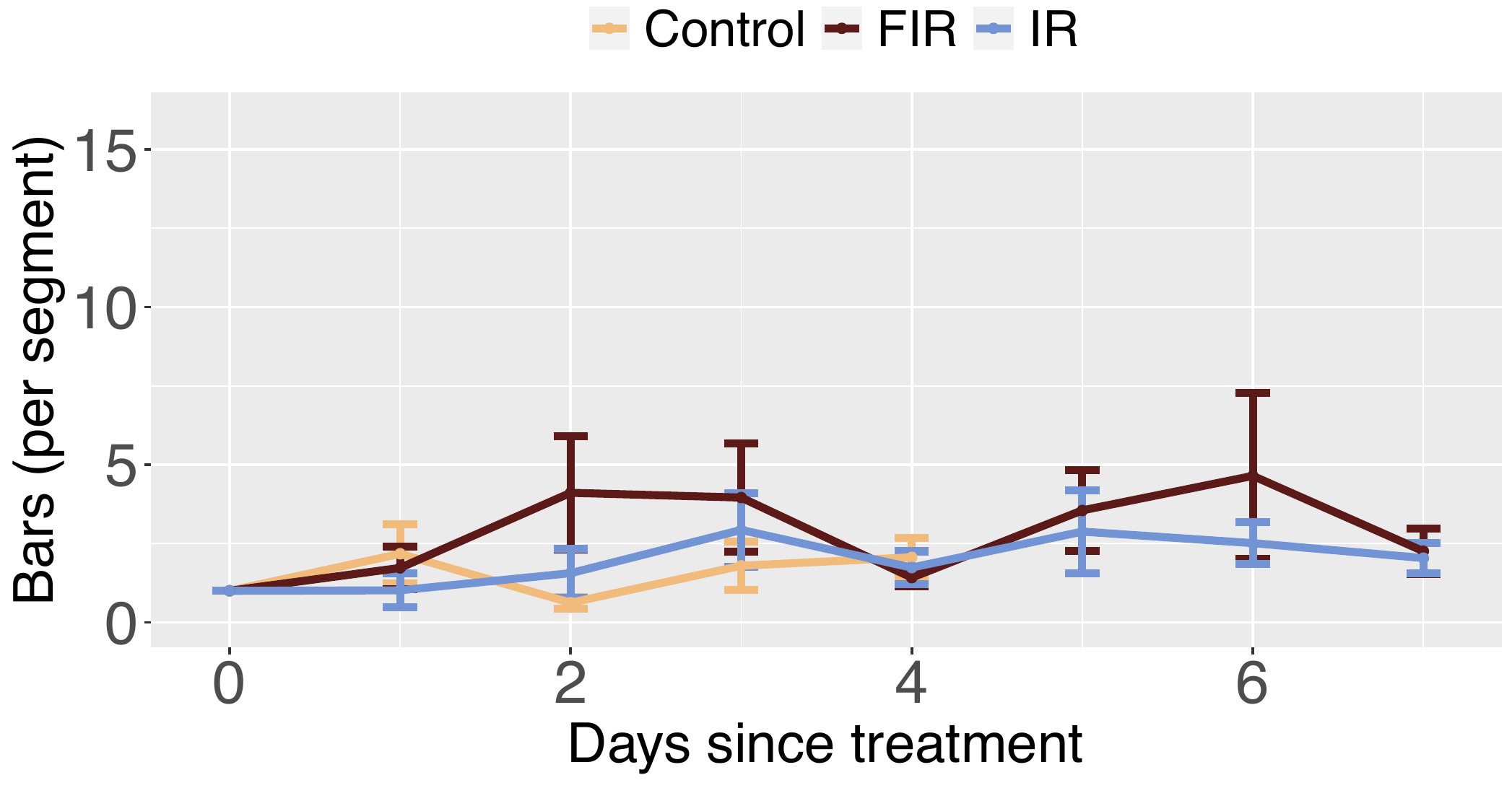}}%
\caption{{\bf Mean time series of the normalised number of loops per vessel segment for different filtration intervals in the intravital dataset.} We show the mean of the number of loops normalised by day 0 and standard error of the mean in different radial intervals. Interval I corresponds to the radial region closest to the tumour centre, while Interval V represents parts of the vessel network that are farthest away from the tumour centre. We separate the treatments into two groups to facilitate the distinction of the trends.}
\label{fig:MC38AverageRadialDim1LoopsIntervals}
\end{figure}

\noindent In Fig.~\ref{fig:MC38AverageRadialDim1LoopsIntervals} we present time-series of the spatio-temporal resolution of the intravital data.

\clearpage

\subsection*{Additional results and statistical analysis}\label{Sec:MoreResStats}

\begin{figure}[ht!]
\centering \includegraphics[width=.9\textwidth]{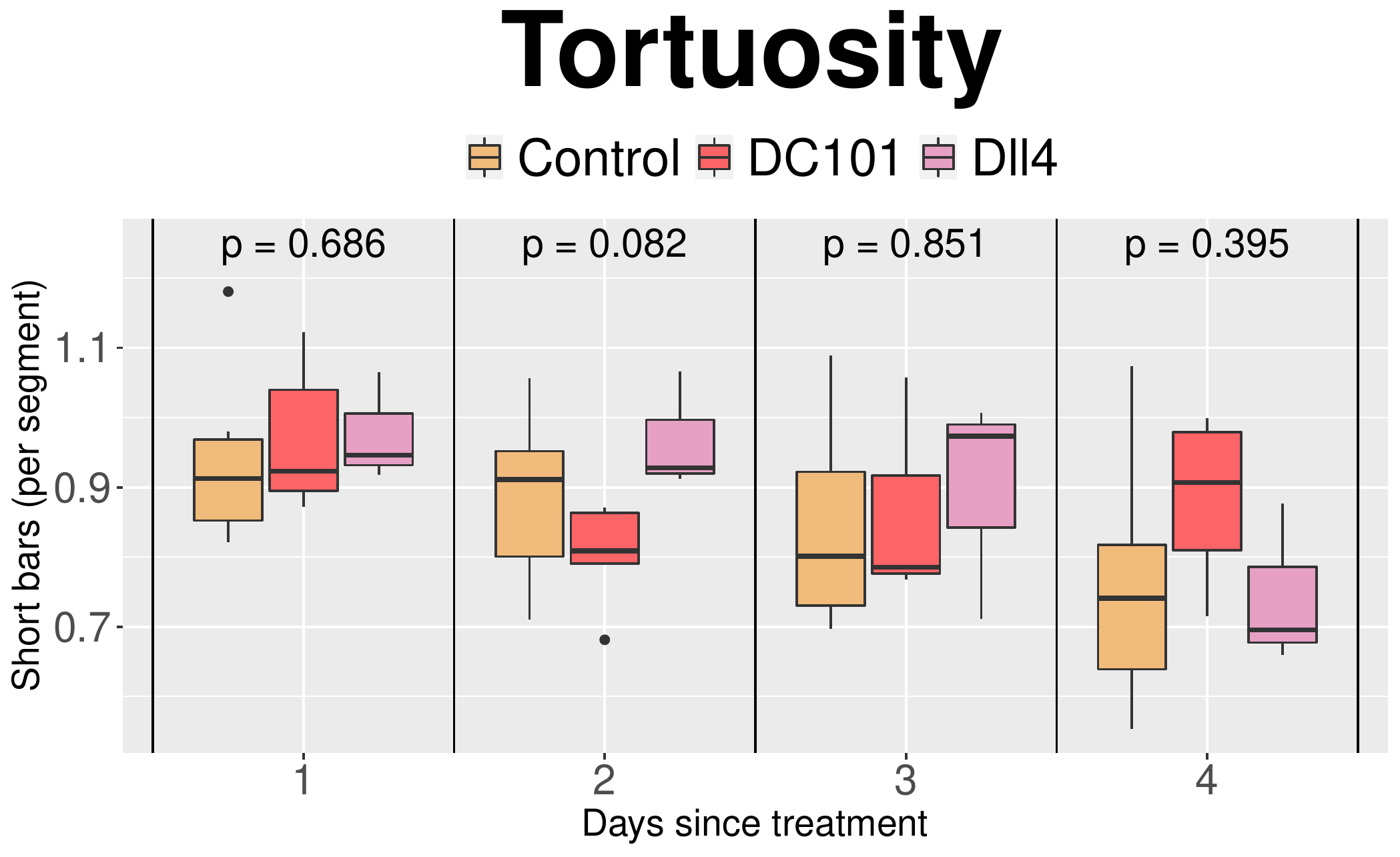}
\caption{{\bf Box plot showing the number of short bars in the dimension 0 barcode of the radial filtration divided by the number of vessel segments.} The values are normalised by day 0 of initial treatment for all treatment regimes to facilitate comparisons of trends over time. We show group level $p$-values according to the Kruskal-Wallis test.}\label{Fig:KruskalMC38TortuosityDC101Dll4}
\end{figure}

\begin{figure}[ht!]
\centering \includegraphics[width=.9\textwidth]{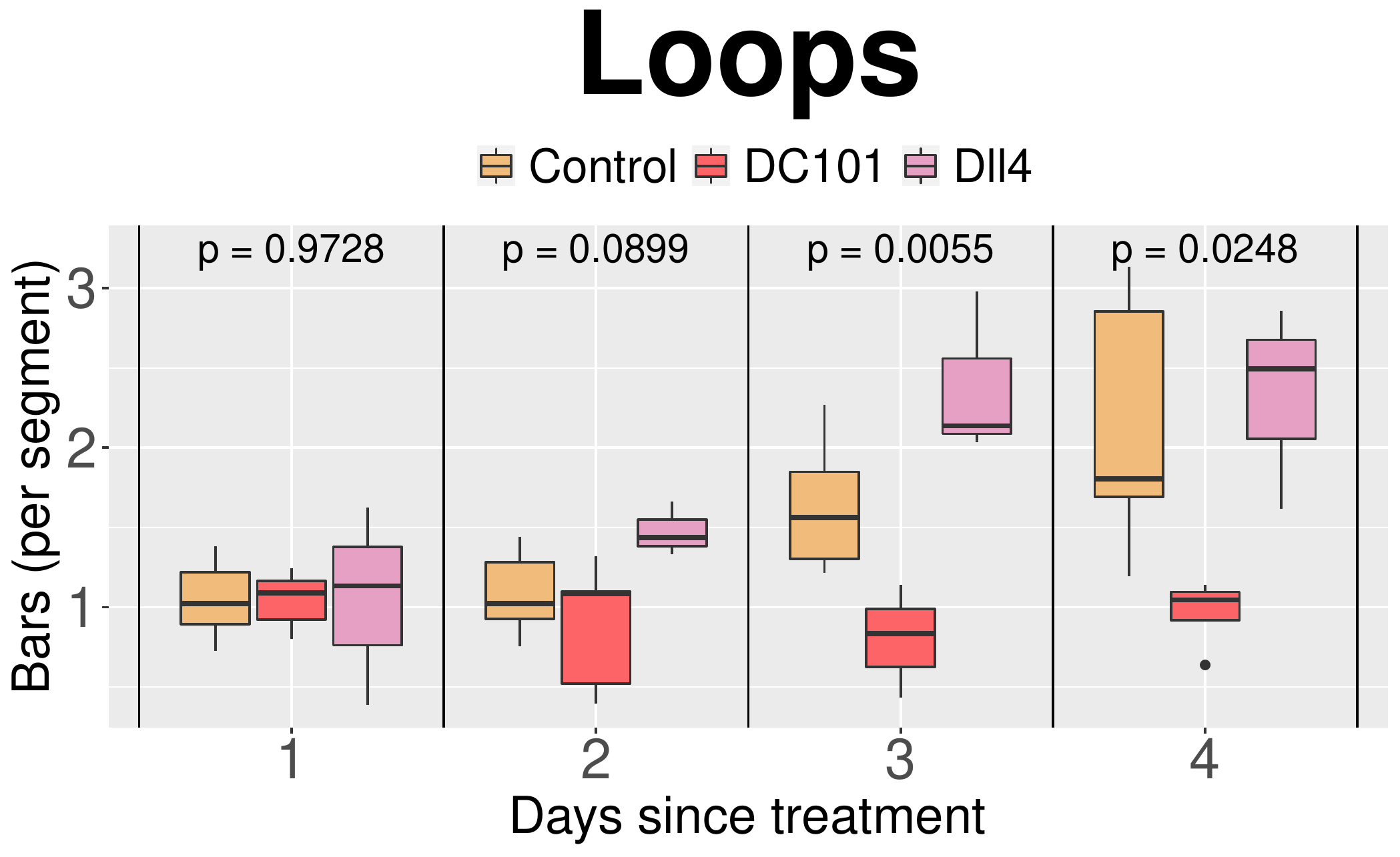}
\caption{{\bf Box plot showing the number of bars in the dimension 1 barcode of the radial filtration divided by the number of vessel segments.} The values are normalised by day 0 of initial treatment for all treatment regimes to facilitate comparisons of trends over time. We show group level $p$-values according to the Kruskal-Wallis test.}\label{Fig:KruskalMC38LoopsDC101Dll4}
\end{figure}

\begin{figure}[ht!]
\centering \includegraphics[width=.9\textwidth]{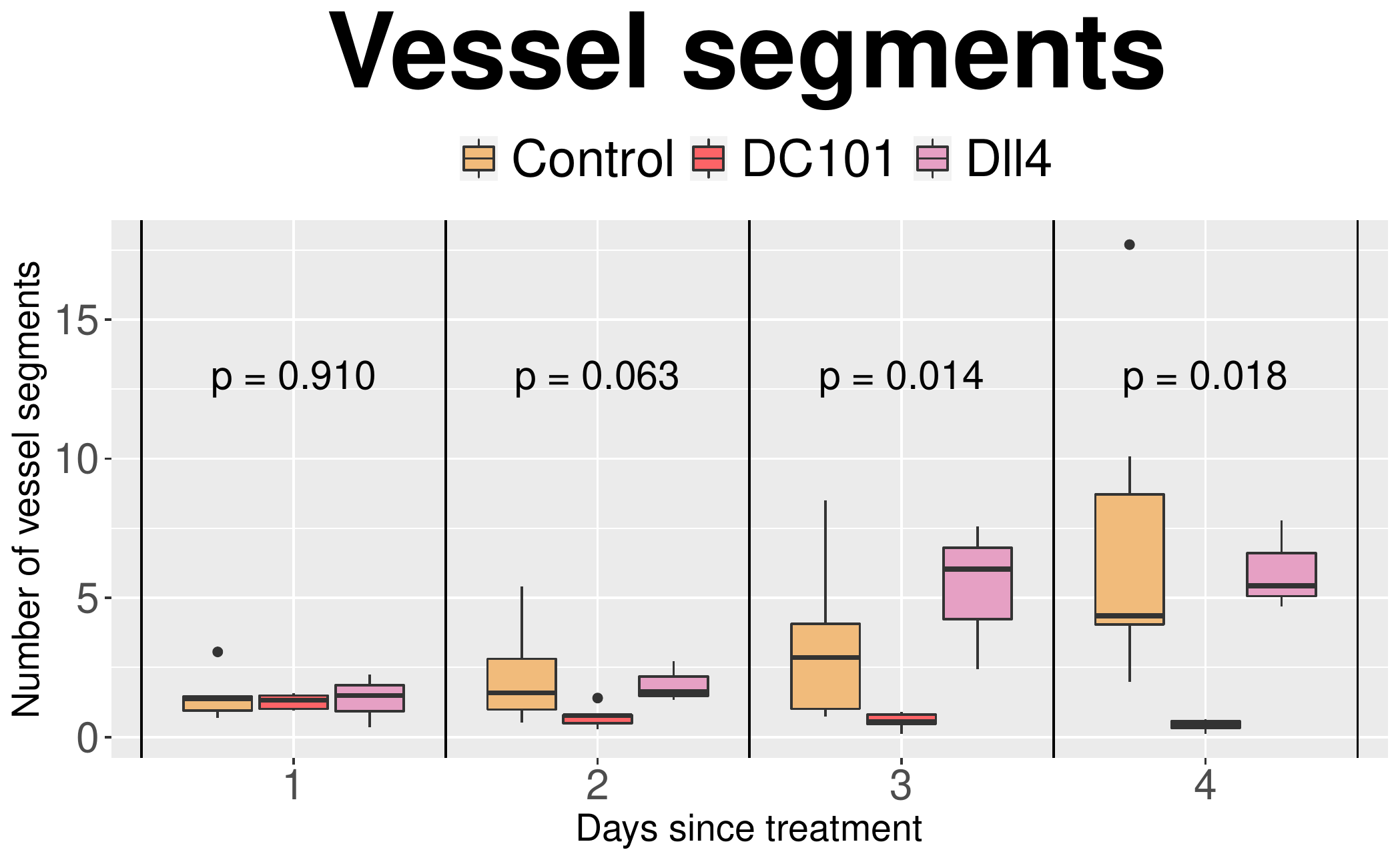}
\caption{{\bf Box plot showing the number of vessel segments.} The values are normalised by day 0 of initial treatment for all treatment regimes to facilitate comparisons of trends over time. We show group level $p$-values according to the Kruskal-Wallis test.}\label{Fig:KruskalMC38SegmentsDC101Dll4}
\end{figure}

\begin{figure}[ht!]
\centering \includegraphics[width=.9\textwidth]{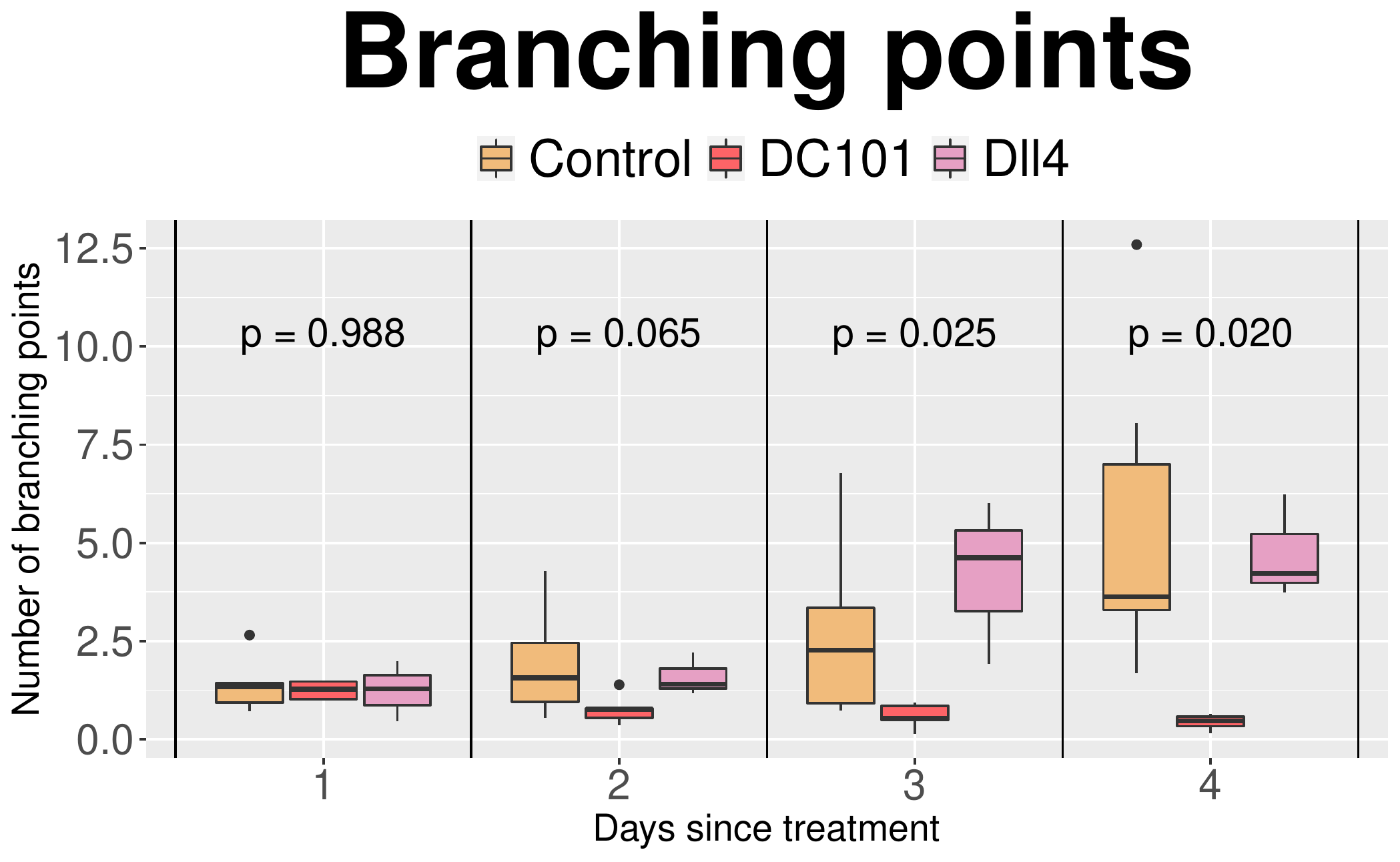}
\caption{{\bf Box plot showing the number of branching points.} The values are normalised by day 0 of initial treatment for all treatment regimes to facilitate comparisons of trends over time. We show group level $p$-values according to the Kruskal-Wallis test.}\label{Fig:KruskalMC38BranchingDC101Dll4}
\end{figure}

\begin{figure}[ht!]
\centering \includegraphics[width=.9\textwidth]{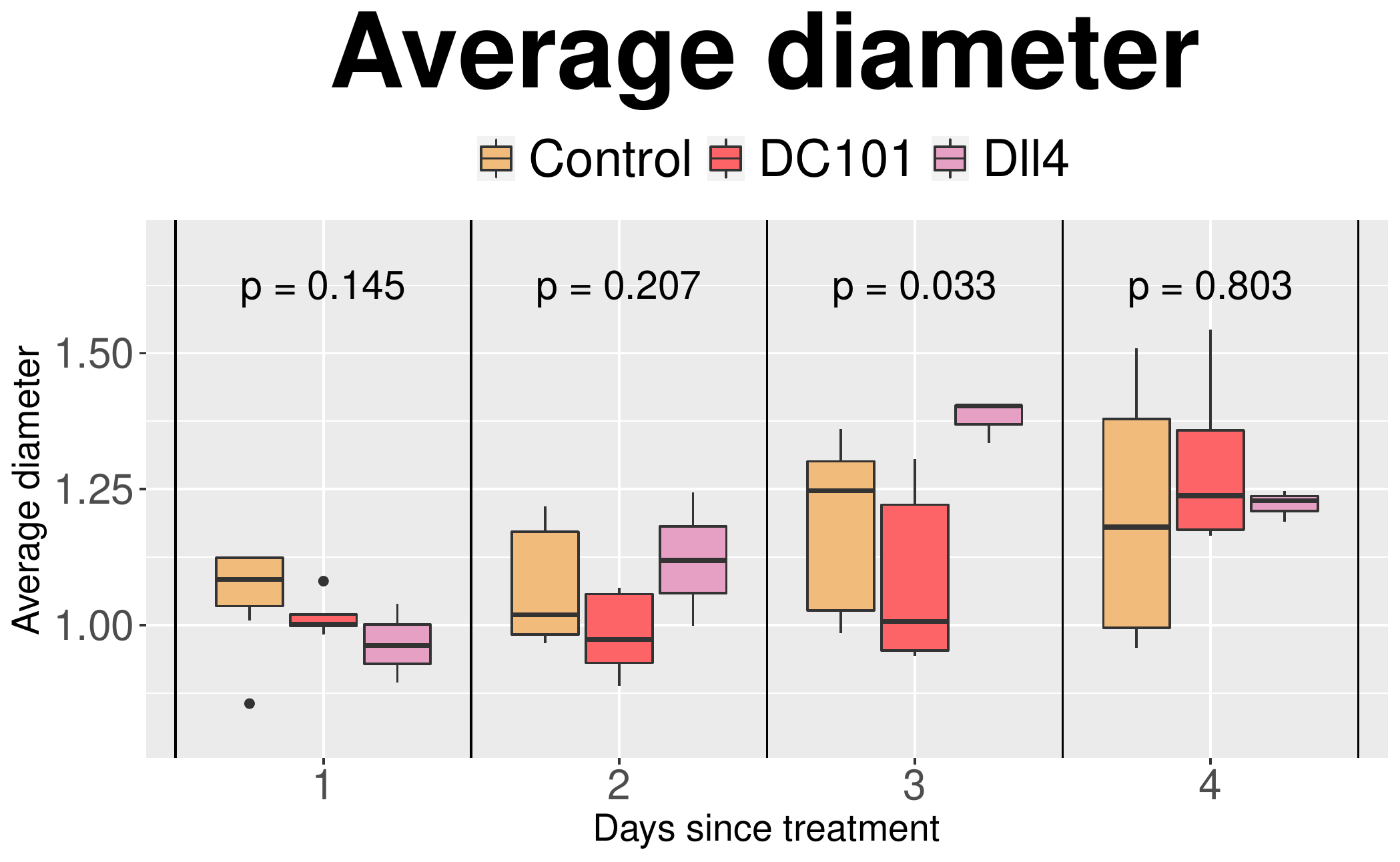}
\caption{{\bf Box plot showing the average (mean) vessel diameter.} The values are normalised by day 0 of initial treatment for all treatment regimes to facilitate comparisons of trends over time. We show group level $p$-values according to the Kruskal-Wallis test.}\label{Fig:KruskalMC38AvgDiamDC101Dll4}
\end{figure}

\begin{figure}[ht!]
\centering \includegraphics[width=.9\textwidth]{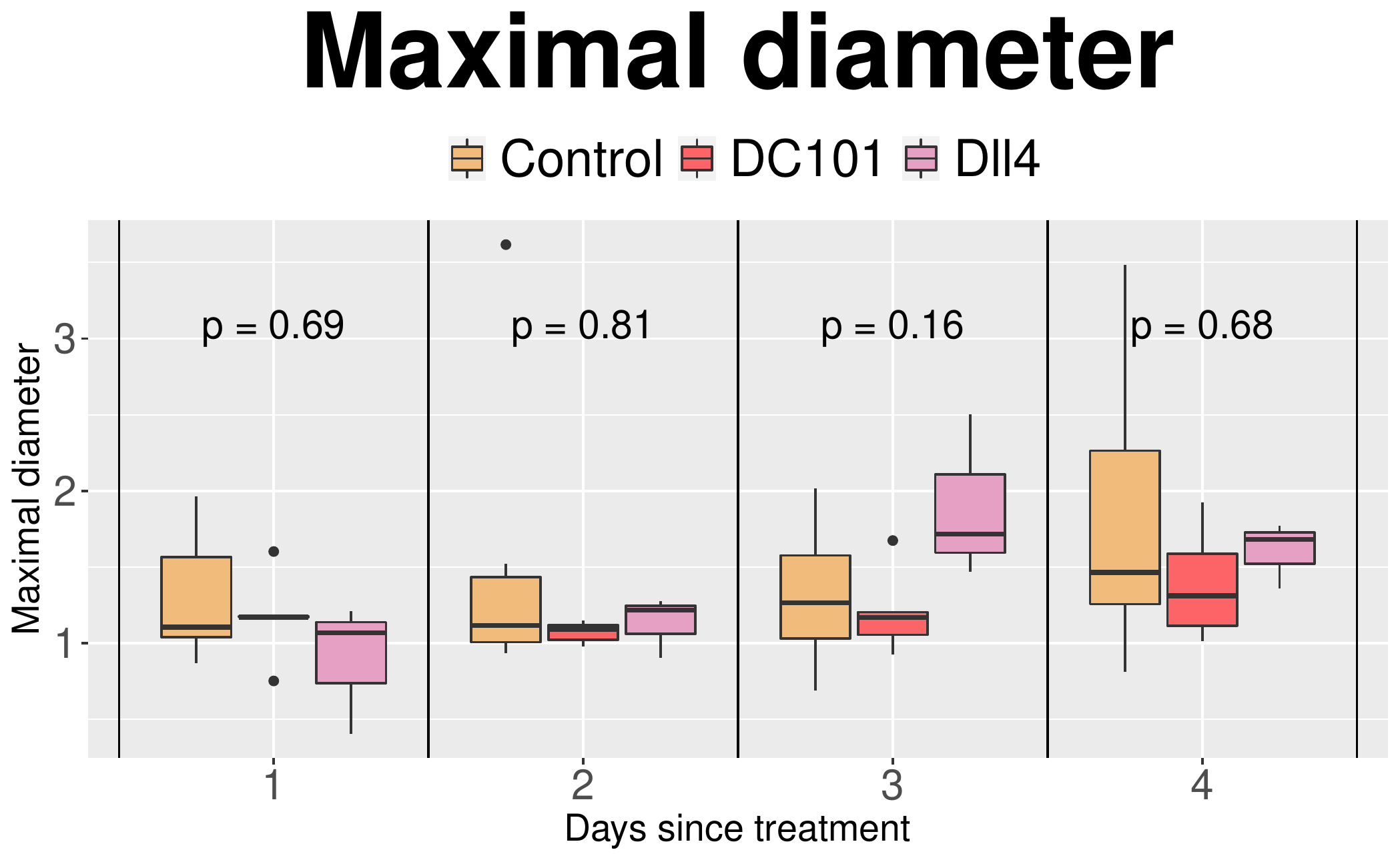}
\caption{{\bf Box plot showing the maximal vessel diameter.} The values are normalised by day 0 of initial treatment for all treatment regimes to facilitate comparisons of trends over time. We show group level $p$-values according to the Kruskal-Wallis test.}\label{Fig:KruskalMC38MaxDiamDC101Dll4}
\end{figure}

\begin{figure}[ht!]
\centering \includegraphics[width=.9\textwidth]{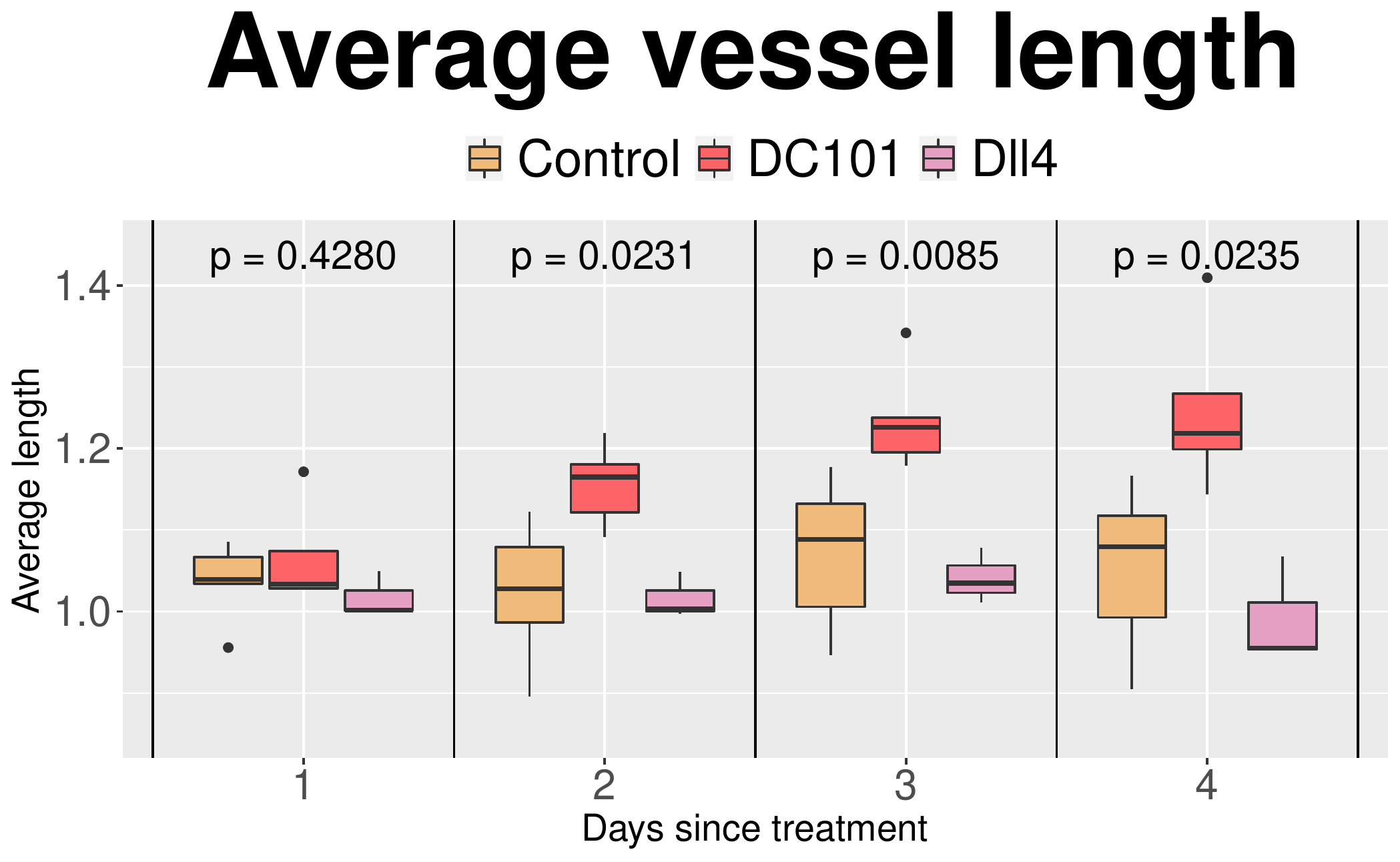}
\caption{{\bf Box plot showing the average (mean) vessel length.} The values are normalised by day 0 of initial treatment for all treatment regimes to facilitate comparisons of trends over time. We show group level $p$-values according to the Kruskal-Wallis test.}\label{Fig:KruskalMC38AvgLengthDC101Dll4}
\end{figure}

\begin{figure}[ht!]
\centering \includegraphics[width=.9\textwidth]{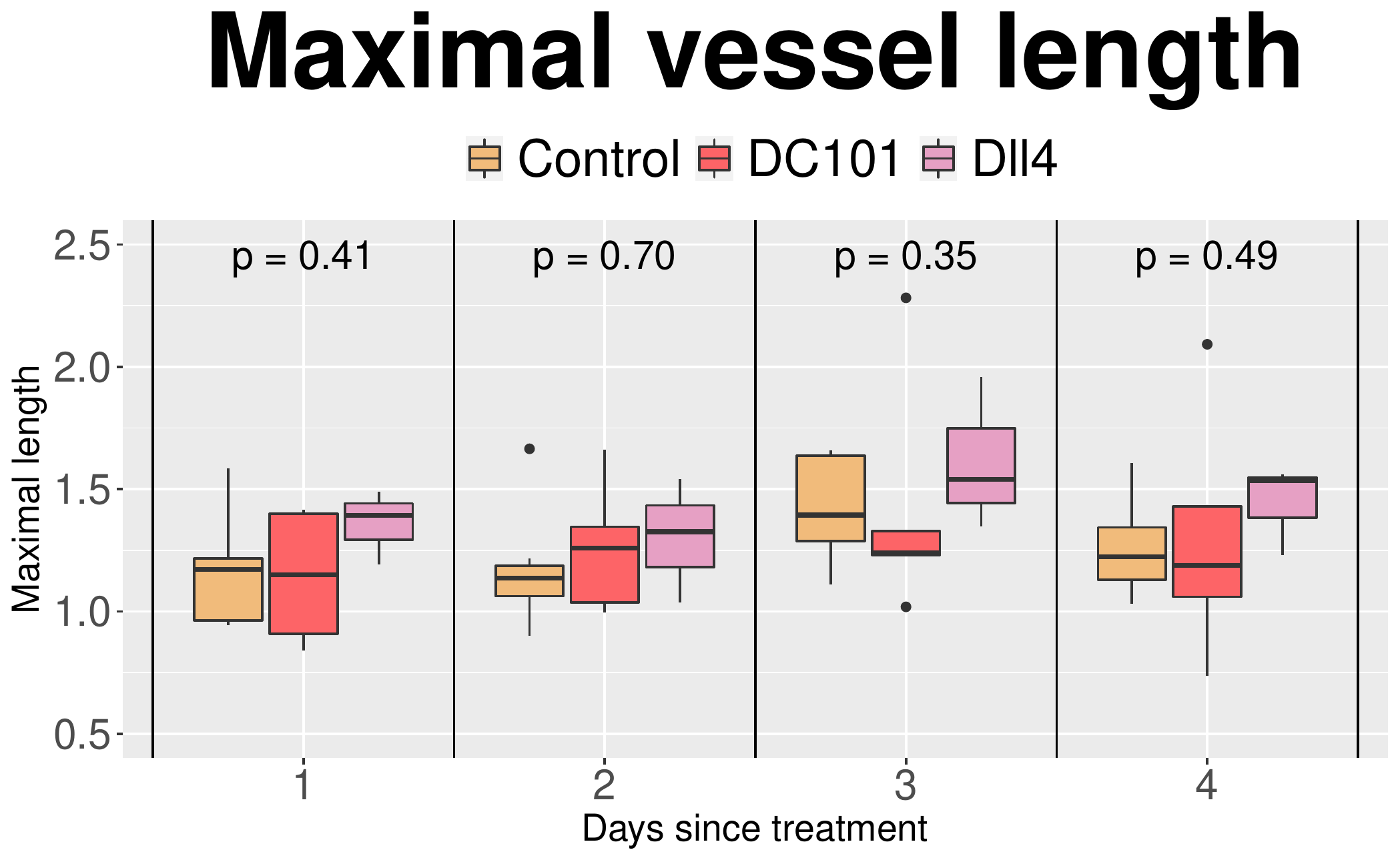}
\caption{{\bf Box plot showing the average (mean) vessel length.} The values are normalised by day 0 of initial treatment for all treatment regimes to facilitate comparisons of trends over time. We show group level $p$-values according to the Kruskal-Wallis test.}\label{Fig:KruskalMC38MaxLengthDC101Dll4}
\end{figure}

\begin{figure}[ht!]
\centering \includegraphics[width=.9\textwidth]{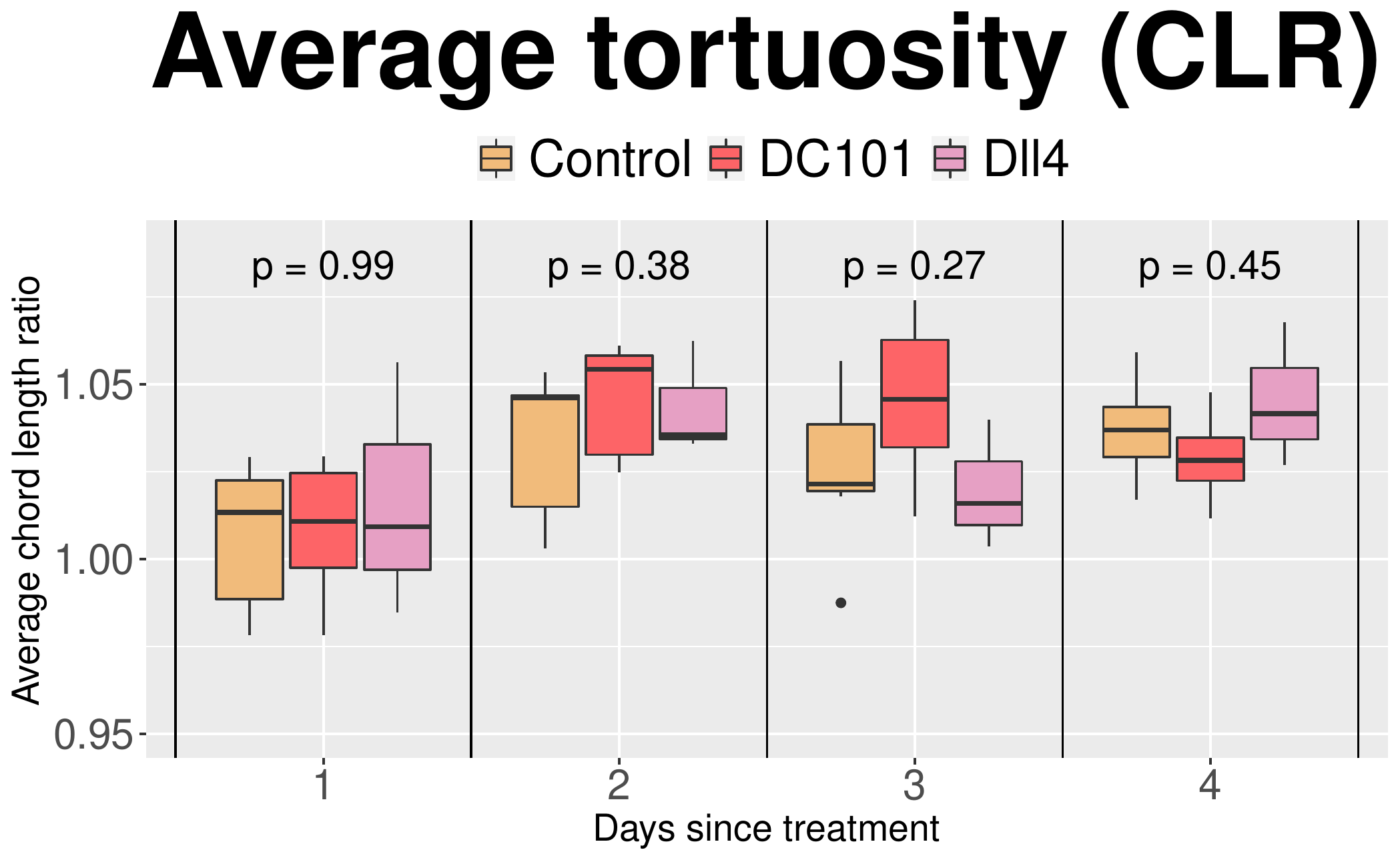}
\caption{{\bf Box plot showing the average (mean) chord length ratio.} The values are normalised by day 0 of initial treatment for all treatment regimes to facilitate comparisons of trends over time. We show group level $p$-values according to the Kruskal-Wallis test.}\label{Fig:KruskalMC38CLRDC101Dll4}
\end{figure}

\begin{figure}[ht!]
\centering \includegraphics[width=.9\textwidth]{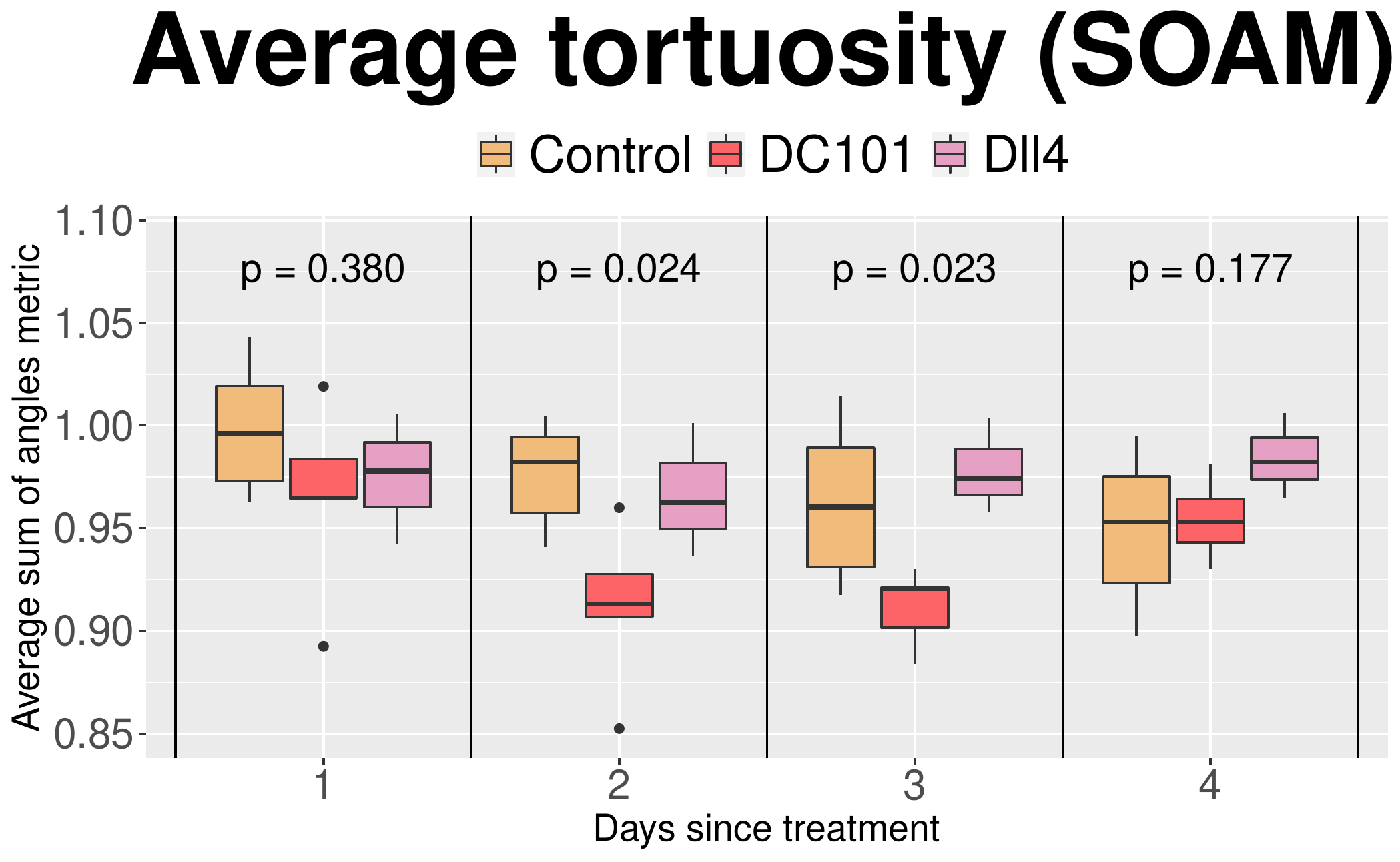}
\caption{{\bf Box plot showing the average (mean) sum of angles metric.} The values are normalised by day 0 of initial treatment for all treatment regimes to facilitate comparisons of trends over time. We show group level $p$-values according to the Kruskal-Wallis test.}\label{Fig:KruskalMC38SOAMDC101Dll4}
\end{figure}

\begin{figure}[ht!]
\centering \includegraphics[width=.9\textwidth]{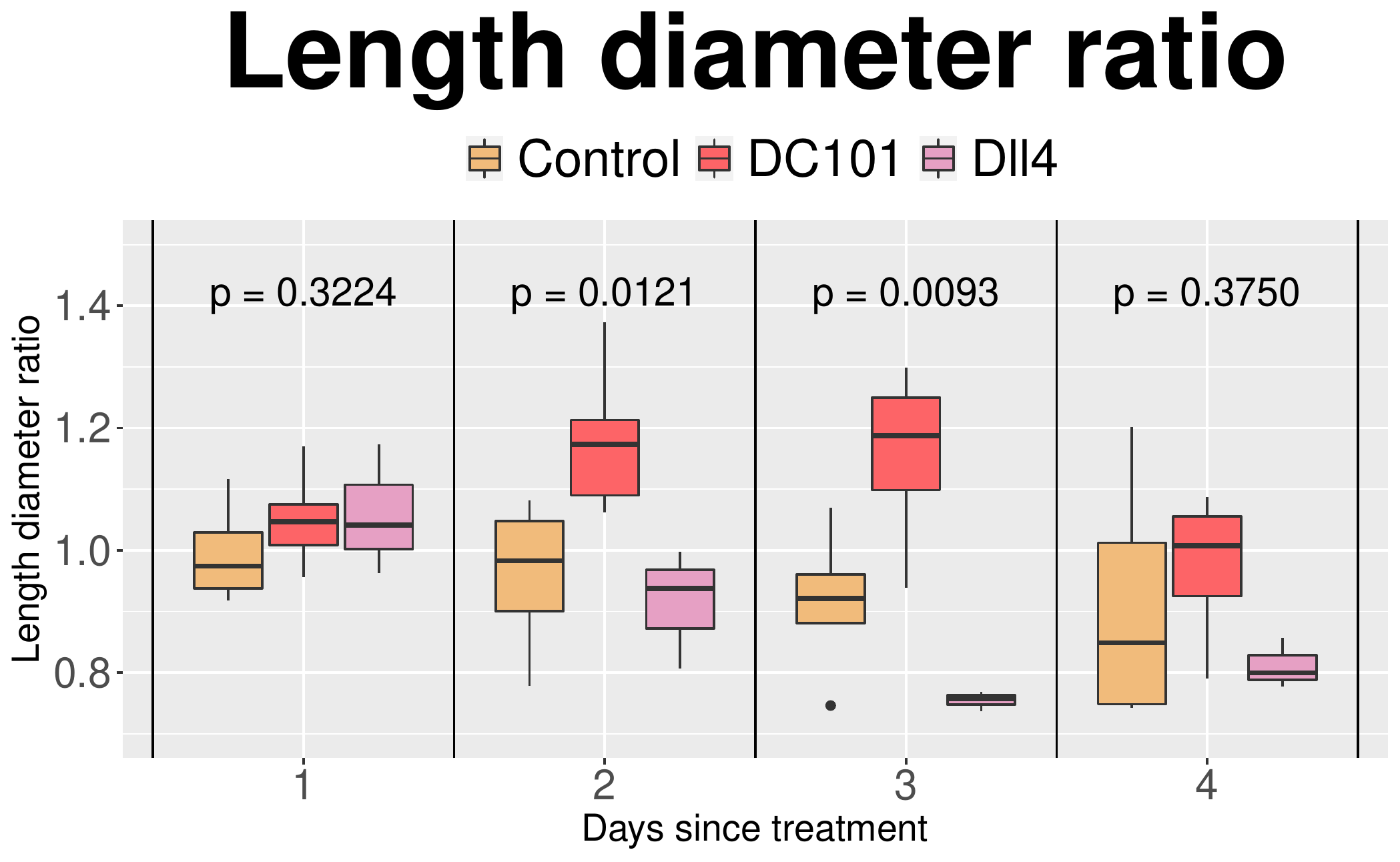}
\caption{{\bf Box plot showing the length to diameter ratio.} The values are normalised by day 0 of initial treatment for all treatment regimes to facilitate comparisons of trends over time. We show group level $p$-values according to the Kruskal-Wallis test.}\label{Fig:KruskalMC38LambdaDC101Dll4}
\end{figure}

\paragraph*{Intravital data: DC101 versus anti-Dll4.} We present statistical analysis on the control group and treatment groups DC101 and anti-Dll4 in the intravital data. We use the function \\ {\tt stat\_compare\_means()} from the library {\tt ggpubr} to compute Kruskal-Wallis test $p$-values for tortuosity (see Fig.~\ref{Fig:KruskalMC38TortuosityDC101Dll4}), number of loops per vessel segment (see Fig.~\ref{Fig:KruskalMC38LoopsDC101Dll4}) as well as the following  standard measures for vascular networks: number of vessel segments (see Fig.~\ref{Fig:KruskalMC38SegmentsDC101Dll4}), number of branching points (see Fig.~\ref{Fig:KruskalMC38BranchingDC101Dll4}), average vessel diameter (see Fig.~\ref{Fig:KruskalMC38AvgDiamDC101Dll4}), maximal diameter (see Fig.~\ref{Fig:KruskalMC38MaxDiamDC101Dll4}), average (mean) vessel length (see Fig.~\ref{Fig:KruskalMC38AvgLengthDC101Dll4}), maximal vessel length (see Fig.~\ref{Fig:KruskalMC38MaxLengthDC101Dll4}), average (mean) chord length ratio (see Fig.~\ref{Fig:KruskalMC38CLRDC101Dll4}), average (mean) sum of angles metric (see Fig.~\ref{Fig:KruskalMC38SOAMDC101Dll4}), and length-diameter ratio (see Fig.~\ref{Fig:KruskalMC38LambdaDC101Dll4}). All values are normalised by day 0 of observation/treatment and were obtained from the {\sc python} code package {\sc unet-core}~\cite{unetRuss}.

\clearpage

\begin{figure}[ht!]
\centering \includegraphics[width=\textwidth]{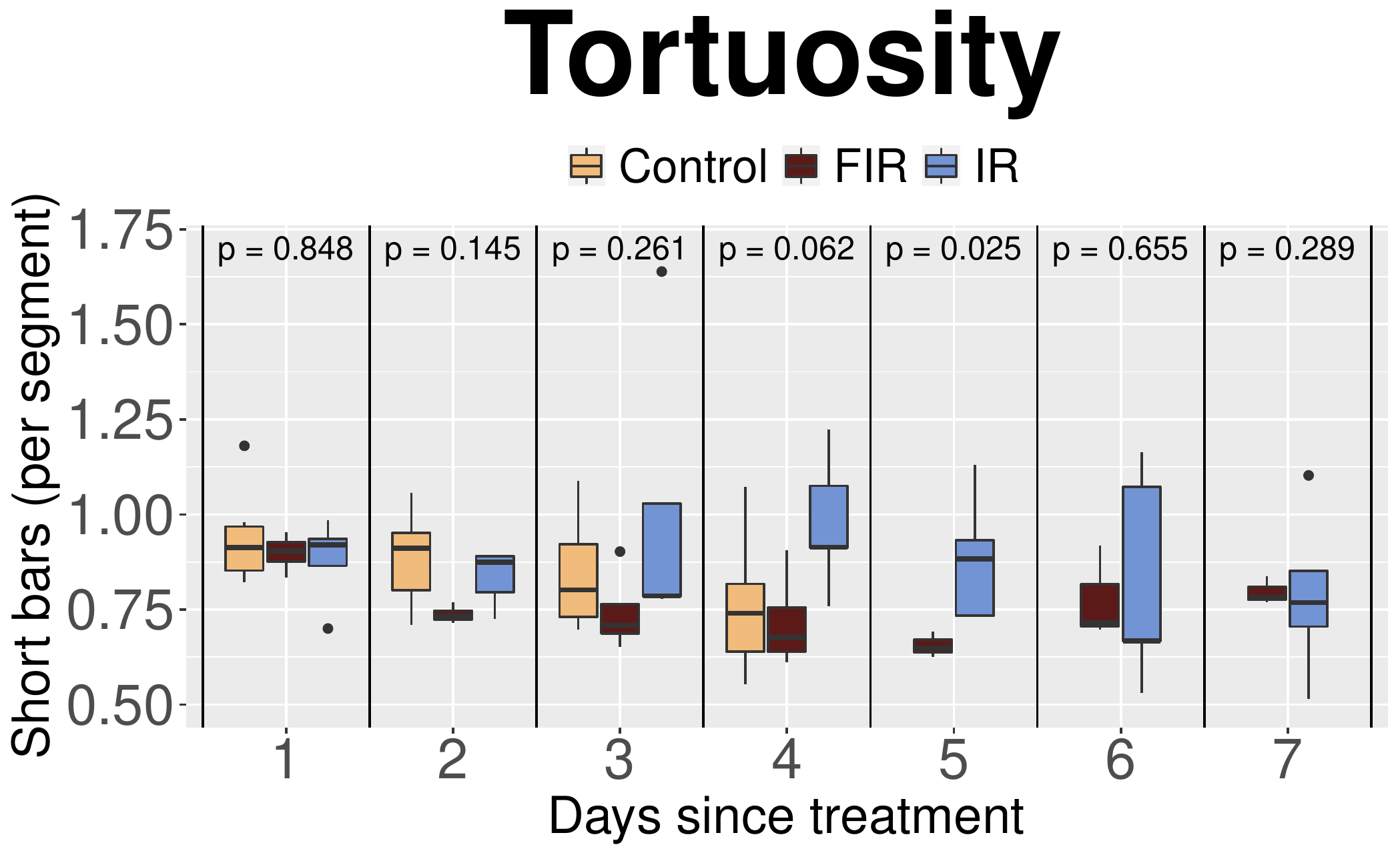}
\caption{{\bf Box plot showing the number of short bars in the dimension 0 barcode of the radial filtration divided by the number of vessel segments.} The values are normalised by day 0 of initial treatment for all treatment regimes to facilitate comparisons of trends over time. We show group level $p$-values according to the Kruskal-Wallis test.}\label{Fig:KruskalMC38TortuosityIRFIR}
\end{figure}

\begin{figure}[ht!]
\centering \includegraphics[width=\textwidth]{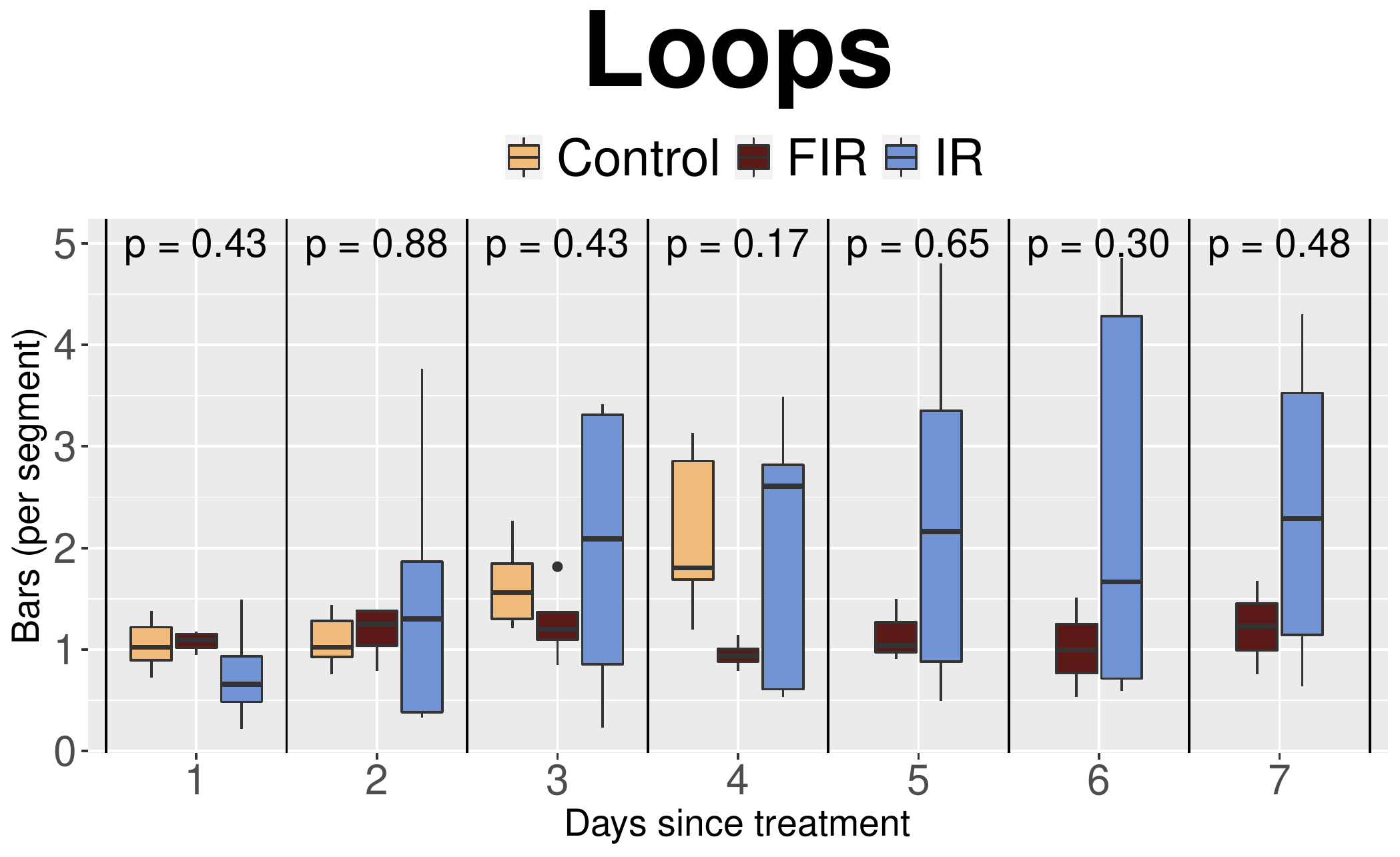}
\caption{{\bf Box plot showing the number of bars in the dimension 1 barcode of the radial filtration divided by the number of vessel segments.} The values are normalised by day 0 of initial treatment for all treatment regimes to facilitate comparisons of trends over time. We show group level $p$-values according to the Kruskal-Wallis test.}\label{Fig:KruskalMC38LoopsIRFIR}
\end{figure}

\paragraph*{Intravital data: single dose irradiation versus fractionated dose irradiation.} We present statistical analysis on the control group and radiation treatment groups IR (single-dose irradiation) and FIR (fractionated-dose irradiation) in the intravital data. We use the function {\tt stat\_compare\_means()} from the library {\tt ggpubr} to compute Kruskal-Wallis test $p$-values for tortuosity (see Fig.~\ref{Fig:KruskalMC38TortuosityIRFIR}) and number of loops per vessel segment (see Fig.~\ref{Fig:KruskalMC38LoopsIRFIR}). All values are normalised by day 0 of observation/treatment.

\clearpage

\begin{figure}[ht!]
\centering \includegraphics[width=\textwidth]{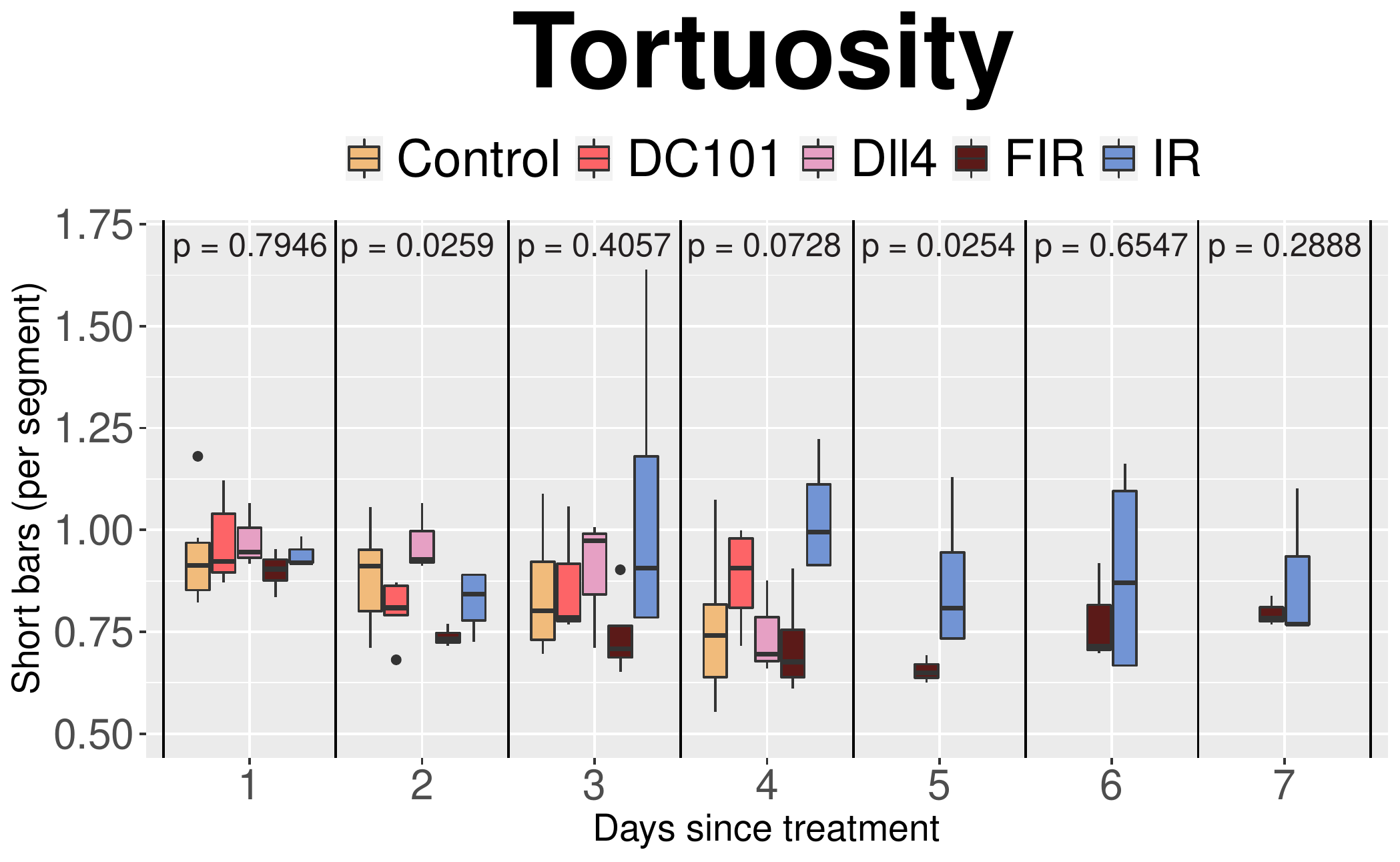}
\caption{{\bf Box plot showing the number of short bars in the dimension 0 barcode of the radial filtration divided by the number of vessel segments.} The values are normalised by day 0 of initial treatment for all treatment regimes to facilitate comparisons of trends over time. We show group level $p$-values according to the Kruskal-Wallis test.}\label{Fig:KruskalMC38Tortuosity}
\end{figure}

\begin{figure}[ht!]
\centering \includegraphics[width=\textwidth]{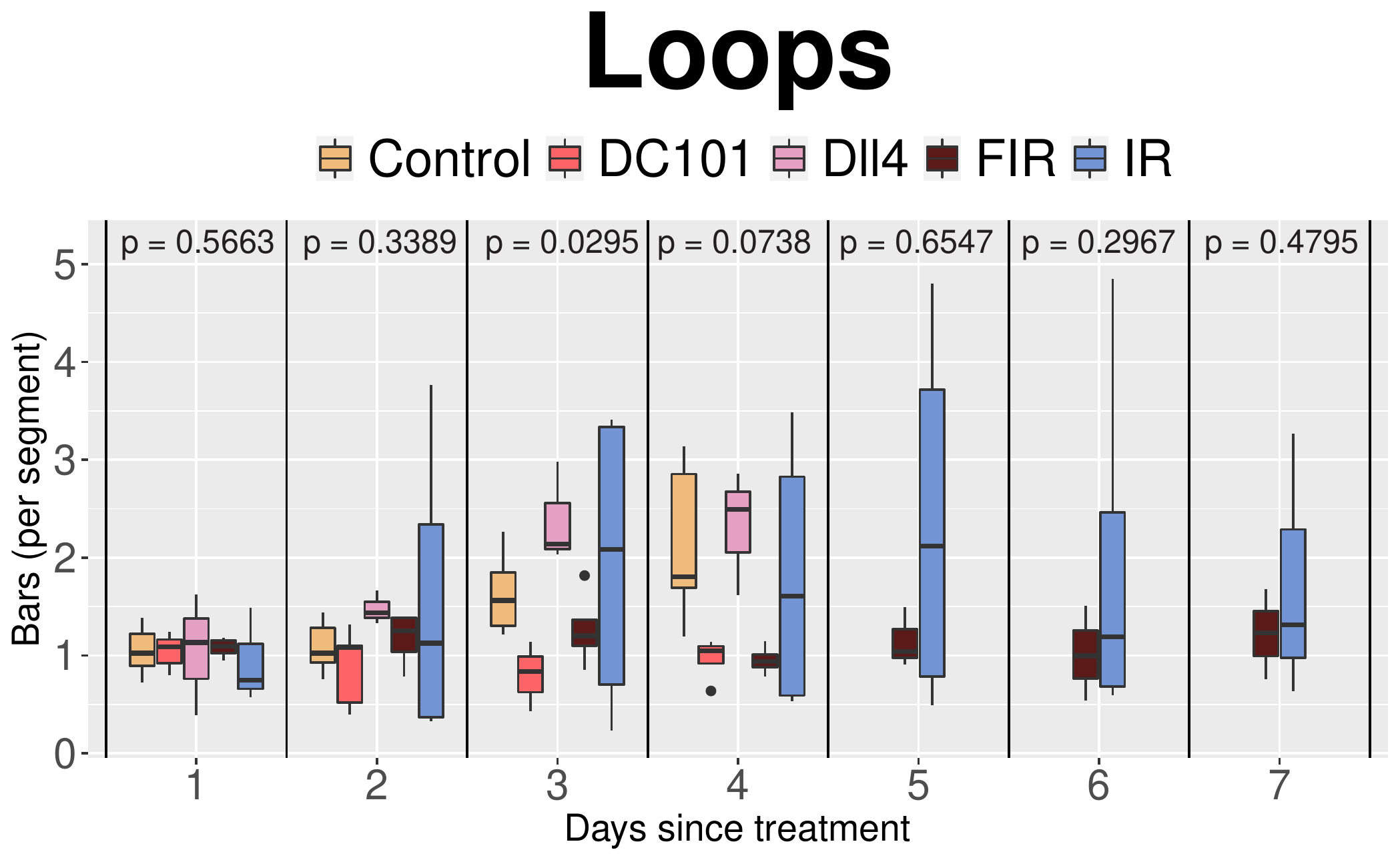}
\caption{{\bf Box plot showing the number of bars in the dimension 1 barcode of the radial filtration divided by the number of vessel segments. } The values are normalised by day 0 of initial treatment for all treatment regimes to facilitate comparisons of trends over time. We show group level $p$-values according to the Kruskal-Wallis test.}\label{Fig:KruskalMC38Loops}
\end{figure}

\begin{figure}[ht!]
\centering \includegraphics[width=\textwidth]{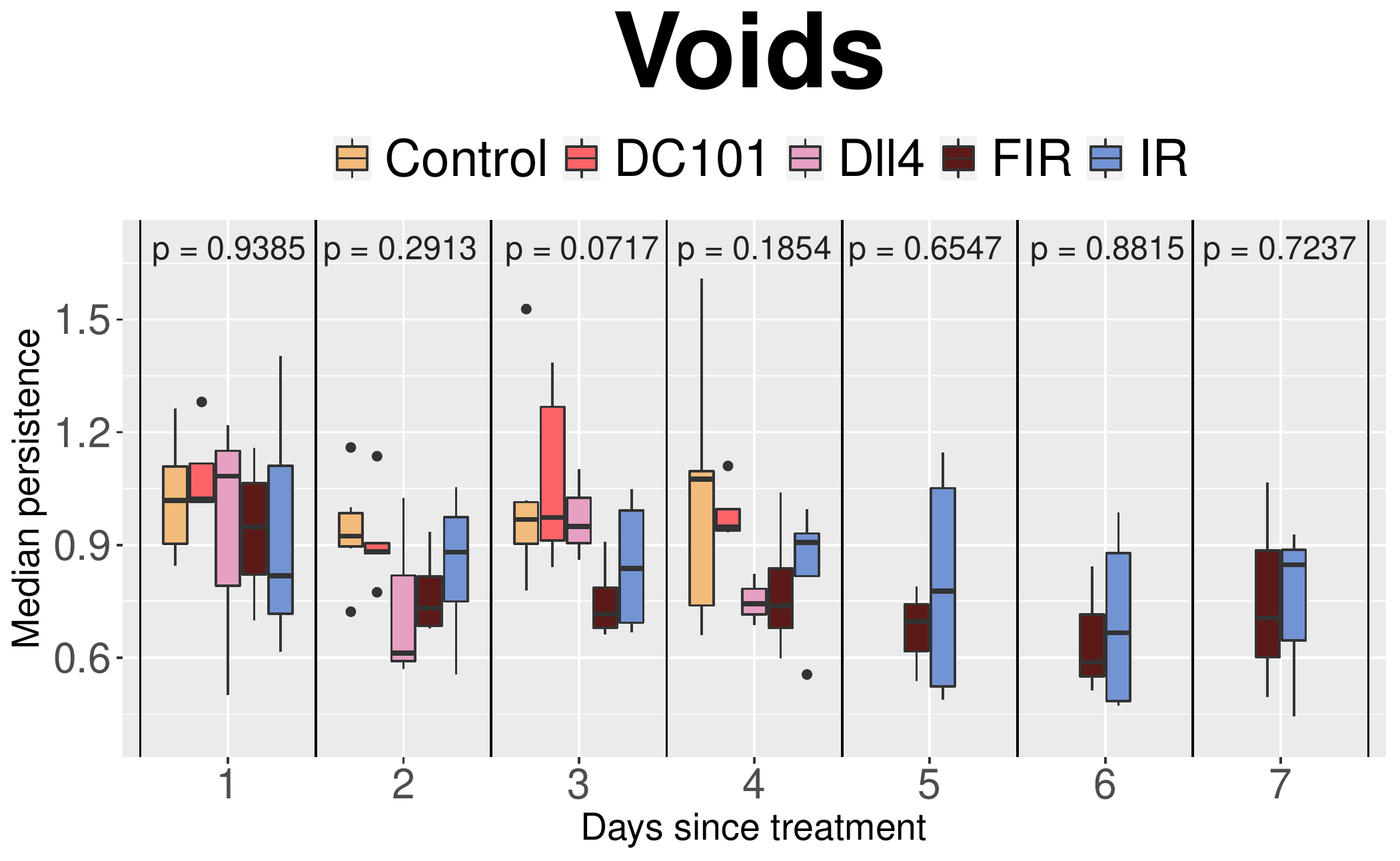}
\caption{{\bf Box plot showing median persistence of bars in the dimension 2 barcode of the $\alpha$-complex filtration.} The values are normalised by day 0 of initial treatment for all treatment regimes to facilitate comparisons of trends over time. We show $p$-values according to the Kruskal-Wallis test.}\label{Fig:KruskalMC38Voids}
\end{figure}

\begin{figure}[ht!]
\centering \includegraphics[width=\textwidth]{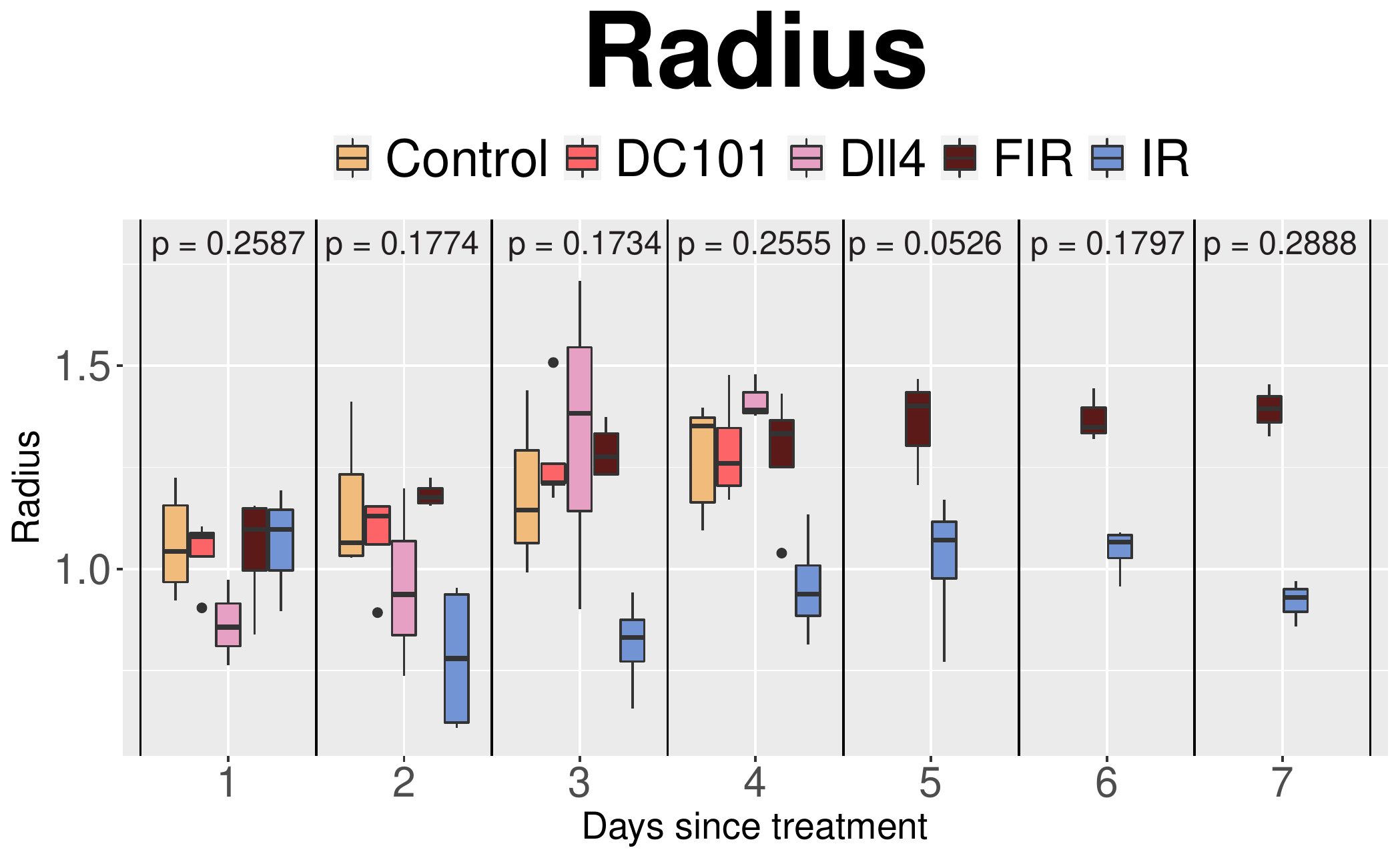}
\caption{{\bf Box plot showing the maximal radius attained in the radial filtration. } The values are normalised by day 0 of initial treatment for all treatment regimes to facilitate comparisons of trends over time. We show $p$-values according to the Kruskal-Wallis test.}\label{Fig:KruskalMC38Radii}
\end{figure}

\paragraph*{Intravital data: all treatment groups.} We present statistical analysis to determine whether at least one of the treatment groups in the intravital data behaves significantly differently to the others in Fig.~\ref{Fig:KruskalMC38Tortuosity} for our extracted tortuosity measure and in Fig.~\ref{Fig:KruskalMC38Loops} for the number of loops per vessel segment.  All values are normalised by day 0 of observation/treatment.
We compute the (non-exact) $p$-values for the using the {\sc R} function {\tt kruskal.test()} to compute Kruskal-Wallis in {\sc RStudio}~\cite{RStudio-Team:2016aa}. 
We further present the same analysis for parameters not shown in the main text, i.e. for voids in Fig.~\ref{Fig:KruskalMC38Voids} and maximal radii used in the radial filtration (i.e. an approximation of the tumour radii) in Fig.~\ref{Fig:KruskalMC38Radii}. Again, all values are normalised by day 0 of observation/treatment. We note that both of these parameters do not show significant differences between treatment groups. In the case of the voids in the intravital dataset this can be explained by the the low penetration depth of the imaging.

\clearpage

\begin{figure}[ht!]
\centering \includegraphics[width=\textwidth]{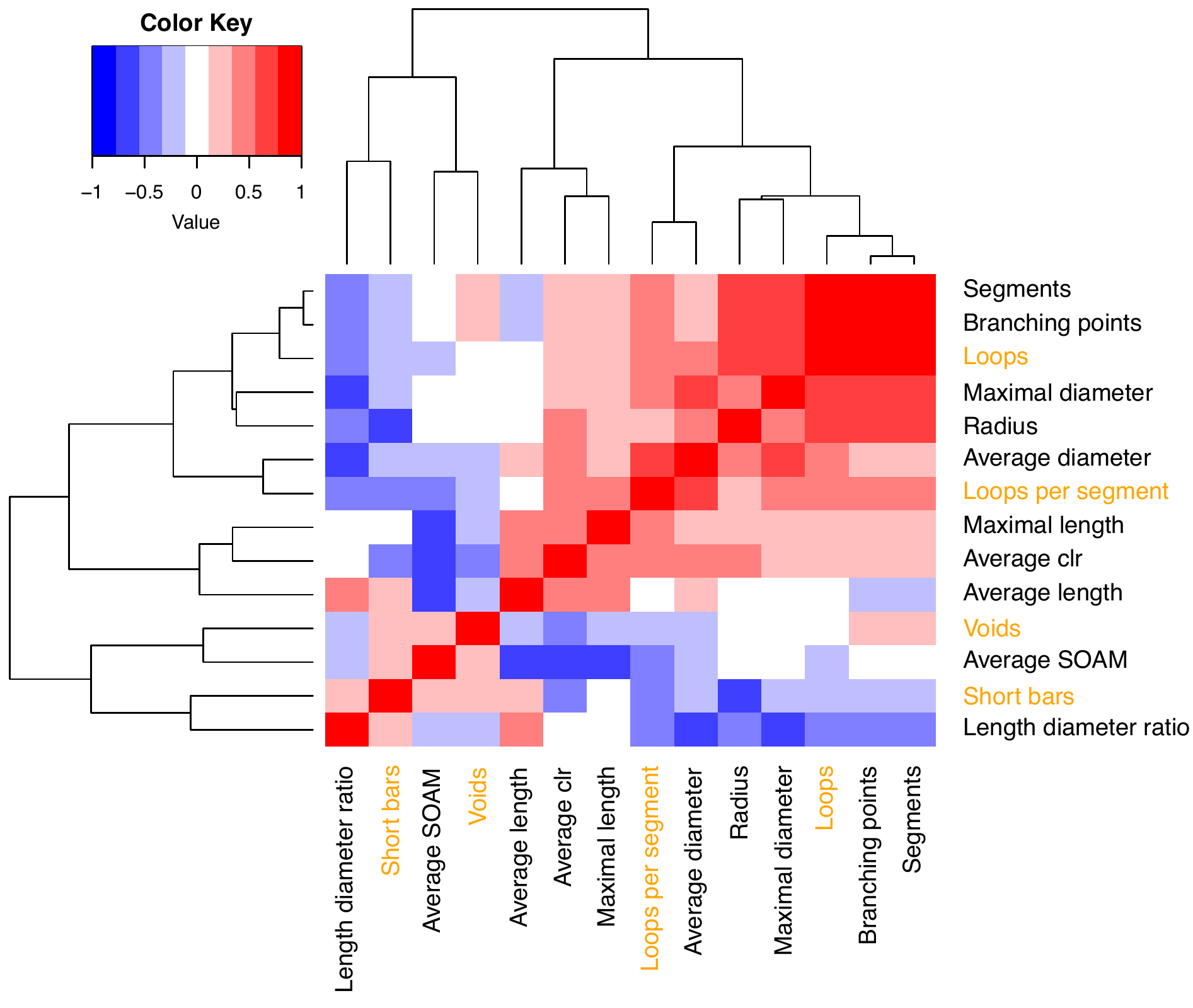}
\caption{{\bf Heatmap displaying the pairwise Pearson correlation coefficients between different vascular characteristics derived from the intravital data. } The dendrogramme represents complete linkage clustering using the Euclidean distance measure. We consider the following vascular characteristics: number of vessel segments (i.e. number of edges), number of branching points (i.e. number of nodes), number of vessel loops, maximal vessel diameter, maximal radius used in the radial filtration, average vessel diameter, number of vessel loops per vessel segment, maximal vessel length, average chord length ratio (clr), average vessel length, median persistence of bars in dimension 2 barcodes (voids), average sum of angles measure (SOAM), number of short bars per vessel segment in the dimension 0 barcodes, vessel length/diameter ratio.
We highlight the topological measures in orange including both the number of loops and number of loops per vessel segment to highlight the effect of the normalisation. }\label{Fig:MC38Cor}
\end{figure}

\noindent Finally, we present a correlation analysis between parameters that are conventionally extracted from vascular networks and our topological parameters in Fig.~\ref{Fig:MC38Cor}. We compute pairwise Pearson correlation using the library {\tt hmisc} and plot our results including a complete linkage clustering dendrogramme of the parameters using the library {\tt corrplot} in {\sc RStudio}~\cite{RStudio-Team:2016aa}.

\clearpage

\begin{figure}[ht!]
\centering \includegraphics[width=.75\textwidth]{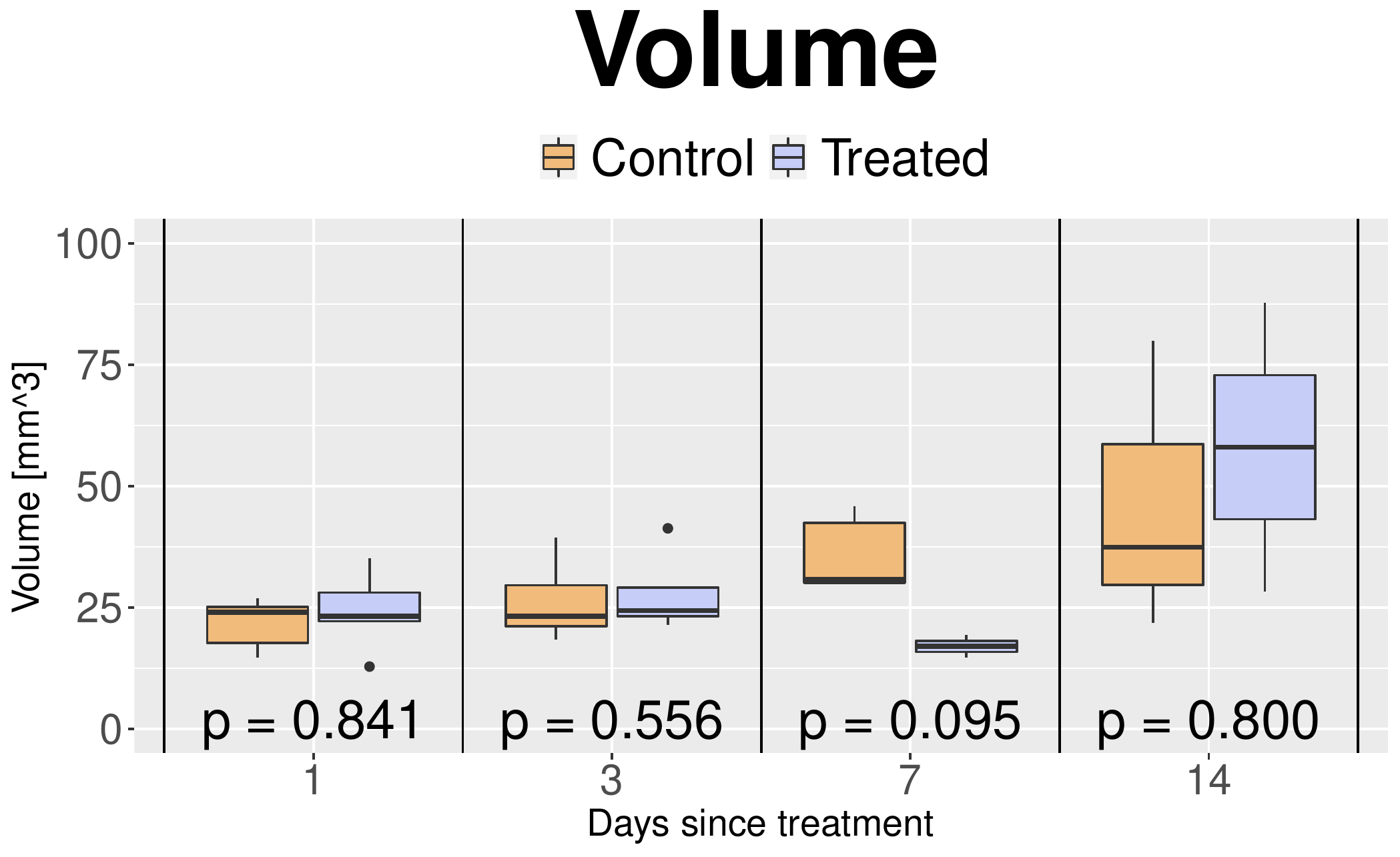}
\caption{{\bf Box plot showing tumour volume as determined by Dobosz \emph{et al.}~\cite{Dobosz2014}.} We show $p$-values according to Wilcoxon's rank sum test.}\label{Fig:RocheVolume}
\end{figure}

\begin{figure}[ht!]
\centering \includegraphics[width=.75\textwidth]{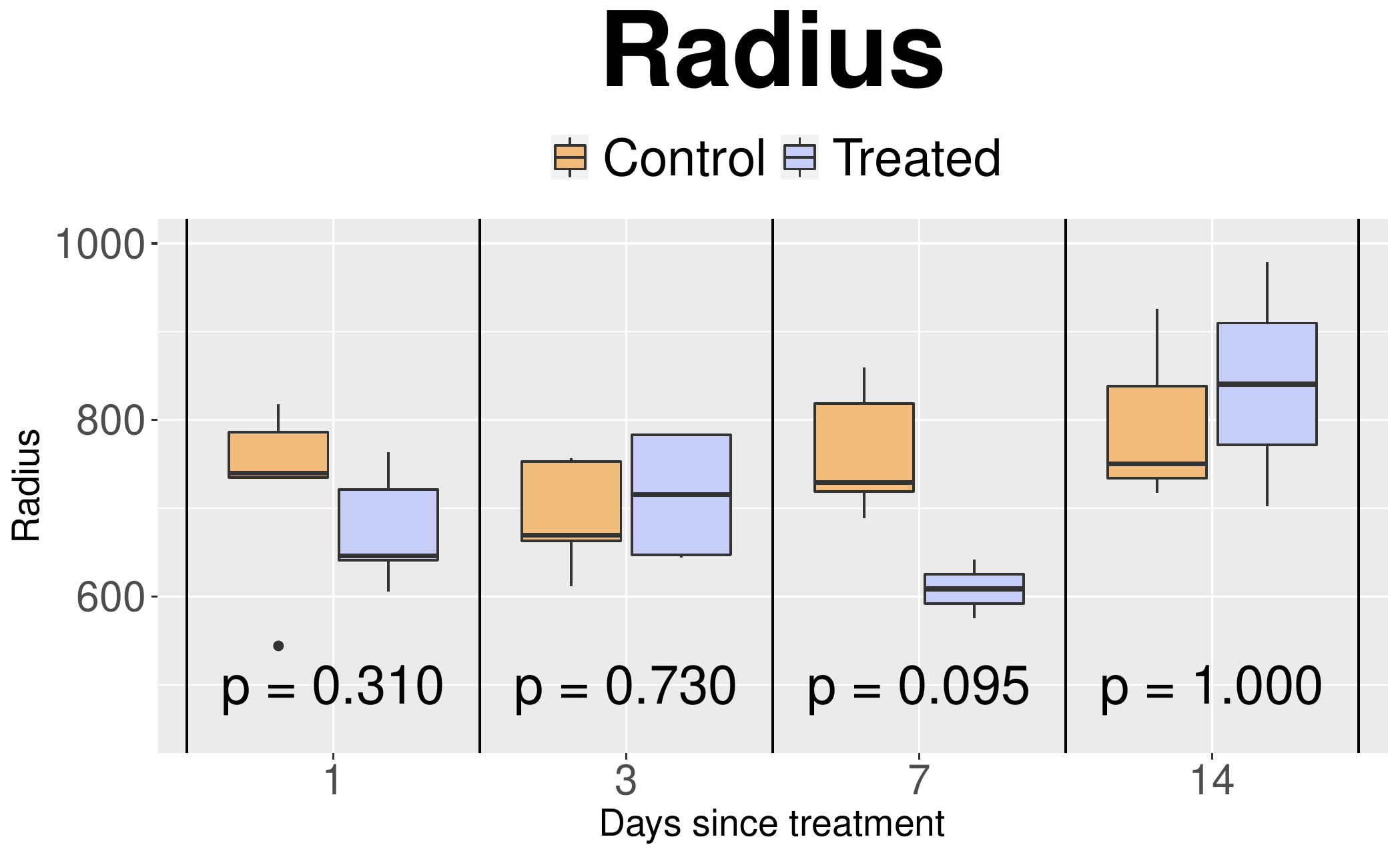}
\caption{{\bf Box plot showing the maximal radius attained in the radial filtration.} We show $p$-values according to Wilcoxon's rank sum test.}\label{Fig:RocheRadius}
\end{figure}

\begin{figure}[ht!]
\centering \subcaptionbox{Radial interval I.}{\centering\includegraphics[width=.49\textwidth]{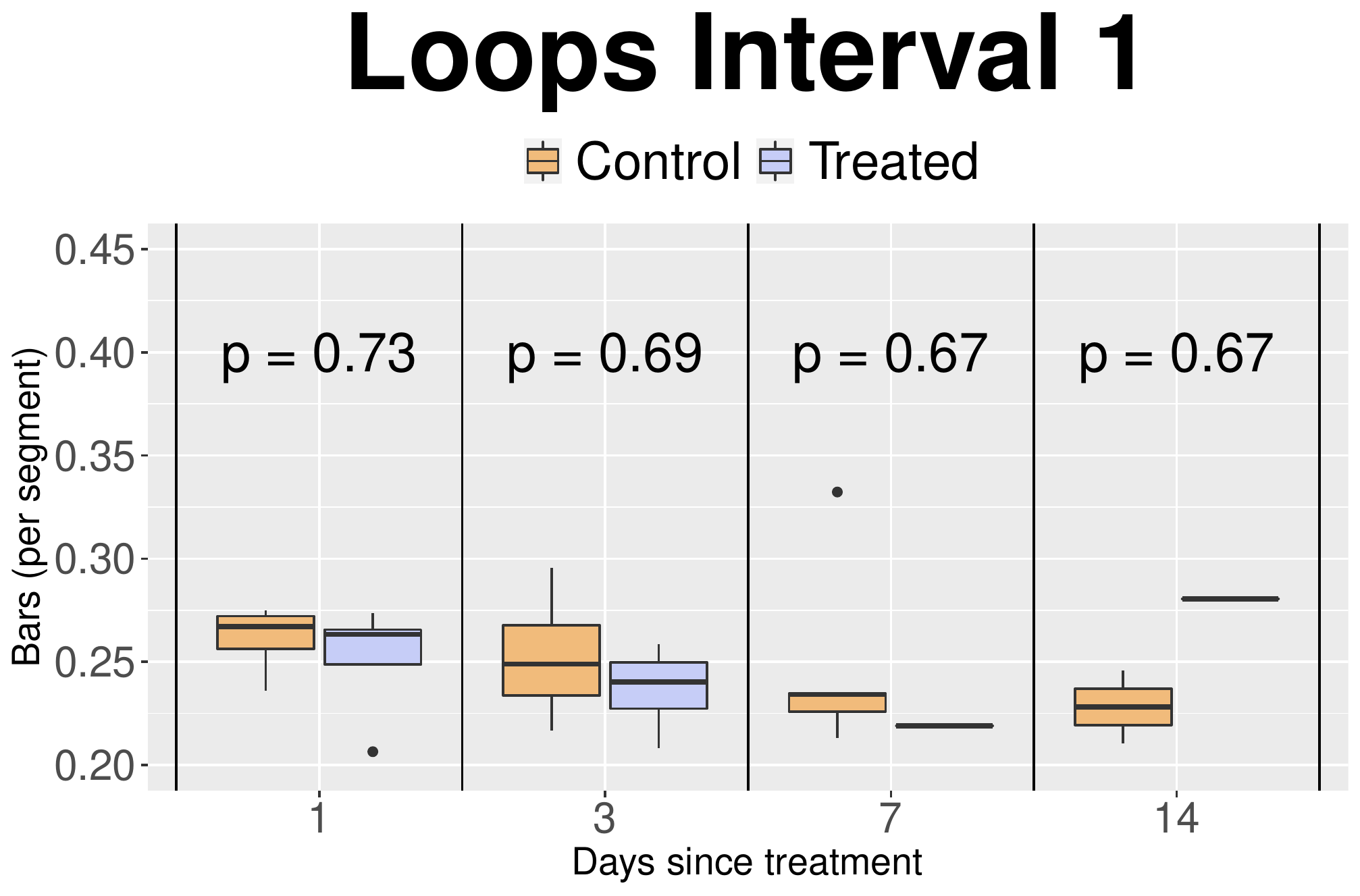}}%
\vspace{0.01\textheight}
\subcaptionbox{Radial interval II.}{\centering\includegraphics[width=.49\textwidth]{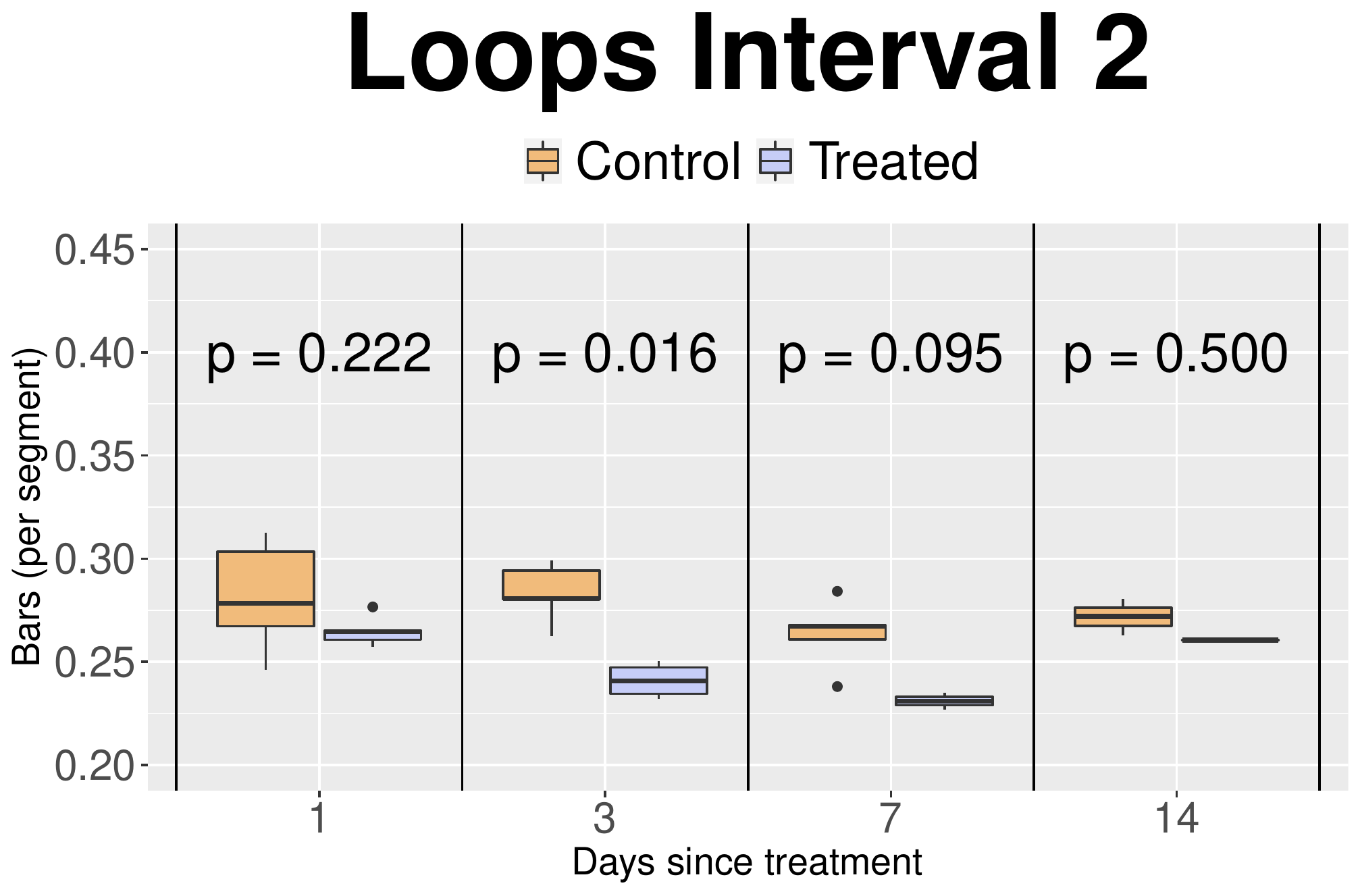}}%
\vspace{0.01\textheight}
\subcaptionbox{Radial interval III.}{\centering\includegraphics[width=.49\textwidth]{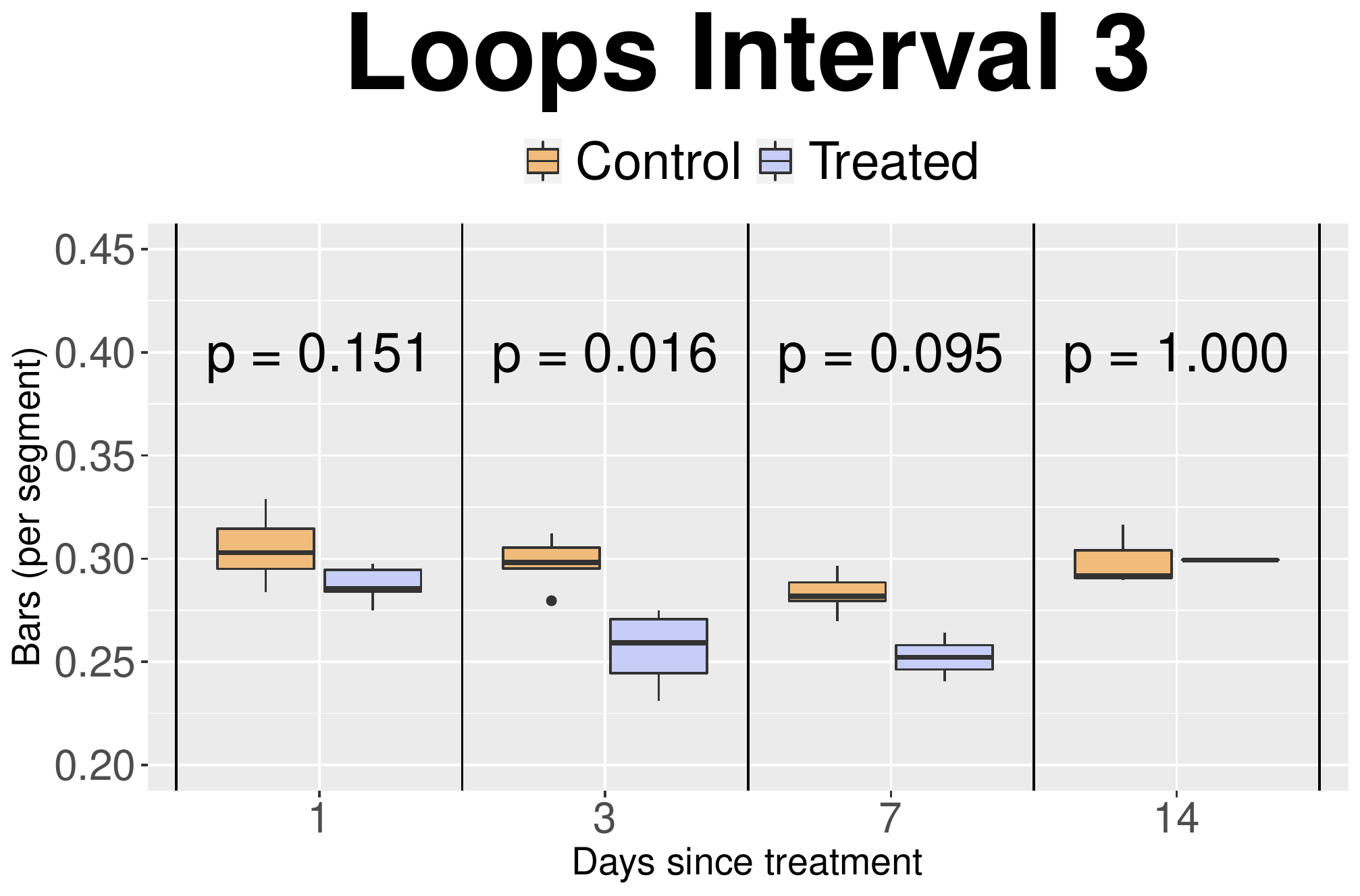}}%
\vspace{0.01\textheight}
\subcaptionbox{Radial interval IV.}{\centering\includegraphics[width=.49\textwidth]{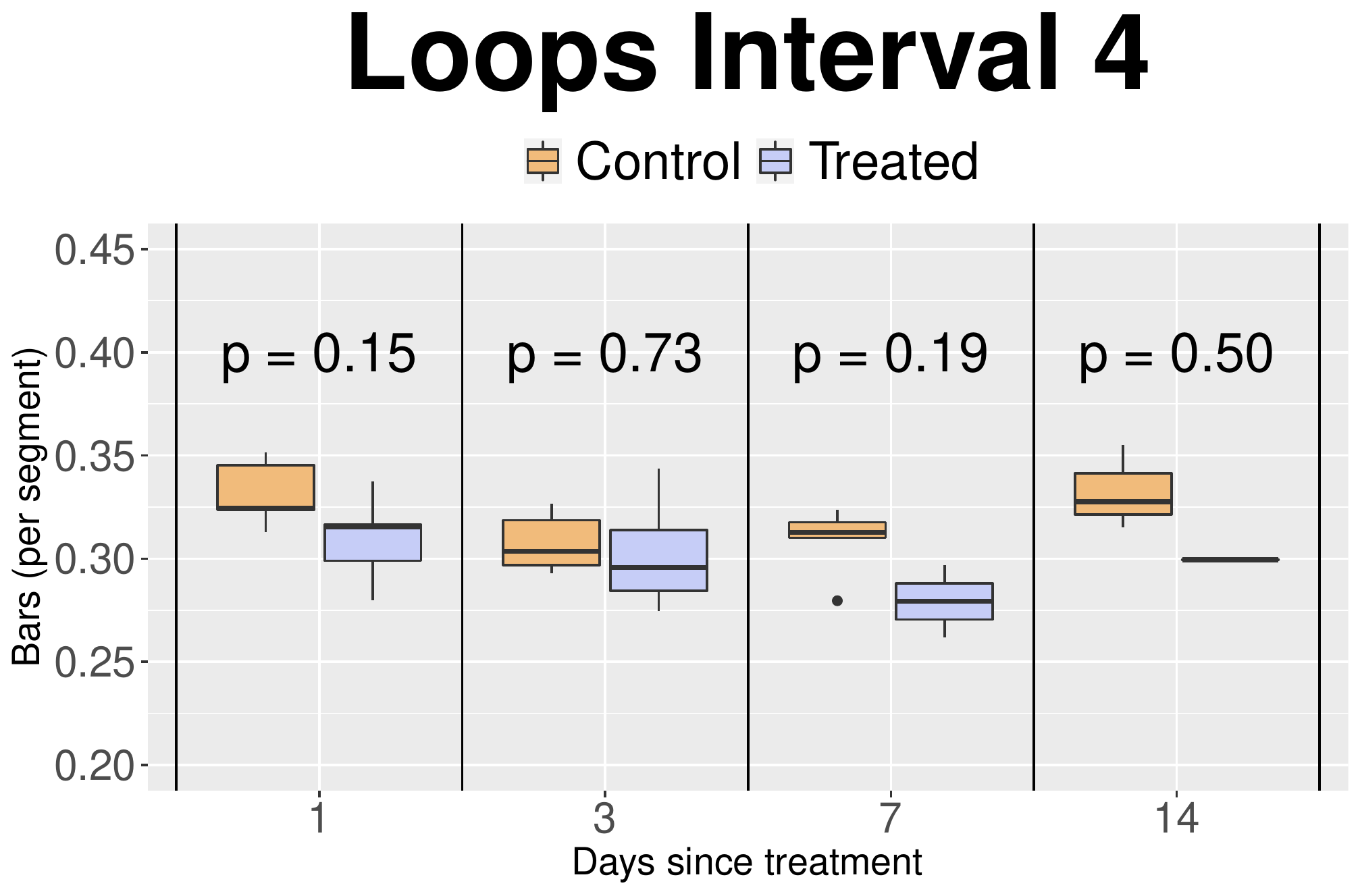}}%
\vspace{0.01\textheight}
\subcaptionbox{Radial interval V.}{\centering\includegraphics[width=.49\textwidth]{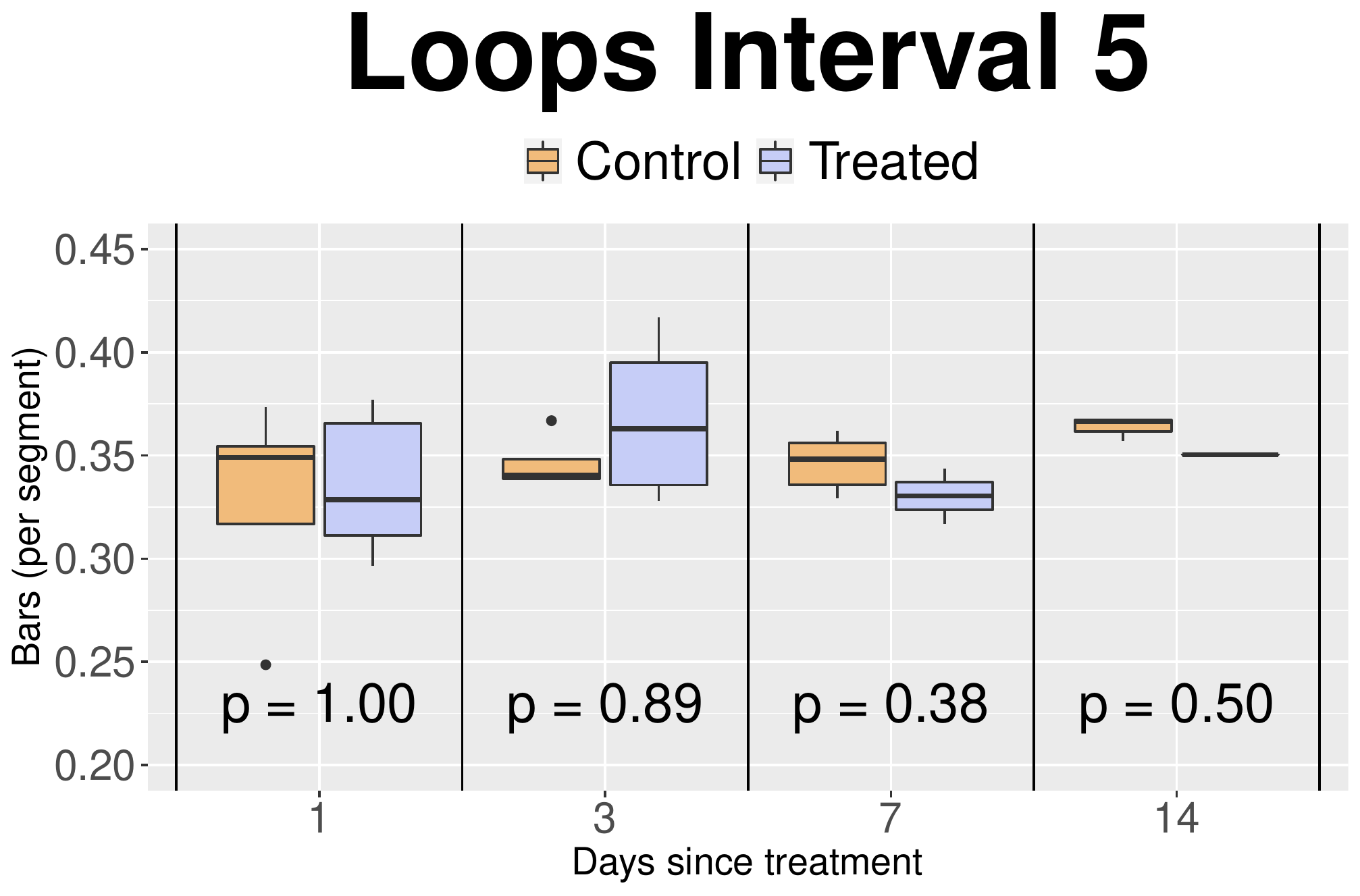}}%
\caption{{\bf Number of loops per vessel segment for different filtration intervals in the ultramicroscopy dataset.} We show box plots of the number of loops per vessel segment. Interval I corresponds to the radial region closest to the tumour centre, while Interval V represents parts of the vessel network that are farthest away from the tumour centre. We show $p$-values according to Wilcoxon's rank sum test.}
\label{Fig:RocheIntervals}
\end{figure}

\paragraph*{Ultramicroscopy data.} We present box plots of the tumour volume as determined by Dobosz \emph{et al.}~\cite{Dobosz2014} in Fig.~\ref{Fig:RocheVolume} and the maximal radii used in the radial filtration in Fig.~\ref{Fig:RocheRadius}. We compute the (non-exact) $p$-values using function {\tt stat\_compare\_means} from the library {\tt ggpubr} in {\sc RStudio}~\cite{RStudio-Team:2016aa} to perform a pairwise Wilcoxon's rank sum test between the control group and the treatment group. All our tests are by default two-sided.
We further show the spatio-temporal resolution of the number of loops in the ultramicroscopy data in Fig.~\ref{Fig:RocheIntervals}. We do not find any marked differences in either treatment group in different spherical shells around the tumour centres.

\clearpage

\clearpage

\begin{figure}[ht!]
\centering \includegraphics[width=\textwidth]{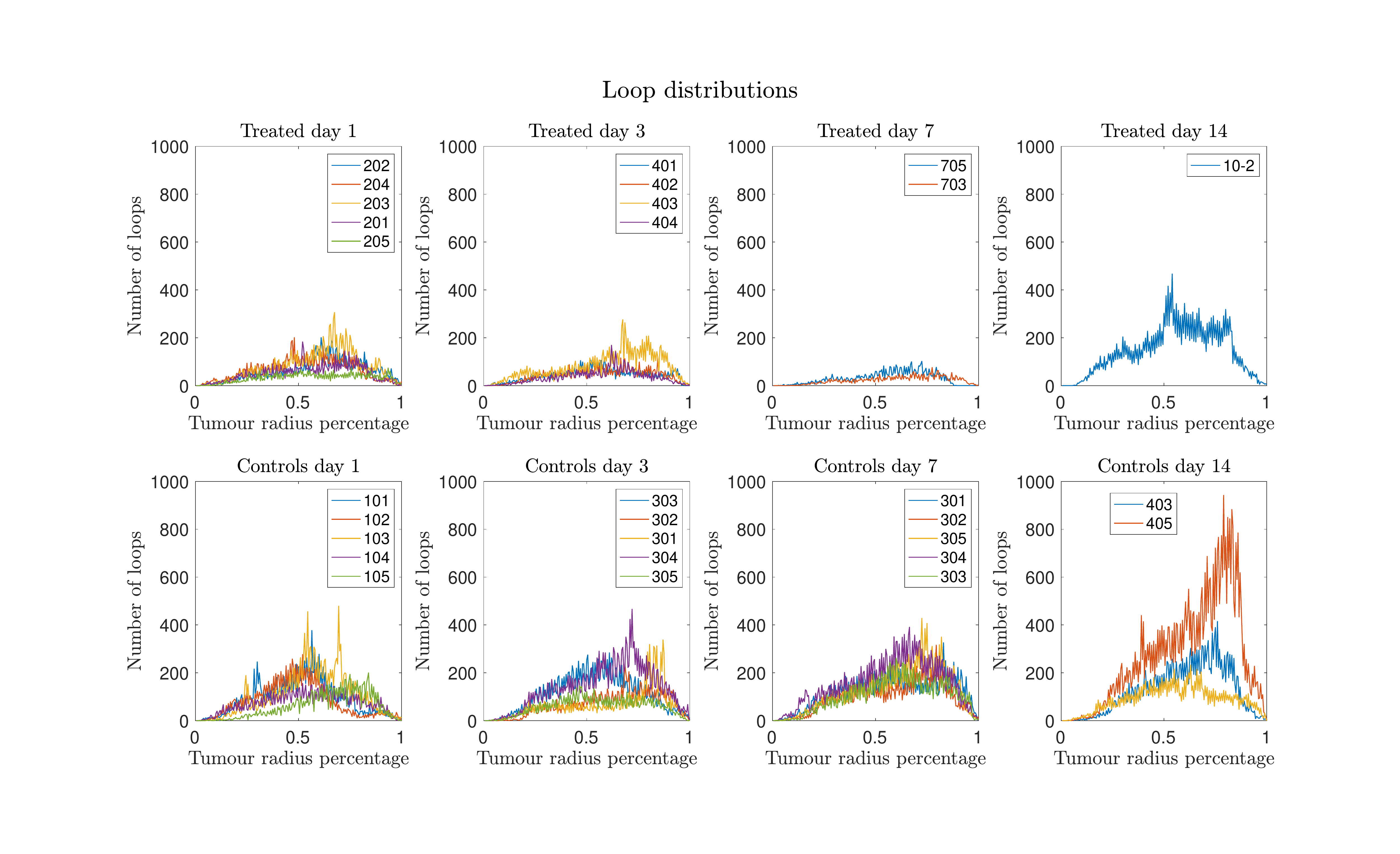}
\caption{{\bf Spatial distribution of the number of loops in the ultramicroscopy data}. We show the distribution of loops in individual tumours grouped by treatment regime (top row: bevacicumab treated tumours; bottom row: control tumours) and time points (column 1: day 1 after treatment; column 2: day 2 after treatment; column 3: day 3 after treatment; column 4: day 4 after treatment). The horizontal axis represents the radial distance to the tumour centre normalised by tumour radius.}\label{Fig:RocheDistributions}
\end{figure}

\begin{figure}[ht!]
\centering \includegraphics[width=\textwidth]{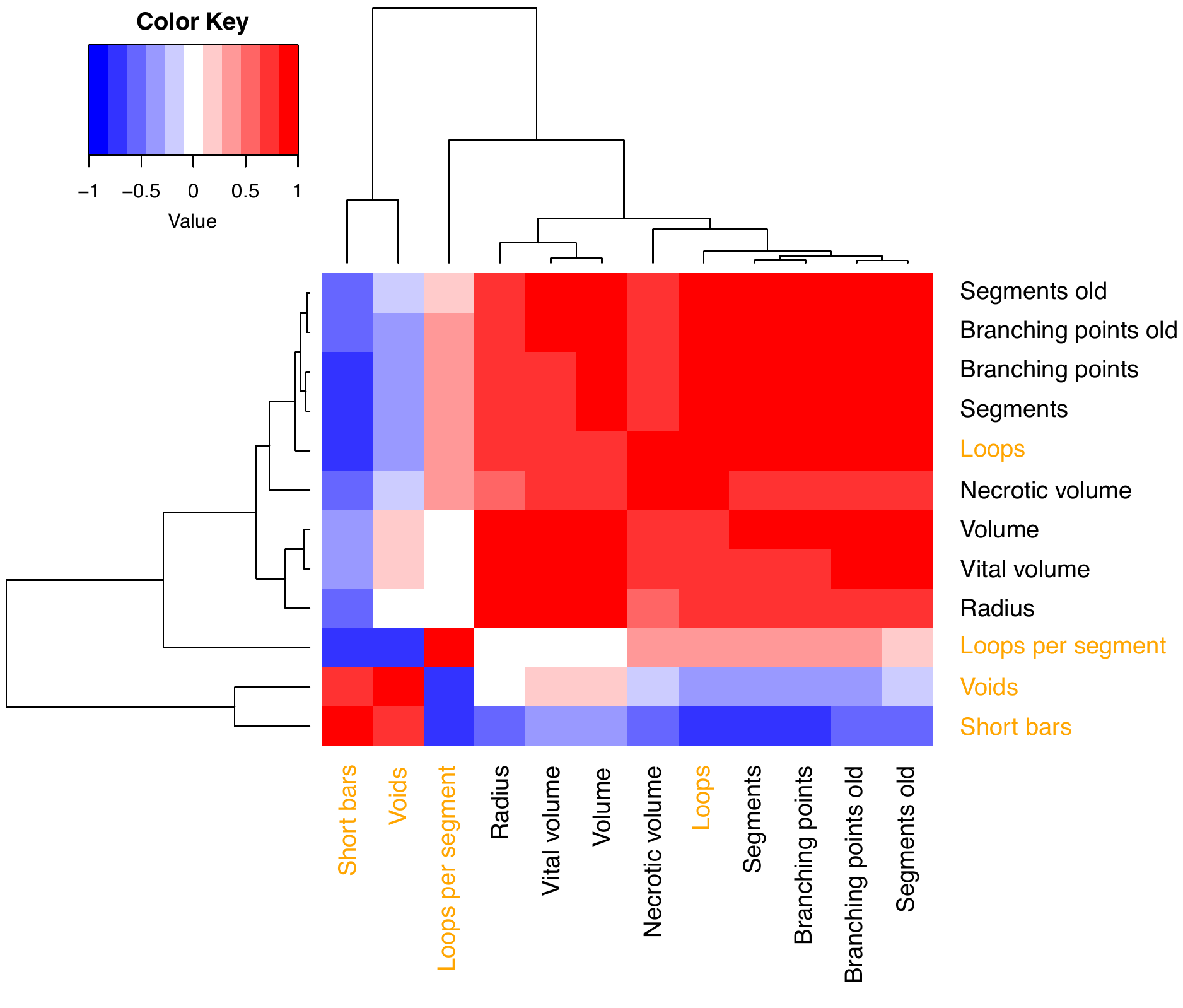}
    \caption{{\bf Heatmap displaying the pairwise Pearson correlation coefficients between different vascular characteristics derived from the ultramicroscopy data.} The dendrogramme represents complete linkage clustering using the Euclidean distance measure. We consider the following vascular characteristics: number of vessel segments as computed by \cite{Dobosz2014} (segments old), number of branching points as computed by \cite{Dobosz2014} (branching points old),  number of branching points as computed by \emph{unet}, number of vessel segments as computed by \emph{unet}, number of vessel loops, necrotic tumour volume as computed by \cite{Dobosz2014}, tumour volume as computed by \cite{Dobosz2014}, vital tumour volume as computed by \cite{Dobosz2014}, maximal radius used in the radial filtration, number of vessel loops per vessel segment, median persistence of bars in dimension 2 barcodes (voids), number of short bars per vessel segment in the dimension 0 barcodes.
We highlight the topological measures in orange including both the number of loops and number of loops per vessel segment to highlight the effect of the normalisation. } \label{Fig:RocheCor}
\end{figure}

We present the distribution of loops in the ultramicroscopy data relative to the tumour radii in Fig.~\ref{Fig:RocheDistributions}. We apply a Anderson-Darling test using the function {\tt ad.test()} from the library {\tt ksamples} in {\sc RStudio}~\cite{RStudio-Team:2016aa} to the different time points and treatment groups to determine whether the samples within one groups come from a common  (unspecified)  distribution.
We do not find this to be the case in any of the groups for any time point.

\clearpage

\noindent Finally, we present a correlation analysis between parameters that were extracted by Dobosz\emph{et al.}~\cite{Dobosz2014} and our topological parameters in Fig.~\ref{Fig:MC38Cor}. We also include the number of segments and branching points determined by our extraction of the vessel networks with {\sc unet-core}~\cite{unetRuss}. Both of these standard parameters correlate strongly with the same parameters extracted by Dobosz\emph{et al.}~\cite{Dobosz2014}.
We compute pairwise Pearson correlation using the library {\tt hmisc} and plot our results including a complete linkage clustering dendrogramme of the parameters using the library {\tt corrplot} in {\sc RStudio}~\cite{RStudio-Team:2016aa}.

\clearpage

\subsection*{Example images from the ultramicroscopy data}

\begin{figure}[ht!]
\subcaptionbox{Control tumour, day 3.}{\centering\includegraphics[height=5.9cm]{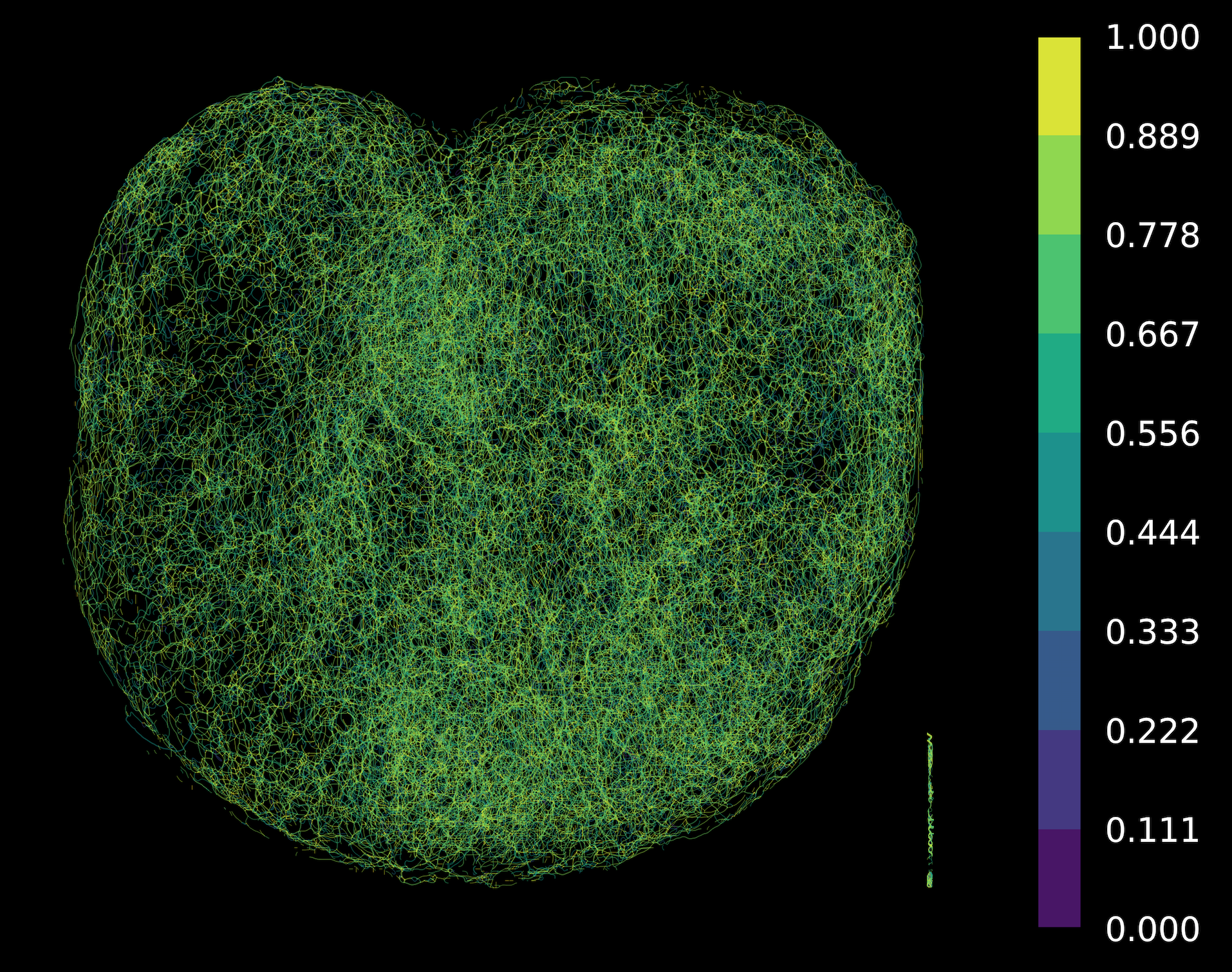}}%
\hspace{0.02\textwidth}
\subcaptionbox{Control tumour, day 7.}{\centering\includegraphics[height=5.9cm]{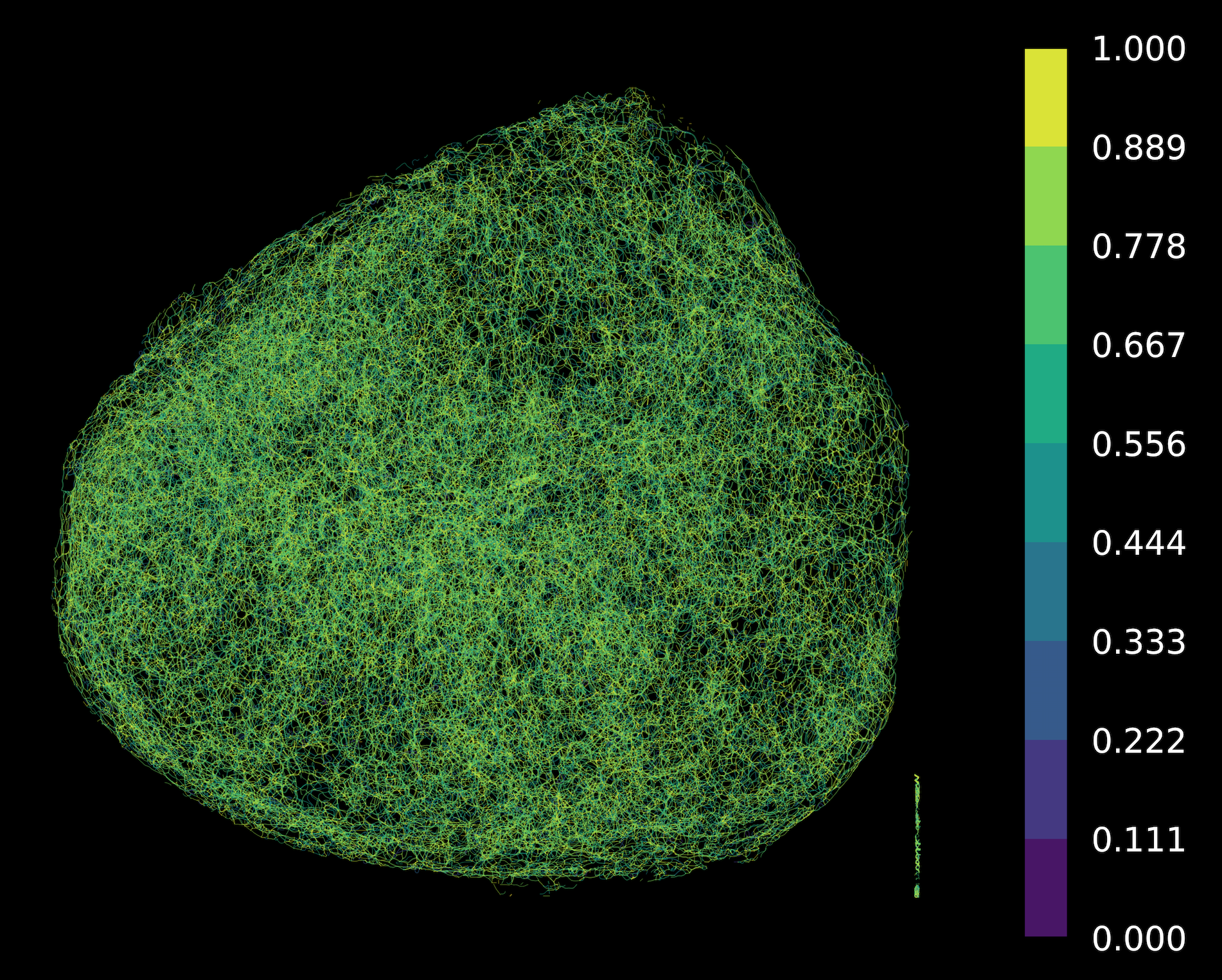} }%
\\
\subcaptionbox{Anti-VEGF-A treated tumour, day 3.}{\centering\includegraphics[height=5.9cm]{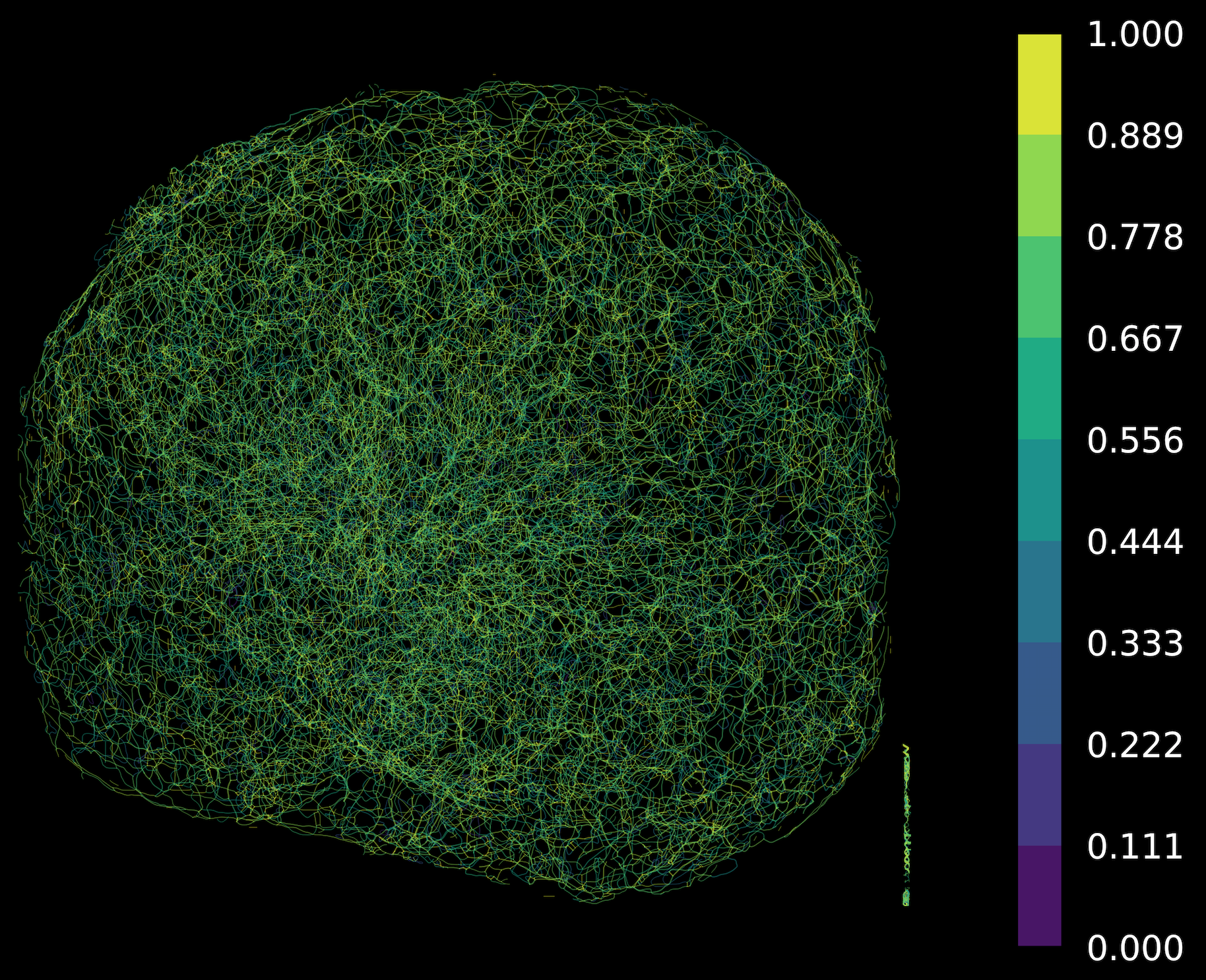}}%
\hspace{0.02\textwidth}
\subcaptionbox{Anti-VEGF-A treated tumour, day 7.}{\centering\includegraphics[height=5.9cm]{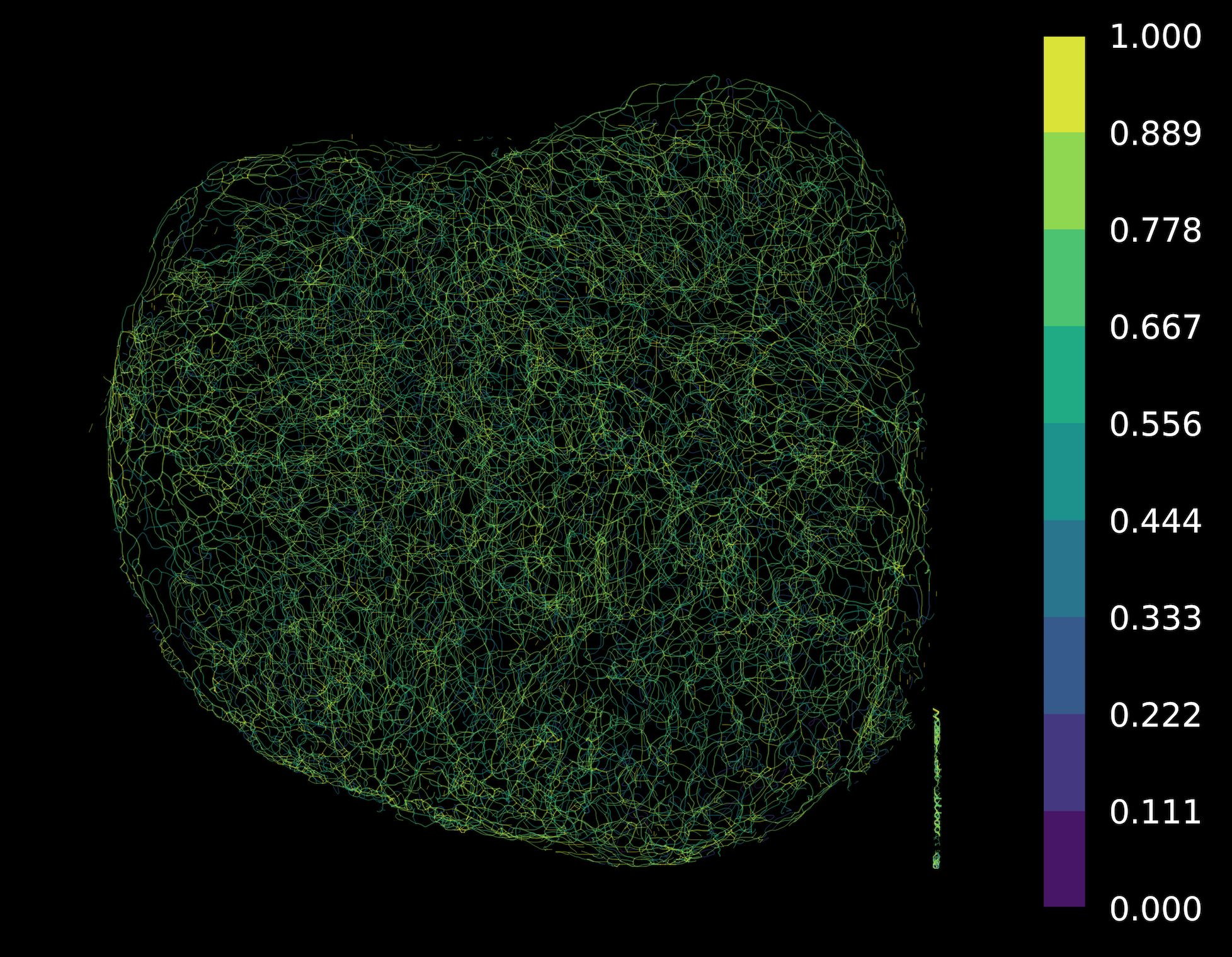} }%
\caption[Example images of extracted vessel networks from multispectral fluorescence ultramicroscopy data.]{{\bf Example images of extracted vessel networks from multispectral fluorescence ultramicroscopy data coloured according to chord-length-ratio (clr) values.} We can see a clear difference between the vessel networks of the treated versus the untreated tumour on both day 3 and day 7 after treatment. Note that the collection of lines in the bottom right corner of the images corresponds to text that was present in the skeleton images in the dataset. We removed these artefacts from our extracted point clouds manually.}\label{fig:ExampleImagesRoche}
\end{figure}

We show example images of the vessel networks extracted from the ultramicroscopy dataset using {\sc unet-core}~\cite{unetRuss} in Fig.~\ref{fig:ExampleImagesRoche}.

\end{document}